\documentclass[a4paper,onecolumn,superscriptaddress,12pt,nofootinbib,twoside,raggedfooter,notitlepage,floatfix]{revtex4-1}
\usepackage{graphicx}

\usepackage{amsbsy}
\usepackage{amsmath}
\usepackage{amssymb}
\usepackage[english]{babel}

\usepackage{graphicx}

\usepackage{wrapfig}
\usepackage{sidecap}

\usepackage{multirow}
\usepackage{color}
\usepackage[normalem]{ulem}

\usepackage{chemarr}

\newcommand{\boson}{\protect\raisebox{-1ex}{\protect\includegraphics[height=3ex,width=0.4ex]{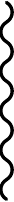}}}
\newcommand{\corfermion}{\protect\raisebox{-0.9ex}{\protect\includegraphics[height=3ex]{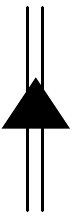}}}
\newcommand{\fermion}{\protect\raisebox{-0.9ex}{\protect\includegraphics[height=3ex,width=1ex]{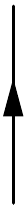}}}
\newcommand{\contfermion}{\protect\raisebox{-0.9ex}{\protect\includegraphics[height=3ex]{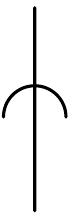}}}
\newcommand{\conttritium}{\protect\raisebox{-0.9ex}{\protect\includegraphics[height=3ex]{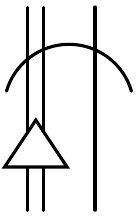}}}
\begin{document}


\title{Nuclear Field Theory predictions for ${}^{11}$Li and ${}^{12}$Be: shedding light on the origin of pairing in nuclei}
\author{G. Potel}
\affiliation{Departamento de Fisica Atomica, Molecular y Nuclear, Universidad de Sevilla, Facultad de Fisica, Avda. Reina Mercedes s/n, Spain.}
\affiliation{CEA-Saclay, DSM/IRFU/SPhN, F-91191 Gif-sur-Yvette, France (Present Address)}
\author{A. Idini}
\affiliation{Department of Physics, University of Milano, via Celoria 16, 20133 Milan, Italy}
\affiliation{INFN, Milan Section, via Celoria 16, 20133 Milan, Italy}
\author{F. Barranco}
\affiliation{Departamento de Fisica Aplicada III, Universidad de Sevilla, Escuela Superior de Ingenieros, Sevilla, 41092 Camino de los Descubrimientos s/n, Spain.}
\author{E. Vigezzi}
\affiliation{INFN, Milan Section, via Celoria 16, 20133 Milan, Italy}
\author{R.A. Broglia}
\affiliation{Department of Physics, University of Milano, via Celoria 16, 20133 Milan, Italy}
\affiliation{INFN, Milan Section, via Celoria 16, 20133 Milan, Italy}
\affiliation{The Niels Bohr Institute, University of Copenhagen, Blegdamsvej 17, DK-2100 Copenhagen, Denmark}


\begin{abstract}
\begin{center}
(\today)
\end{center}

Recent data resulting from studies of two--nucleon transfer reaction  on ${}^{11}$Li, analyzed through a unified nuclear--structure--direct--reaction theory 
have provided strong direct as well as indirect confirmation, through the population of the first excited state of $^9$Li and of the observation of a strongly quenched ground state 
transition, of the prediction that   phonon mediated pairing interaction is the main mechanism  binding the neutron halo of the 8.5 ms--lived ${}^{11}$Li nucleus. In other words, the ground state of $^{11}$Li can be viewed as a neutron Cooper pair bound to the $^9$Li core, mainly through the exchange of collective vibration of the core and of the pigmy resonance arizing from the sloshing back and forth of the neutron halo against the protons of the core, the mean field leading to unbound two particle states, a situation essentially not altered by the bare nucleon-nucleon  interaction acting between the halo neutrons.  Two-neutron pick-up data, together with  $(t,p)$ data on ${}^7$Li, suggest the existence of a pairing vibrational band based on ${}^9$Li, whose members can be excited with the help of inverse kinematic experiments as was done in the case of 
${}^{11}\textrm{Li} (p,t) {}^9\textrm{Li}$ reaction. The deviation from harmonicity can provide insight  into the workings of medium polarization effects on Cooper pair nuclear pairing, let alone specific information concering the ``rigidity'' of the N=6 shell closure. Further information concerning 
these questions is provided by the predicted absolute differential cross sections  $\sigma_{abs}$ associated with  the reactions 
 ${}^{12}\textrm{Be}(p,t) {}^{10}\textrm{Be}$(gs) and ${}^{12}\textrm{Be}(p,t) {}^{10}\textrm{Be}$($pv$)($\approx ^{10}${\textrm Be}(p,t)$^8{\textrm Be} $(gs)). 
In particular, concerning this last reaction,
predictions of $\sigma_{abs}$ can change by an order of magnitude depending on 
whether the halo properties  associated with the $d_{5/2}$ orbital  are treated selfconsistently in calculating the  ground state correlations of
the (pair removal) mode, or not.
\end{abstract}

\maketitle


\section{Introduction}

At the basis of BCS theory of superconductivity one finds the condensation of strongly overlapping Cooper pairs. This model can be extended to the atomic nucleus, provided one takes pair fluctuations into account, fluctuations which renormalize in an important way the different quantities entering the theory, in particular the pairing gap ($\Delta = (\Delta^2_\textrm{BCS} + \Delta^2_\textrm{Fluct})^{1/2}$, see e.g. \cite{Brink:05} Fig. 6.2 p. 152 and connected discussion). Not only pairing vibrations are more collective in the nuclear case than in condensed matter (see e.g. \cite{Anderson:58}) and in liquid $^{3}$He \cite{Wolfle:78}, where they 
 essentially correspond to 
a two quasiparticle--like state lying on top of twice the pairing gap. They also develop, in the nuclear case,  as well defined vibrational bands around closed shell nuclei, where one can 
clearly distinguish between particle and hole degrees of freedom, in keeping with the relatively  large gap existing between occupied and unoccupied single--particle states. Systematic evidence exists of the correlation and stability of the pair addition and pair subtraction modes (single Cooper pairs) in medium heavy nuclei (see e.g. \cite{Brink:05} and \cite{Broglia:73} and references therein). Making use of these building blocks a number of pairing vibrational bands have been identified, providing important insight into the mechanism in which nuclear superfluidity emerges from the condensation of many phonon pairing modes. 
In medium heavy nuclei,  a large fraction of the pairing correlation energy ($\gtrsim$ 50--70\%) is due to the strong ${}^1\textrm{S}_0$ $N$--$N$ bare interaction, medium polarization effects being important although not dominant (see \cite{Barranco:99} as well as \cite{Brink:05}, Ch. 10 and refs. therein). This is, among other things, one of the reasons why the multi--phonon pairing spectra seen to date are rather harmonic (see e.g. \cite{Flynn:72}). In other words, the interaction between pairing modes in heavy stable nuclei is weak and anharmonicities effects are, 
although very revealing of the nuclear structure around closed shell nuclei, not very important (note however Pauli principle effects \cite{Bortignon:78}).

We expect the situation to be rather different in the case of weakly bound, strongly polarizable light halo nuclei like ${}^{11}$Li and, to some extent also ${}^{12}$Be. In 
both cases, the N--N bare interaction 
provides a small contribution ($\approx 100$ keV) to the two-neutron binding energy. Thus a major fraction of the glue holding the neutron Cooper pair (halo) to the ${}^9$Li and to the ${}^{10}$Be cores arizes from the exchange of collective vibrations \cite{Barranco:01,Gori:04}. To further clarify the origin of pairing in nuclei, 
it would then be important to look for the pairing vibrational band associated with the $N=6$ closed shell systems ${}^9$Li and ${}^{10}$Be. Larger or smaller anharmonicities as compared with other closed shell systems (like e.g. $^{208}$Pb) would provide a welcome input to shed light on the relative contributions to the nuclear pairing force, both in terms of the strength as well as of the radial shape (form-factors), of the bare $N$--$N$ force and of the effective pairing force arising from the exchange of collective vibrations between pairs of nucleons moving in time reversal states close to the Fermi energy. In fact, if all of the ${}^1\textrm{S}_0$ correlations came from the short range, bare interaction, 
many processes leading to anharmonic effects    would  not be present in the pairing vibrational nuclear spectrum. In particular, a special group of those related to Pauli principle corrections reflecting the overcompleteness of the NFT basis of elementary excitations, i.e. single-particle, particle-hole-like and  pair vibrational modes.

\section{Nuclear Field Theory}

Nuclear Field Theory (NFT) provides a systematic method for describing both nuclear structure \cite{Bes:76a,Bes:76b,Bes:76c,Bes:75,Mottelson:76,Bortignon:77,Broglia:76} and nuclear reactions \cite{Broglia:05c}. In particular, 
it allows to deal with the problem of  overcompleteness of the degrees of freedom arising from the identity of particles appearing explicitly and the particles participating in the collective motion. 
The non--orthogonality of the variety of degrees of freedom must be expected
 quite generally in descriptions that: a) from the nuclear structure point of view 
exploit simultaneously  the single-particle degrees of freedom, as well as the quanta of particle--hole type together with  those involving pairing vibrations \cite{Bes:66} made out of two correlated particles (pair addition modes) or of two correlated holes (pair subtraction modes); b) from the reaction point of view, make use of mean fields of target and of projectile which are assumed to be independent of each other, with the added possibility, in the case of the study of pairing correlations through two--particle transfer reactions, that a member of a Cooper pair \cite{Cooper:56}, \cite{Leggett:06} is in one system (e.g. target) and the second partner in the other one (i.e. projectile). This is in keeping with the fact that 
the typical  value of the pairing gap in superfluid nuclei  (or of correlation energies in the case of normal systems) is $\Delta \approx $ 1 MeV, implying  
that Cooper pair partners are correlated over distances of the order of the associated coherence length, $\xi \approx $ 20-30 fm.
This result, together with the fact that $\Delta/U(r \approx R_0)\approx 4\times10^{-2}$, $U(r)$ being the average mean field potential, 
implies, as a rule, that successive transfer, where $U$ acts twice,  is dominant as compared to the simultaneous transfer of  two nucleons, a process in 
which $U$ acts once, the second nucleon being transferred  partially making use of the interaction energy between the partners of the Cooper pair, and partially 
exploiting the non--orthogonality of the single--particle wavefunctions associated with the mean field of target and projectile. It is of notice that this argumentation is at the basis of one of the cornerstones of superconductivity in metals, namely the Josephson effect 
\cite{Josephson:62,Cohen:62} (see also \cite{Potel:09a} and references therein).

\begin{figure}
\centerline{\includegraphics*[width=.75\textwidth,angle=0]{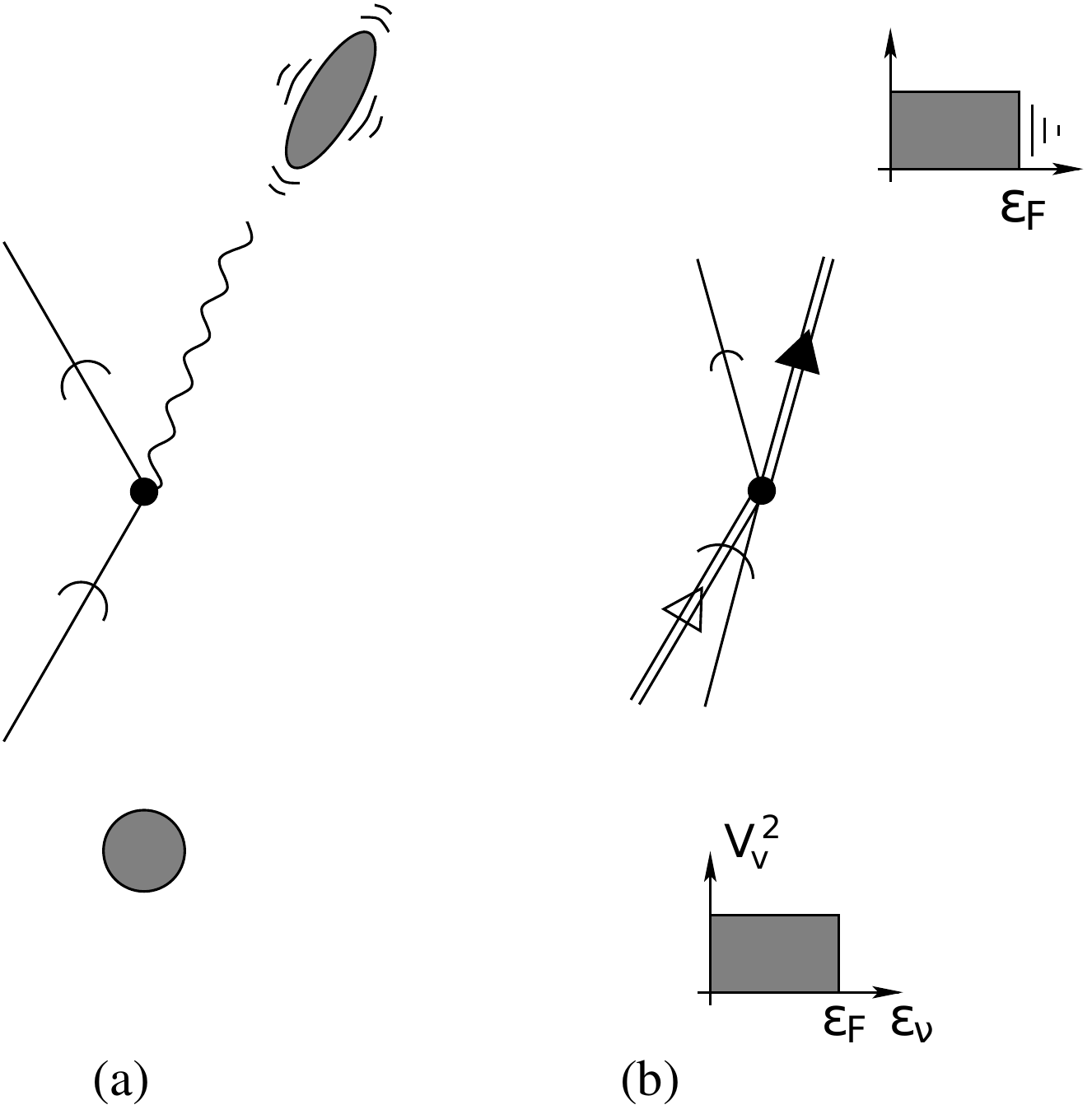}}
\caption{\begin{footnotesize}Diagramatic representation of: a) inelastic scattering process (transfer quantum number $\beta=0$) of a collective particle--hole like (quadrupole surface vibration) induced by the nonhomogeneous time dependent field created by a nucleon (projectile, curved arrowed line) passing by the target nucleus, b) two--particle transfer process in which a dineutron is stripped from the incoming triton exciting the pair addition mode (transfer quantum number $\beta=+2$) of e.g. a closed shell system (double lines labeled with a conventionally empty arrow corresponds to two neutrons bound in the projectile (triton); when labeled by a solid arrow, it corresponds to a pair addition mode, i.e. to two nucleons bound to the core nucleus). The solid dot indicates the particle vibration coupling vertex measured by the strength $\Lambda_\alpha(\beta)$, where the subindex $\alpha$ indicate the angular momentum and eventually, if the mode is fragmented, whether the state is the lowest, next to lowest, etc. in energy.\end{footnotesize}}\label{fig1}
\end{figure}

Tailored after Feynman's version of Quantum Electrodynamics (QED) \cite{Feynman:61}, that is in terms of Feynman diagrams, NFT is essentially only restricted by the limitations
in the (experimental)  knowledge one has of the particle--vibration coupling vertices $\Lambda_\alpha(\beta)$, where $\alpha$ indicates the quantum numbers of the phonon but associated with gauge space, as well as whether the mode in consideration is the first, second, etc. in energy, while $\beta = 0,\pm 2$, is the transfer quantum number ($\beta=0$ implies particle-hole excitations, while $\beta=+2$ pair-addition and $\beta=-2$ pair-removal modes respectively, see e.g. \cite{Broglia:73}, \cite{Bortignon:77} and \cite{Bohr:75} and refs. therein).
 At variance with QED for which an essentially definitive value of the fine structure constant $e^2/\hbar c$ exists, uncertainties are to be ascribed to the variety of $\Lambda_\alpha(\beta)$ values. This is due to the combined effect of the fragmentation of the modes, as well as the fact that to extract $\Lambda_\alpha(\beta)$ from a reaction measurement one needs to use optical potentials, with all the associated limitations of such an approach (see Fig. \ref{fig1}).


In any case, progress which has taken place during recent years has ushered studies of nuclear pairing with two-nucleon transfer reactions into the quantitative era, with uncertainties well within experimental errors (see e.g. \cite{Potel:10,Potel:11a,Potel:11b,Potel:12a} and refs. therein).

\section{The Cooper pair problem and ${}^{11}$\textrm{Li}}

The basis for the understanding of superconductivity in metals worked out by Bardeen, Cooper and Schrieffer (BCS theory \cite{Bardeen:57a}) was provided by Cooper \cite{Cooper:56} who solved the problem of a pair of electrons interacting above a noninteracting Fermi sea of electrons via a two--body potential. ${}^{11}$Li can be viewed as a rather accurate realization of Cooper's model, within the framework of nuclear physics: a pair of  neutrons weakly bound to the  ${}^9$Li core.

\begin{figure}
\centerline{\includegraphics*[width=.9\textwidth,angle=0]{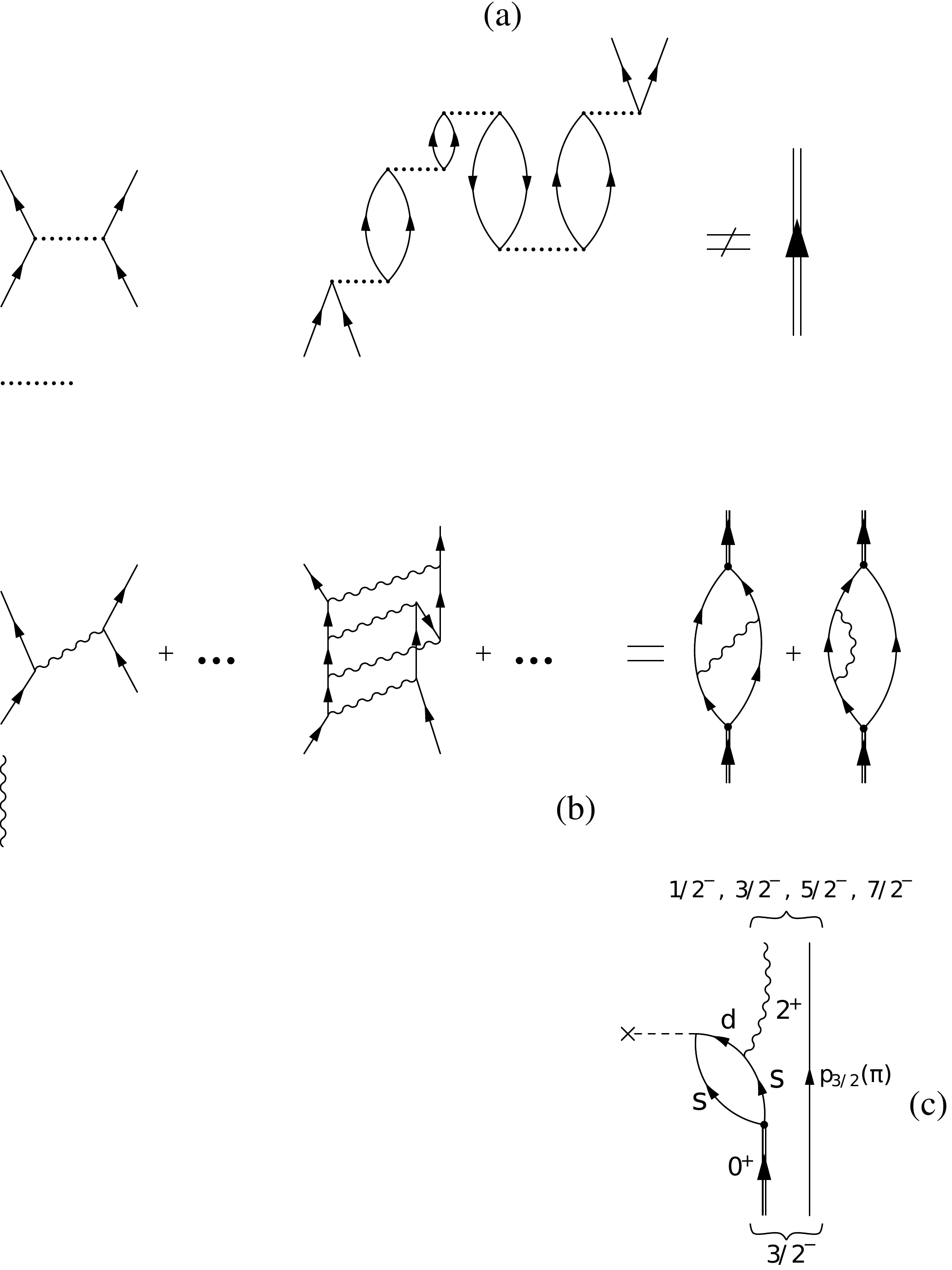}}
\caption{\begin{footnotesize}In the case of ${}^{11}$Li the bare interaction (e.g. Argonne $v_{14}$  $N-N$ potential; horizontal dotted line) acting to infinite order 
between the halo neutrons (see (a)) does not lead to a bound Cooper pair (double solid arrowed line), but merely shifts downwards (lowers the energy) of the resonant configurations $s_{1/2}^2(0)$ and  $p_{1/2}^2(0)$ by about 100 keV individually, without leading to any appreciable mixing \cite{Barranco:01}. The exchange of a dipole pigmy resonance of the halo field and of a quadrupole vibration of the $^{9}$Li core (wavy lines, see (b)) 
provides essentially all of the glue for the Cooper pair to become bound ($S_{2n}=330$ keV, theory \cite{Barranco:01} as compared to the experimental data of 378$\pm$5 keV \cite{Bachelet:08}; 369.15$\pm$0.65 keV \cite{Smith:08}). In (c) a schematic representation of the ${}^{11}\textrm{Li}(p,t) {}^9\textrm{Li}(1/2^{-})$ process (see \cite{Tanihata:08,Potel:10}) is displayed (see also Figs. \ref{fig3} and \ref{fig4}).\end{footnotesize}}\label{fig2}
\end{figure}

Now, it was found in a NFT study of this halo nucleus \cite{Barranco:01}, that the bare nucleon--nucleon interaction does not bind the Cooper pair (see Fig. \ref{fig2} (a)). It is the exchange of the 
vibrations of the halo field and of the core (dipole and quadrupole) 
between the halo neutrons which, similar to phonons in metals, 
binds the Cooper pair (see Fig. \ref{fig2} (b)). The NFT results of ref. \cite{Barranco:01} provide a rather accurate description of the nuclear structure properties of $^{11}$Li.
The calculations were carried out following all what is well known about NFT and the particle--vibration coupling mechanism (see, e.g. \cite{Bes:76a,Bes:76b,Bes:76c,Bes:75,Mottelson:76,Bortignon:77,Broglia:76}, \cite{Bohr:75} and refs. therein). 
A single mean field potential (Saxon--Woods) was used for all single--particle ($s-,p-,d-,$\ldots) states (cf. \cite{Bohr:69} pp. 236-240).
Parity inversion of the $s_{1/2}$ and $p_{1/2}$ states of ${}^{10}$Li results from the coupling of these resonances to   vibrations of the ${}^9$Li core \footnote{Of notice that this is tantamaunt to saying of the $^{8}$He core, in keeping with the fact that the $p_{3/2}$ proton state is considered throughout as a spectator.},
when the coupling to the continuum is taken properly into account. Making use of the resulting single-particle  (dressed) states,
of the bare N-N interaction  and of the interaction induced by the exchange of collective modes between the dressed single-particle halo states, 
including  the $1^{-}$ pigmy resonance of $^{11}$Li
(cf. Fig. \ref{fig2} (b)), 
a quantitative account of the experimental findings was obtained, in particular concerning the extremely low two--neutron separation energy  $S_{2n} \approx 0.38$ MeV 
\cite{Bachelet:08,Smith:08} (see also Caption to Fig. \ref{fig2}). Notice that if the $N$--$N$ interaction is not 
included, the absolute value of $S_{2n}$ is decreased by only about 100 keV.

\begin{figure}
\centerline{\includegraphics*[width=.90\textwidth,angle=0]{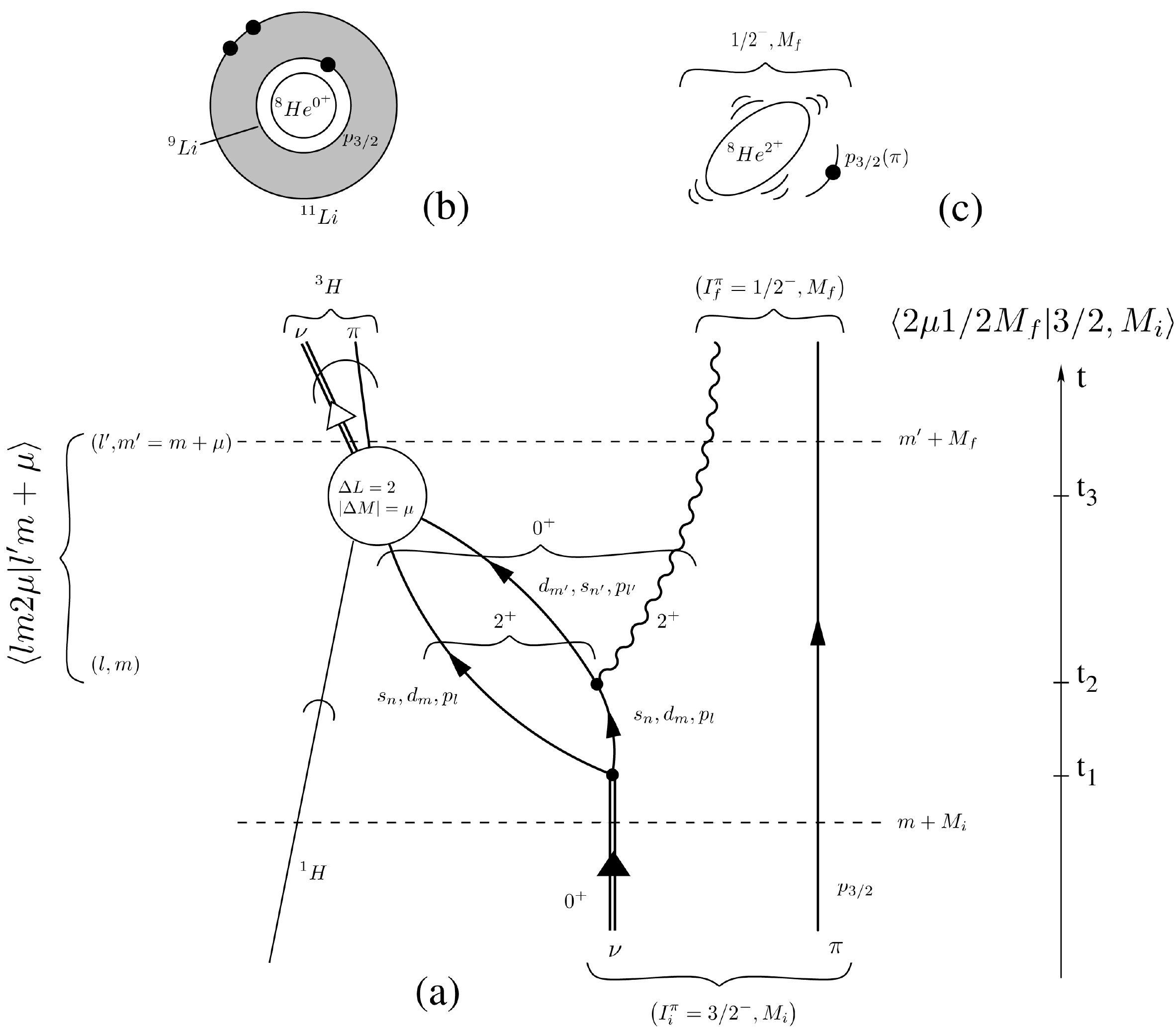}}
\caption{\begin{footnotesize}(a) NFT--Feynman diagram associated with the process \mbox{${}^1$H (${}^{11}$Li(gs),${}^9$Li($1/2^-$;2.69 MeV)${}^3$H)}, which treats on equal footing the nuclear structure \end{footnotesize} \normalsize ($\corfermion^{2^+},\boson^{2+},\fermion$)\begin{footnotesize} and the reaction mechanism \end{footnotesize}\normalsize($\contfermion, \conttritium$).\begin{footnotesize} The neutron correlated Cooper pair (pair addition mode), has a somewhat different structure when bound to ${}^9$Li to make ${}^{11}$Li, than to the proton to make the triton (${}^3$H) (depicted with a solid and with an open arrow respectively). This is the reason why there is a finite overlap, called $\Omega_n(n=0,1,\ldots)$ between the corresponding relative motion wavefunctions displaying zero,one, etc. nodes (cf. e.g. \cite{Broglia:73} and references therein). In keeping with this fact one can posit that although ($p,t$) reactions are quite specific to probe pairing correlations in nuclei, in particular in the case of the ${}^{11}$Li neutron Cooper pair, they display limitations. In particular the specific probe in the case under discussion is the ${}^{11}$Li(${}^9$Li,${}^{11}$Li)${}^9$Li reaction ($\Omega_n=\delta(n,0)$), although obviously much more demanding experimentally. Curly brackets indicate angular momentum coupling, while horizontal dashed lines indicate magnetic quantum number conservation. In (b) and (c) a schematic representation of the initial (${}^{11}$Li) and final (${}^9$Li($1/2^-,M_f$; 2.69 MeV)) nuclear states are given, respectively. The ordinate to the right indicates time. Following NFT of both structure 
\cite{Bes:76a,Bes:76b,Bes:76c,Bes:75,Mottelson:76,Bortignon:77,Broglia:76}
and reaction \cite{Broglia:05c} processes, all possible time orderings are to be considered in the calculations.It is of notice that curved arrowed lines indicate continuum scattering states, while standard arrowed lines correspond to bound states. \end{footnotesize}}\label{fig3}
\end{figure}

\begin{figure}
\centerline{\includegraphics*[width=.98\textwidth,angle=0]{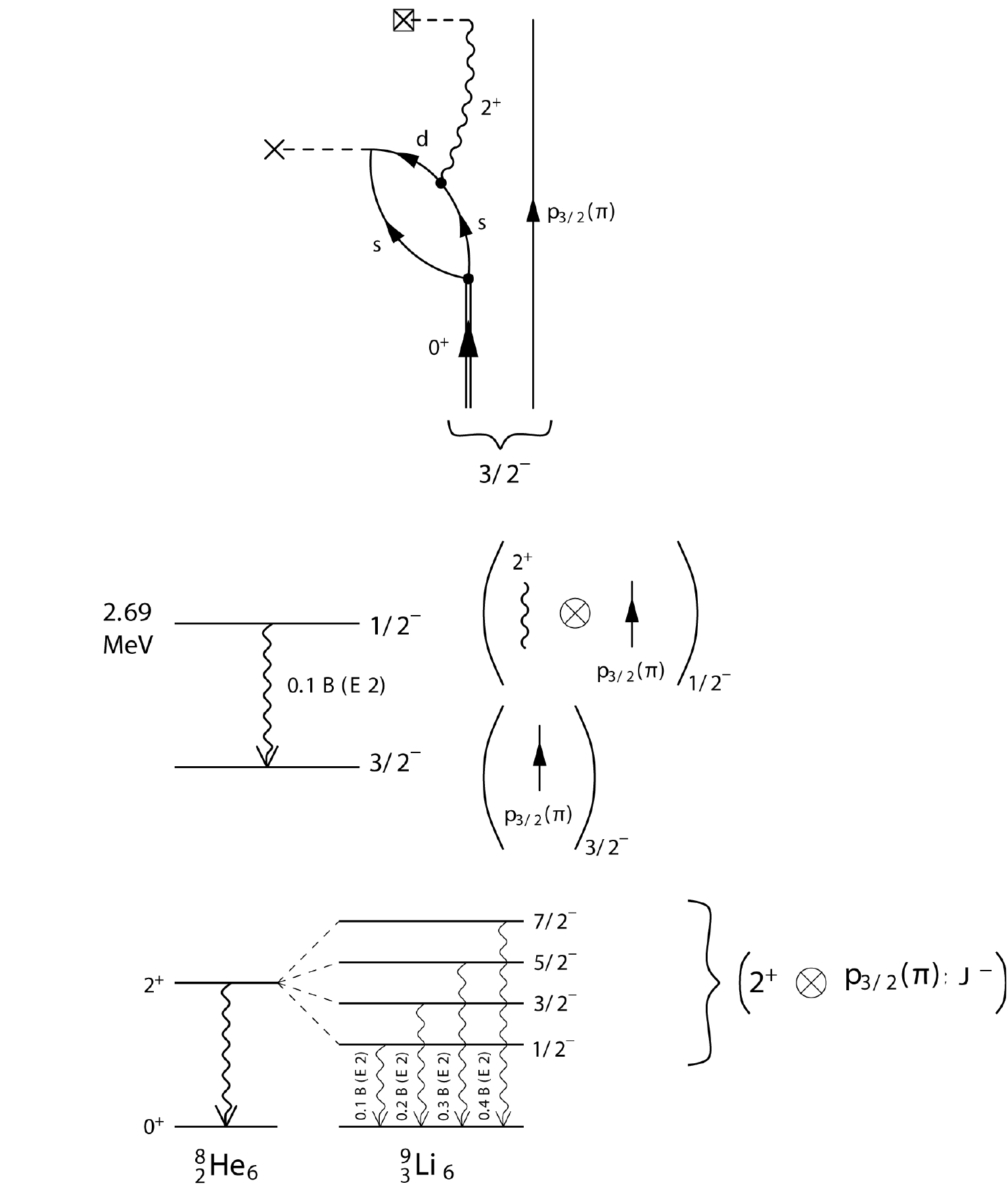}}
\caption{\textit{Gedanken} (two--particle transfer)--($\gamma$ decay) coincidence experiments aimed at better individuating the couplings involved in the neutron halo Cooper pair correlations in $^{11}$Li and of the $1/2^-$ member of $^9$Li excited in the $^1$H($^{11}$Li,$^9$Li)$^{3}$H reaction \cite{Tanihata:08,Potel:10}.}\label{fig4}
\end{figure}

The NFT description of the structure of $^{11}$Li 
also contained precise predictions (see e.g. \cite{Brink:05} Ch. 11, in particular Fig. 11.6) concerning how to force virtual correlation processes, to become real. Namely, through two--neutron pick--up reactions, like e.g. ${}^{11}\textrm{Li}(p,t) {}^9\textrm{Li}$. In this way it was found that one can excite, among others, the multiplet of states $(2^+\otimes p_{3/2}(\pi))_{1/2^-,3/2^-,5/2^-,7/2^-}$ (see Fig. \ref{fig2} (c), see also Figs. \ref{fig3} and \ref{fig4}). These predictions had to wait short of ten years to be tested 
(see ref. \cite{Tanihata:08}), and proved not to be wrong.
\footnote{First we guess it. Then we compute the consequences of the guess to see what would be implied if the law we guess is right. Then we compare the result of the computation to nature, with experiment or experience, compare it directly with observation, to see if it works. If it disagrees with experiment it is wrong. In that simple statement is the key to science. It does not make any difference how beautiful your guess is. It does not make any difference how smart you are, who made the guess, or what your name is. If it disagrees with experiment it is wrong. That is all there is to it. Richard Feynman} In fact, making use of a detailed nuclear reaction theory of two-particle transfer which considers successive, simultaneous and non--orthogonality processes, together with inelastic and break up channels and thus accurately treats the coupling to the continuum concerning both the structure and reaction aspects of the calculations, one has been able \cite{Potel:10} to check the validity of such predictions, and reproduce 
in a quantitative fashion, the experimental findings of ref. \cite{Tanihata:08}. In particular the observation of the $1/2^-$ member of the ($2^+ \otimes p_{3/2}(\pi)$) multiplet of ${}^9$Li which is the first excited state of this nucleus (see Fig. \ref{fig4}). This result provided the first, direct confirmation of the central role played by vibrational states in nuclear pairing.


\section{Pairing vibrational band of ${}^{9}${\textrm Li}}

\begin{figure}
\centerline{\includegraphics*[width=.99\textwidth,angle=0]{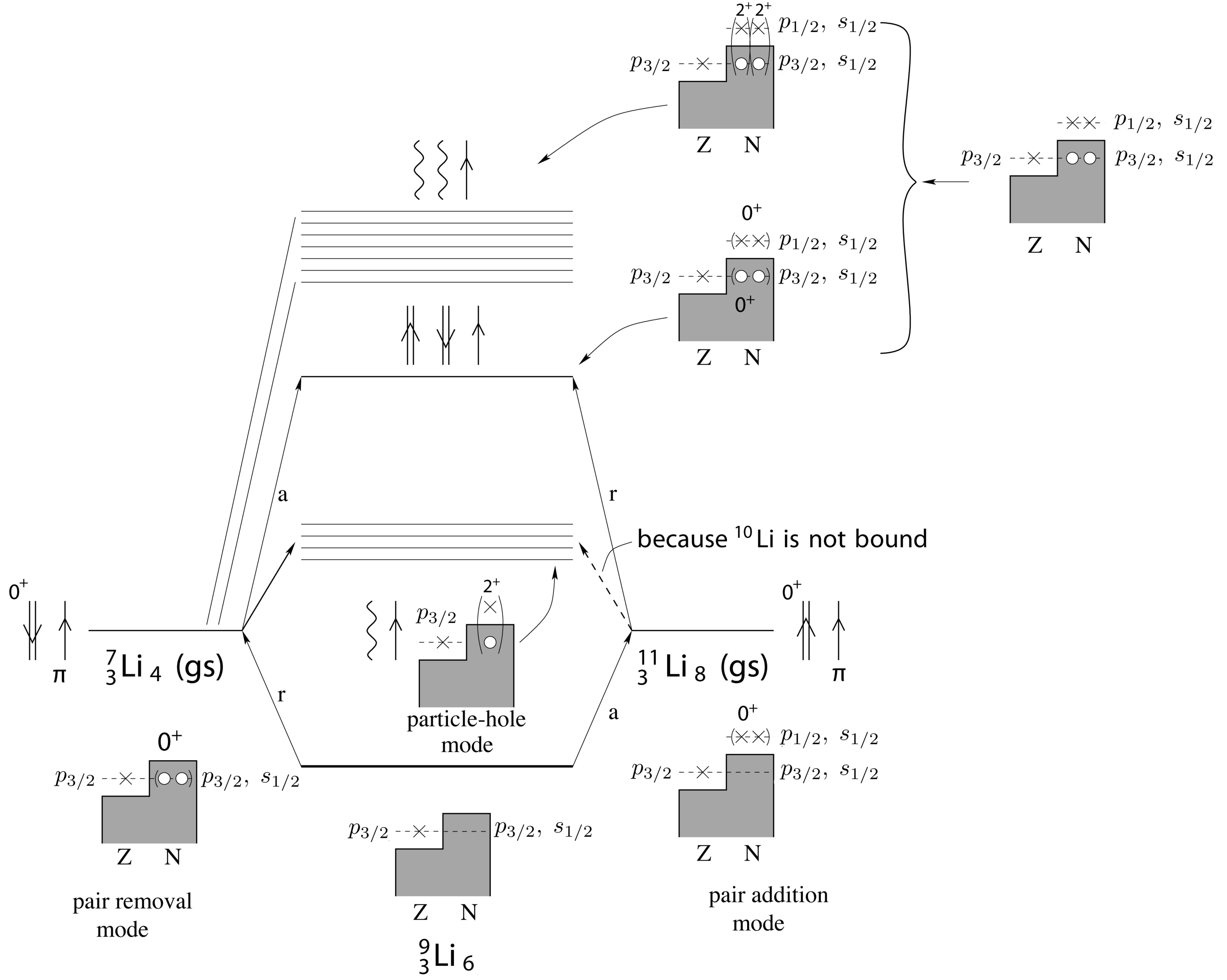}}
\caption{\begin{footnotesize}Schematic representation of the phonon pairing ($\beta=\pm 2$) and density vibrational ($\beta=0$) spectrum based on  ${}^9_3$Li$_6$ (harmonic approximation), constructed making an analogy with studies of two--particle transfer reactions carried out around ${}^{208}$Pb (cf. e.g. \cite{Brink:05} p. 109 Fig. 5.5 and refs. therein; see also \cite{Barranco:01}). The Fermi representation of occupied proton ($Z$) and neutron ($N$) states is schematically shown in terms of grey areas, where holes are shown as empty circles, and particles are indicated in terms of crosses. A single--arrowed line pointing upwards (downwards) represent a particle (hole) neutron or proton state. A double arrow indicates the correlated neutron pair addition (subtraction) modes. The predicted two--particle transfer cross sections $a,r$  and excitation energy $E_{pv}$ associated with the two--phonon (pair--addition), (pair--subtraction) and  ($2p$--$2h$ like excitation) pairing vibrational (pv) state expected in ${}^9$Li are, in the harmonic approximation: $E_{pv}=5.7$ MeV, 
$a =\textrm{d}\sigma \left( {}^{11}\textrm{Li}(p,t) \; {}^9\textrm{Li}(gs); \; \theta=60^\circ \right)/\textrm{d}\Omega = 0.7$ mb/sr \cite{Tanihata:08}, and 
$r =\textrm{d}\sigma \left( {}^{7}\textrm{Li}(t,p) \; {}^9\textrm{Li}(gs); \; \theta=60^\circ \right)/\textrm{d}\Omega = 3.0$ mb/sr \cite{Young:71}, (see also Fig. \ref{LiPairVibr}). \end{footnotesize}}\label{fig5}
\end{figure}

The unified NFT description of structure and reactions used in refs. \cite{Barranco:01,Potel:10}, provides a number of suggestions concerning the consequences 
the findings reported in \cite{Tanihata:08} has. 
In particular, the existence of a many--phonon pair vibrational spectrum (see Fig. \ref{fig5}), similar to that observed in $(p,t)$ and $(t,p)$ reactions in the Pb region\footnote{The main expected difference concerns the relative importance of the bare and induced interaction which binds the Cooper neutron pair to the core, namely to ${}^{208}$Pb in one case and ${}^9$Li in the other. An important consequence of this difference is to be related with the expected anharmonicities of the halo pairing vibrational spectrum.} (see e.g. \cite{Brink:05,Broglia:73,Flynn:72}).

The energy of the two--phonon state ($n_a=1,\; n_r=1$) mode of ${}^9$Li, where $n_i$ indicates the number of pair addition ($i=a$; ground state of ${}^{11}$Li) and pair removal ($i = r$; ground state of ${}^7$Li) modes is expected to be, in the harmonic approximation \cite{Bes:66} (see also App. A),
\begin{eqnarray}
	W(n_r=1;\;n_a=1) = \left( B(9) - B(7) \right) - \left( B(11) - B(9) \right)  \\ \nonumber
	  = (45.34-39.25)-(45.71-45.34) \textrm{ MeV} = ( 6.1 - 0.4 ) \textrm{ MeV} = 5.7 \textrm{ MeV} \;\\ \nonumber
         = W_1 (\beta= 2) + W_1 (\beta= -2),
\label{eq_1}
\end{eqnarray}
where $B(A)$ is the binding energy of the Li--isotope with mass number $A$ (see Fig.\ref{LiPairVibr}, see also App. \ref{Appendix.A} and Fig. \ref{fig21} below).

The excitation energy of a state of the vibrational band (again in the harmonic approximation), can be written as
\begin{equation}
	W(n_r,n_a) = n_r W_1 (\beta = -2) + n_a W_1 (\beta= 2) \;.
\end{equation}
The absolute value of the different transitions between these states can be expressed in terms of the basic cross--sections
\begin{equation}
	a = \sigma \left( {}^9\textrm{Li}(gs) \leftrightarrow {}^{11}\textrm{Li}(gs) \right) \;\;\textrm{and}\;\; r = \sigma \left( {}^9\textrm{Li}(gs) \leftrightarrow {}^7\textrm{Li}(gs) \right) \;.
\end{equation}
An embodiment of the above definitions is provided by the values \cite{Tanihata:08,Young:71} (see also \cite{AjzenbergSelove:78}),
\begin{equation}
	a = \textrm{d}\sigma \left( {}^{11}\textrm{Li}(p,t) {}^9\textrm{Li}(gs); \; \theta = 60^\circ \right)/ \textrm{d}\Omega = 0.7 \textrm{ mb/sr},
	\label{eq:4}
\end{equation}
and
\begin{equation}
	r = \textrm{d}\sigma \left( {}^{7}\textrm{Li}(t,p) {}^9\textrm{Li}(gs); \; \theta = 60^\circ \right)/ \textrm{d}\Omega = 3 \textrm{ mb/sr},
	\label{eq:5}
\end{equation}
and the principle of detailed balance. 
It is of notice that while $W_1(\beta=2)$ and $W_1(\beta=-2)$ are intrinsic properties of the pairing vibrational modes
(it is reminded that the Fermi energy within the framework of the pairing vibrational model is determined  by the value of the energy associated with the minimum
of the RPA dispersion relation, see  Apps. A and C),
(\ref{eq:4}) and (\ref{eq:5}) depend also on the bombarding conditions (bombarding energy, scattering angle, etc), as well as on the reaction ($(p,t)$, (${}^{18}$O,${}^{16}$O), etc). Within this context it is of notice that the values reported in (\ref{eq:4}) and (\ref{eq:5}) corresponds to tritons of $\sim 3$ MeV and of 5 MeV per nucleon respectively (see also App. D).

With such caveat in mind one expects that (see also \cite{Flynn:72}), at equal bombarding conditions as those corresponding to (\ref{eq:4}) and (\ref{eq:5}) (see \cite{Tanihata:08,Young:71})
\begin{equation}
	\textrm{d}\sigma \left( {}^{11}\textrm{Li}(t,p) {}^{13}\textrm{Li}(gs); \; \theta = 60^\circ \right)/ \textrm{d}\Omega \approx 1.4 \textrm{ mb/sr}
	\label{eq:6},
\end{equation}
and
\begin{equation}
	\textrm{d}\sigma \left( {}^{7}\textrm{Li}(p,t) {}^5\textrm{Li}(gs); \; \theta = 60^\circ \right)/ \textrm{d}\Omega \approx 6 \textrm{ mb/sr},
	\label{eq:7}
\end{equation}
in a similar way in which the transition 
$\left( \boson^{2^+} \hspace{-0.2cm}\otimes \boson^{2^+} \right) \rightarrow \left( \boson^{2^+} \right)$ \normalsize is expected to display a $B(E2)$--value twice as large as that associated with the transition of the one phonon state to the ground state i.e.  \Large $\left( \boson^{2^+} \right) \rightarrow$ \normalsize $(gs)$.

In Fig. \ref{fig5} the one-- and two--phonon quadrupole spectrum of ${}^9$Li expected 
in the harmonic approximation is also shown.
In Fig. \ref{LiPairVibr} we display the pairing vibrational spectrum of $^{9}$Li and the associated absolute two--particle transfer differential cross sections of the pair addition and pair 
removal modes, in comparison with the experimental data \cite{Tanihata:08},\cite{Young:71}. 

In the case of the reaction $^{11}$Li$(p,t)^9$Li(gs) the absolute differential cross section  was calculated making use of the two-nucleon spectroscopic amplitudes obtained from the $|0>$ term of the neutron component of the $^{11}$Li ground state 
($\equiv |\tilde{0}\rangle_\nu \otimes |p_{3/2}(\pi)\rangle$, cf. \cite{Potel:11b}),
\begin{equation}
|\tilde{0}>_\nu =  |0> + \alpha |(p_{1/2},s_{1/2})_{1^-} \otimes 1^-; 0> + \beta |(s_{1/2},d_{5/2})_{2^+} \otimes 2^+ ; 0 >,
\end{equation}
with $\alpha =0.7$ and $\beta \approx 0.1$,
and 
\begin{equation}
|0> =   0.45 |s_{1/2}^2(0)> +  0.55 |p_{1/2}^2(0)> + 0.04 |d_{5/2}^2(0)>,
\label{component}
\end{equation}
describing the motion of the two halo neutrons around $^9$Li. Successive, simultaneous and non-orthogonality contributions to the two-particle transfer process
were taken into account (see \cite{Potel:10,Potel:12a,Tanihata:08}  and refs. therein).  The optical parameters used to describe  the $^{11}$Li+p, $^{10}$Li+d and the $^9$Li+t channels were taken from refs. \cite{Tanihata:08},
\cite{An:06} 
.

In the case of $^7$Li(t,p)$^9$Li(gs) reaction, the optical parameters  were taken from refs. \cite{Young:71},\cite{An:06} while the two-neutron spectroscopic amplitudes, reported in Table I, were calculated 
within the RPA (cf. e.g. \cite{Brink:05} and \cite{Broglia:73}, see App. A), making use of a Saxon-Woods potential
with standard parametrization  (see  \cite{Bohr:69} pp. 236-240) to describe the single-particle energies, and 
of the experimental correlation energy of the two-neutron holes in $^{7}$Li$(gs)$ (as an example (in this case regarding Be-isotopes), cf. Caption to Fig. \ref{fig22}). 
As seen from Fig. \ref{LiPairVibr} and Table II, theory provides an overall account of the experimental findings. While it is the component (\ref{component})  which is directly involved in the 
$^{11}$Li$(p,t)^9$Li(gs)  transfer process, the amplitudes $\alpha$ and $\beta$ play a central role  concerning the absolute value  of the cross section. In fact, 
incorrectly normalizing the state (\ref{component}) to 1, that is $ 0.63 |s_{1/2}^2 (0)> + 0.77 |p^2_{1/2}(0)> +0.06 |d^2_{5/2} (0)> $, leads to an integrated cross section 
in the angular range $20^{\circ}$ - $150^{\circ}$ of 12.1 mb. This result is a factor of 2 larger than that predicted making use of (\ref{component}), and lies outside the errors of the experimental value $\sigma_{exp} = 5.7 \pm 0.9$ 
(see Table II as well as \cite{Potel:10} and \cite{Tanihata:08}); see also Fig. \ref{fig7a}(a), result labeled TDA).

Within this context, one can also play the same game concerning the two-nucleon spectroscopic amplitudes associated with the reaction  $^7$Li(t,p)$^9$Li(gs), by setting 
$Y^{r}_{1}(k) =0$  and normalizing the RPA pair removal wavefunction according to $\sum_i (X^{r}_{1}(i))^2 =1$. In other words, neglecting ground state correlations  (TD approximation) which leads
to $X^{r}_{1} (s_{1/2}) = 0.055$ and $X^{r}_{1} (p_{3/2}) = 0.998$. The resulting absolute cross section for the reaction $^7$Li(t,p)$^9$Li(gs) integrated in the range 
$10^{\circ}-109^{\circ}$ becomes  9.1 mb, a value which lies  outside the experimental errors of the observed value (see Table II and Fig. \ref{fig7a} (b)). 


As shown in Figs. \ref{fig7} and \ref{fig8}, anharmonicities in the pairing vibrational spectrum 
can arise due to a number of physical effects. For example: 1) Overcompletness (see Fig. \ref{fig8} (b) and (c)); 2) Pauli principle correction to multiphonon states (see e.g. Figs. \ref{fig7}(a), \ref{fig7}(b) and \ref{fig8} (c)) leading, in connection     
with phonon mediated pairing (see e.g. Fig. \ref{fig7}(b)), to correlation (CO)-like processes \cite{Mahaux:85} (see inset (a) of Fig. \ref{fig9}). Within this context, it is of notice that the parity inversion observed to take place between the $s_{1/2}$ and $p_{1/2}$ states of a number of 
$N=7$ isotones (see Fig.\ref{fig9}) is due, to a large extent, to CO processes \cite{Barranco:01,Gori:04} (see inset (a) in Fig. 10). This is in keeping with the fact that among the largest components of the $2^+$ vibration of 
${}^8$He and of $^{10}$Be one finds the 
$(1p_{1/2},1p_{3/2}^{-1})$  particle-hole component.

It is then expected that the contributions to the anharmonicity of the pairing vibrational band in light halo nuclei (like ${}^{11}$Li but also ${}^{12}$Be, see Sect. V), can be important, in keeping with the high polarizability displayed by these systems as compared with, for example, $^{208}$Pb.
The energy of the two--phonon pairing vibrational mode given in Fig. \ref{LiPairVibr} for $^{9}$Li (as well as that associated with $^{10}$Be, see Sect. V below)
could, in principle, be strongly modified by such effects, as well as by coupling to states consisting of two (particle-hole-like) phonons (see Fig. \ref{fig8}(c); within this connection see \cite{Broglia:71b} as well as App. B).

\subsection{The optical potential}


In keeping with the fact that structure and reactions are just but two aspects of the same physics and that in the study of light halo nuclei, continuum states are to be treated on, essentially, equal footing in the calculation of the wavefunctions describing bound states (\textbf{structure}) as well as of the asymptotic distorted waves entering in  
the calculation of the absolute two-particle transfer differential cross sections (reaction; see Figs. \ref{fig10} (a) and \ref{fig10} (b)), the calculation of the optical potentials is essentially within reach (\textbf{reaction}, see Fig. \ref{fig10} (c) and Fig.\ref{fig11}, see also e.g. \cite{Fernandez:10a,Fernandez:10b} and refs. therein).

Because the real and imaginary parts of complex functions are related by simple dispersion relations (see e.g. \cite{Mahaux:85} and refs. therein) it is sufficient to calculate only one of the two (real or imaginary) components of the self-energy function to obtain the full scattering, complex, nuclear dielectric function (optical potentials). Now, absorption is controlled by on-the-energy shell contributions. Within this scenario it is likely that the simplest way to proceed is that of calculating the absorptive potential and then obtain the real part by dispersion (see e.g. \cite{Broglia:81b}, \cite{Pollarolo:83}; see also \cite{Broglia:05c}). Of notice that in heavy ion reactions, one is dealing with leptodermous systems. Thus, the real part of the optical potential can, in principle, be obtained by convolution of the nuclear densities and of the surface tension (see e.g. \cite{Broglia:05c} and refs therein).

Within the present context, one can mention  the ambiguities encountered in trying  to properly define a parentage coefficient  relating 
the system of $(A+1)$ nucleons to the system of $A$ nucleons, and thus a spectroscopic amplitude (see e.g. \cite{Dickhoff:05,Jennings:11}, see also \cite{Mahaux:85}).
In other words, a prefactor which allows to express the absolute one-particle transfer differential cross section in terms of the elastic cross section.

Making use of  NFT diagrams like the one shown in Fig. 12, it is possible to calculate, one at a time, 
the variety of contributions leading to one- and two-particle transfer 
processes (in connection with this last one see Fig. 11). Summing up the different contributions, taking also proper care of those 
arising from four-point vertex, tadpole processes, etc. 
(see refs. \cite{Bes:76a,Bes:76b,Bes:76c,Bes:75,Mottelson:76,Bortignon:77,Broglia:76}  and refs. therein,
see also \cite{Dickhoff:05}), a consistent description of the different channels can be worked out, in which the predicted quantities to be directly compared with observables 
are absolute differential cross sections, or, more generally, absolute values of strength functions for different scattering angles.

\section{The pairing  vibrational spectrum of $^{10}$\textrm{Be}}

Calculations similar to the ones discussed in previous sections have been carried out in connection with the expected $N=6$ shell closure pairing vibrational band of $^{10}\textrm{Be}$. In Fig. 
13  we display the associated pairing vibrational spectrum in the harmonic approximation (see App. \ref{Appendix.A}, in particular section $b$ of this Appendix). Also given are the absolute  two-nucleon transfer differential cross sections 
associated with the excitation of the one-phonon pair addition and pair subtraction modes excited in the reactions $^{12}$Be(p,t)$^{10}$Be(gs) and 
$^{10}$Be(p,t)$^{8}$Be(gs) respectively, calculated   for a bombarding energy appropriate for planned studies making use of inverse kinematic techniques
\cite{Kanungo:11}.

The ((2p-2h)-like) two-phonon pairing vibration state of $^{10}$Be is expected, in this approximation,  to lie at 4.8 MeV, equal to the sum of  the energies 
of the pair removal $W_1(\beta = -2)$ = 0.5 MeV and of the pair addition $W_1 (\beta= 2)$ = 4.3 MeV modes. In keeping with the fact that the lowest known $0^+$ excited state of $^{10}$Be appears
at about  6 MeV \cite{Alburger:69}, we have used this excitation energy in the calculation of the $Q-$value associated with the $^{12}$Be(p,t)$^{10}$Be(pv) cross section. The associated shift in energy from the harmonic 
value of 4.8 MeV can, arguably, be connected with anharmonicities of the $^{10}$Be pairing vibrational spectrum, as discussed in the case of $^{11}$Li in connection with Figs. 8 and 9. Medium polarization effects (see e.g. Fig. 8(b)) may also lead to conspicuous anharmonicities in the pairing vibrational spectrum.  


The two-nucleon spectroscopic amplitudes corresponding to the reaction $^{10}$Be(p,t)$^8$Be(gs) and displayed in Table III were obtained solving the RPA 
coupled equations (determinant) associated with the $^{10}$Be(gs) pair-removal  mode, making use, as explained in Sect. b of App. A, of two pairing coupling
constants , to properly deal with the difference in matrix elements (overlaps) between core-core, core-halo and halo-halo two-particle 
configurations. 
In other words with a ``selfconsistent'' treatment of the halo particle states ($\varepsilon_k > \varepsilon_F$), in particular of the $d_{5/2}(0)$ halo state.
The absolute differential cross sections displayed in the figure were calculated making use of the optical parameters of refs. \cite{An:06,Fortune:94}.

The two--nucleon spectroscopic amplitudes  associated with  the reaction 
${}^{12}\textrm{Be} (p,t) {}^{10}\textrm{Be} (gs)$  
correspond to the numerical coefficients appearing in  Eq. (\ref{Eq.waveBe_b}) below, and associated with the wavefunction describing 
the neutron component of the $^{12}$Be ground state (cf. ref. \cite{Gori:04}): 
\begin{equation}
 \vert \widetilde{0} \rangle = \vert 0 \rangle + \alpha \vert (p_{1/2},s_{1/2})_{1^{-}} \otimes 1^{-};0 \rangle + \beta \vert (s_{1/2},d_{5/2})_{2^{+}} \otimes 2^{+};0\rangle + \gamma \vert (p_{1/2},d_{5/2})_{3^{-}} \otimes 3^{-};0 \rangle,
\label{Eq.waveBe}
\end{equation}
with 
\begin{equation}
 \alpha = 0.10,\; \beta = 0.35,\; \textrm{ and }  \gamma= 0.33,
\label{Eq.waveBe_a}
\end{equation}
and
\begin{equation}
\vert 0 \rangle= 0.37 \vert s^{2}_{1/2} (0) \rangle + 0.50 \vert p^{2}_{1/2} (0) \rangle + 0.60 \vert d^{2}_{5/2} (0) \rangle,
\label{Eq.waveBe_b}
\end{equation}
the states $\vert 1^{-} \rangle$, $\vert 2^{+} \rangle$, $\vert 3^{-} \rangle$ being the corresponding lowest states of $^{10}$Be, calculated 
with the help of a multipole separable interaction in the RPA (see e.g. Table IV). It is of notice that a rather similar absolute differential cross section  
to the one displayed in Fig. 13 for the $^{12}$Be$(p,t)^{10}$Be(gs) reaction  is obtained making use of the spectroscopic amplitudes provided by the RPA wavefunction
describing the $^{10}$Be pair addition mode (see Table III and App. A). This can be seen from the results displayed in Fig. 14. 

 To assess the    correctness of the structure description of $|^{12}$Be(gs)$>$ provided by the wavefunction (\ref{Eq.waveBe}-\ref{Eq.waveBe_b})
and of the second order DWBA-reaction mechanism (successive, simultaneous  plus non-orthogonality) employed to calculate the absolute value of the $^{12}$Be$(p,t)^{10}$Be(gs)
differential  cross section, we compare in Fig. 15 the predictions of the model for the reaction $^{10}$Be(p,t)$^{12}$Be(gs) at 17 MeV triton bombarding energy 
with the experimental data. Theory  provides an overall account  of observation within experimental errors.         

It is of notice that the components proportional to $\alpha, \beta$ and $\gamma$ of the state (\ref{Eq.waveBe}) can lead, in a $^{12}$Be$(p,t)$ reaction, to the direct excitation of the $1^-,2^+$ and $3^-$ states of $^{10}$Be. Such results will add to the evidence obtained in the reaction \mbox{${}^1$H (${}^{11}$Li(gs),${}^9$Li($1/2^-$;2.69 MeV))${}^3$H} \cite{Tanihata:08} of phonon mediated pairing \cite{Potel:10}. The role of these components is assessed by the fact that (wrongly) normalizing the state (\ref{Eq.waveBe_b}) to 1, one obtaines a value of $\sigma = 4.5$ mb ($4.4^{\circ} \leq \theta_{CM} \leq 57.4^{\circ}$), a factor 2 larger than the experimental value \cite{Fortune:94} (see Fig. \ref{fig19}).

Let us now return to Fig. 13. \footnote{
In the harmonic approximation, and assuming simultaneous transfer, the cross section $\sigma(r)$ associated with the pair removal mode of a closed shell system ($A_{0}(p,t)(A_{0}-2)(gs)$) coincides by definition with that of the reaction $(A_{0}+2)(p,t)A_{0}(pv)$ ($\sigma(pv)$) exciting the two-phonon pairing vibrational mode starting from the ground state of the $(A_0+2)$ system, this pair addition mode acting as spectator. However, taking properly into account the contribution of successive transfer, $\sigma(r)$ and $\sigma(pv)$ are expected to differ because of the difference in Q-values associated with the corresponding intermediate, virtual, one-particle transfer states. This is the reason why the cross section associated with the reaction $^{10}$Be$(p,t)^{8}$Be$(gs)$ ($\sigma(r)$) shown in Fig. 13, differs from that of $^{12}$Be$(p,t)^{10}$Be$(pv; 6$ MeV) ($\sigma(pv)$) displayed in the lower part of Fig. \ref{fig16} (aside from the fact that the energy of 6 MeV has been used instead of the harmonic value of 4.8 MeV). In spite of this fact, the two cross sections ($\sigma = 16$ mb and $\sigma = 17.4$ mb respecitvely) are rather similar, testifying to the validity of the pairing vibrational scheme.
}
The ratio of the integrated absolute cross section 
at $E_{CM}$= 7 MeV   in the range 10$^{\circ} \leq \theta_{CM} \leq 50^{\circ}$ appropriate for planned experimental studies making use of inverse kinematic
techniques  \cite{Kanungo:11} is,
\begin{equation}
R=\frac{\sigma \left({}^{12}\textrm{Be}(p,t){}^{10}\textrm{Be}(pv;6 \textrm{MeV}) \right)}{\sigma \left({}^{12}\textrm{Be}(p,t){}^{10}\textrm{Be}(gs) \right)}=
{\frac{16.0 \textrm{mb}}{6.9\textrm{mb}} \approx 2.3}, \label{Eq.ratio}
\end{equation}
a result which testifies to the clear distinction between occupied and empty states taking place at $N=6$, 
and thus of the \textit{bona fide} nature of this magic number for halo, drip line nuclei.
The ratio (\ref{Eq.ratio}) reflects the fact that the pairing Zero Point Fluctuations (ZPF in gauge space) displayed by the $|^{10}$Be$(gs)\rangle$ as embodied in the pair addition and pair removal modes, and quantified by the absolute values of the associated two-nucleon transfer cross sections, are of the same order of magnitude.
This is an intrinsic property of the vibrational modes, in the same way in which  e.g. the width (lifetime) of a nuclear state is an intrinsic (nuclear structure) property of such a state. An experiment displaying an energy resolution better than the intrinsic width of the states under study will provide structure information. Otherwise, eventually an upper limit.
Within this scenario and in keeping with the fact that the successive transfer induced by the single-particle potential is the intrinsically (structure) dominant contribution to the absolute two-particle transfer cross section, Q-value (kinematic) effects can strongly distort the picture. In particular in the case in which single-particle transfer channels are closed at the studied bombarding energies.

Let us elaborate on these arguments 
in the case in which the reaction $^{12}$Be(p,t)$^{10}$Be populates the ${}^{10}$Be ground state, where the two correlated nucleons participating in the process are the two valence neutrons of $^{12}$Be$(gs)$. As the first neutron leaves ${}^{12}$Be$(gs)$, it leads to ${}^{11}$Be$(j^{\pi})$ 
where $j^{\pi}$ labels $2s_{1/2}$, $1p_{1/2}$ and $1d_{5/2}$. In Fig. 16 
we report, for concreteness, the $Q-$value scenario where $j^{\pi}=1/2^{+}$, although the calculations of the absolute cross sections displayed 
in Figs. 13 and 14  were carried out making use of the full wavefunction, and thus of all possible ($j^{\pi}=1/2^{+}$, $1/2^{-}$ and $5/2^{+}$) 
intermediate states in the ${}^{11}$Be$(j^{\pi})+d$ channel.
Of notice that, because of ground state correlations,  situations in which $j^{\pi}= 3/2^{-} (1p^{-1}_{3/2})$ and $1/2^{+} (1s^{-1}_{1/2})$ are also possible (as hole states). The corresponding contributions are not negligible (see Table 3), although the most important ones arise from those mentioned above.

The case in which  one populates the excited pairing vibrational state ${}^{10}\textrm{Be}(pv; 6$ MeV)
is illustrated in Fig. 17. The fermions involved correspond to the correlated (valence) holes 
 characterizing the ground state of ${}^{8}$Be, the two correlated (valence) neutrons building the ${}^{12}$Be ground state acting, in the harmonic approximation, only as spectators. Within this scenario, the most important contributions to the ${}^{11}$Be$(j^{\pi})+d$ intermediate states are hole states moving in the $1p_{3/2}$ and 
$1s_{1/2}$ orbitals.
As seen from the figure, at $E_{CM}=5$ MeV, the ${}^{11}Be(j^{\pi})$ channels (e.g. ${}^{11}$Be$(p^{-1}_{3/2}; 4 \textrm{MeV})$) are barely open, thus quenching in a major way the excitation of the ${}^{10}\textrm{Be}(pv; 6$ MeV) state.
The closing or opening of single-particle transfer channels due to Q-value effects, may constitute a unique opportunity to learn about the mechanism which is at the basis of two-nucleon transfer reaction processes (see Figs. 18-20). By properly adjusting the bombarding conditions (excitation function), one may tune on situations in which two-particle transfer switches from successive- to simultaneous-dominated transfer regimes.

Simple mechanical interpretations of the above mentioned Q-value effects can be given within the framework of the semiclassical approximation \cite{Broglia:05c}. The transfer formfactors associated with the excitations of pairing correlated modes are, as a rule, smoothly varying functions along the trajectories of relative motion, displaying none or few nodes, in the neighborhood of the distance of closest approach. To obtain the variety of contributions to the two-nucleon transfer amplitudes the different formfactors are to be weighted with (imaginary) exponentials (i.e. periodic functions), resulting from the mismatch in momentum (recoil effects, Galilean-like transformations), and in stationary state phases (i.e. exp$\{-iEt/\hbar \}$) associated with the different relative orbitals. Large transfer amplitudes and eventually two-nucleon transfer cross sections are obtained when the arguments of the phases appearing in the imaginary exponential are small along the relative motion trajectory (little mismatch). The situation is reversed when the large (stationary) energy differences are not compensated by recoil effects and viceversa. In this case, the harmonic functions change rapidly sign along the relative motion trajectories, thus canceling the (smooth behavior) transfer formfactor contributions (large mismatch).

The extension (of the same arguments) to situations of long wavelength of relative motion (fully quantal scenario) can be made in terms of the WKB approximation (e.g. in terms of Stokes lines), as well as in terms of the Feynman path integral method. In this case all possible trajectories are to be considered. The associated formfactors are weighted with action phases. The different amplitudes thus result from the interweaving of these two types of functions. Depending on their mismatch different bombarding energies, angles, etc. will be privileged over others, as in the simpler (semiclassical) scenario discussed above.


\section{Conclusions}

It is an open question whether (mainly) phonon mediated neutron halo pairing can lead to a well developed (quasi harmonic) pairing vibrational multiphonon band. Inverse kinematics two--particle transfer processes,  as well as standard two--particle transfer reactions can test the validity of the harmonic spectrum discussed above, and thus the actual nature of the $N=6$ shell closure, as well as shed light on the role medium polarization effects play in these fragile, highly polarizable systems. In particular concerning the induced pairing 
interaction. The question of how large the coupling between particle--hole phonons and the ($n_a=1,\;n_r=1$) pairing modes is, could also be tested through the above mentioned two--particle transfer experiments. Last, but not least, the NFT description of structure and reactions 
provides a natural framework for the calculation of the optical potential, a possibility which would bring one step further the concrete realization of the fact that structure and reactions are but two aspects of the same physics, in particular in the case of halo nuclei.

Discussions with I. Tanihata and R. Kanungo  are gratefully acknowledged.

Financial support from the Ministry of Science and Innovation of Spain Grant No. FPA2009-07653 is acknowledged by G. P.

F.B. acknowledges financial support from the Ministry of Science and Innovation of Spain grants FPA2009-07653 and
ACI2009-1056.

\bibliographystyle{unsrt}
\bibliography{nuclear}

\clearpage

\begin{figure}
\centerline{\includegraphics*[width=.99\textwidth,angle=0]{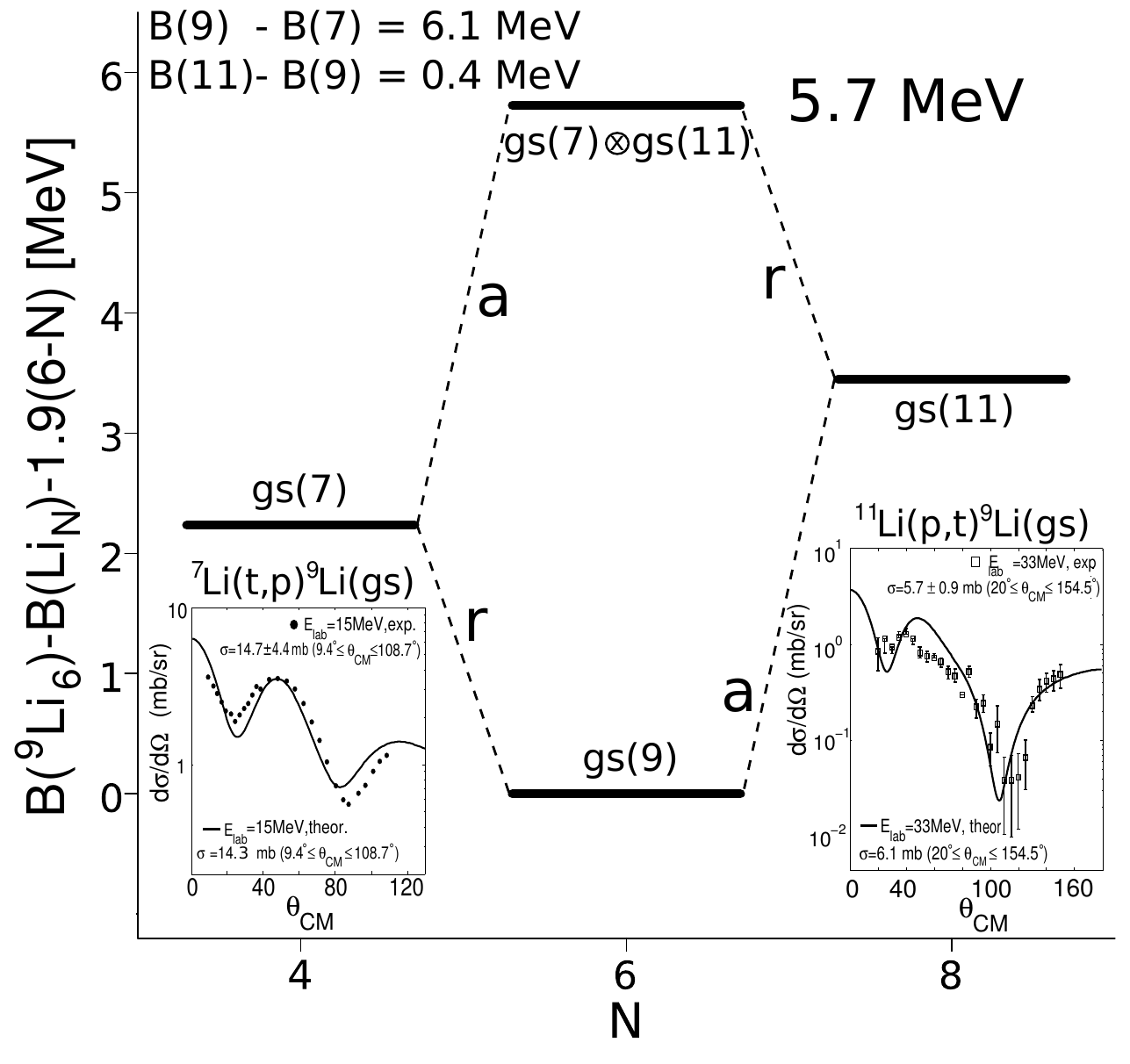}}
\caption{
\begin{footnotesize}
Pairing vibrations around $^{9}$Li and absolute cross sections associated with removal \cite{Young:71} (see also \cite{AjzenbergSelove:78})  and addition \cite{Tanihata:08} modes. The theoretical \textbf{absolute} differential cross sections for ${}^{11}\textrm{Li}(p,t) \; {}^9\textrm{Li}(gs)$ (addition: a) is reported in \cite{Potel:10}. The theoretical \textbf{absolute} differential cross section associated with the reaction ${}^{7}\textrm{Li}(t,p) \; {}^9\textrm{Li}(gs)$ (removal: r) was carried out making use of the wavefunction associated with the RPA solution of the pairing Hamiltonian (see \cite{Broglia:73}, \cite{Bes:66} and App. A as well as Table 1) , adjusting the coupling constant $G$ to reproduce the correlation energy of the two neutron holes in the core of ${}^{9}$Li (i.e. in the ground state of ${}^{7}$Li). 
The optical potential parameters used were taken from ref. \cite{Young:71,An:06}. Of notice that  throughout in this paper, in particular in connection with this figure, we report \textbf{absolute} differential cross sections
(see also Table 2).
\end{footnotesize} 
}
\label{LiPairVibr}
\end{figure}

\begin{figure}
\centerline{\includegraphics*[width=.75\textwidth,angle=0]{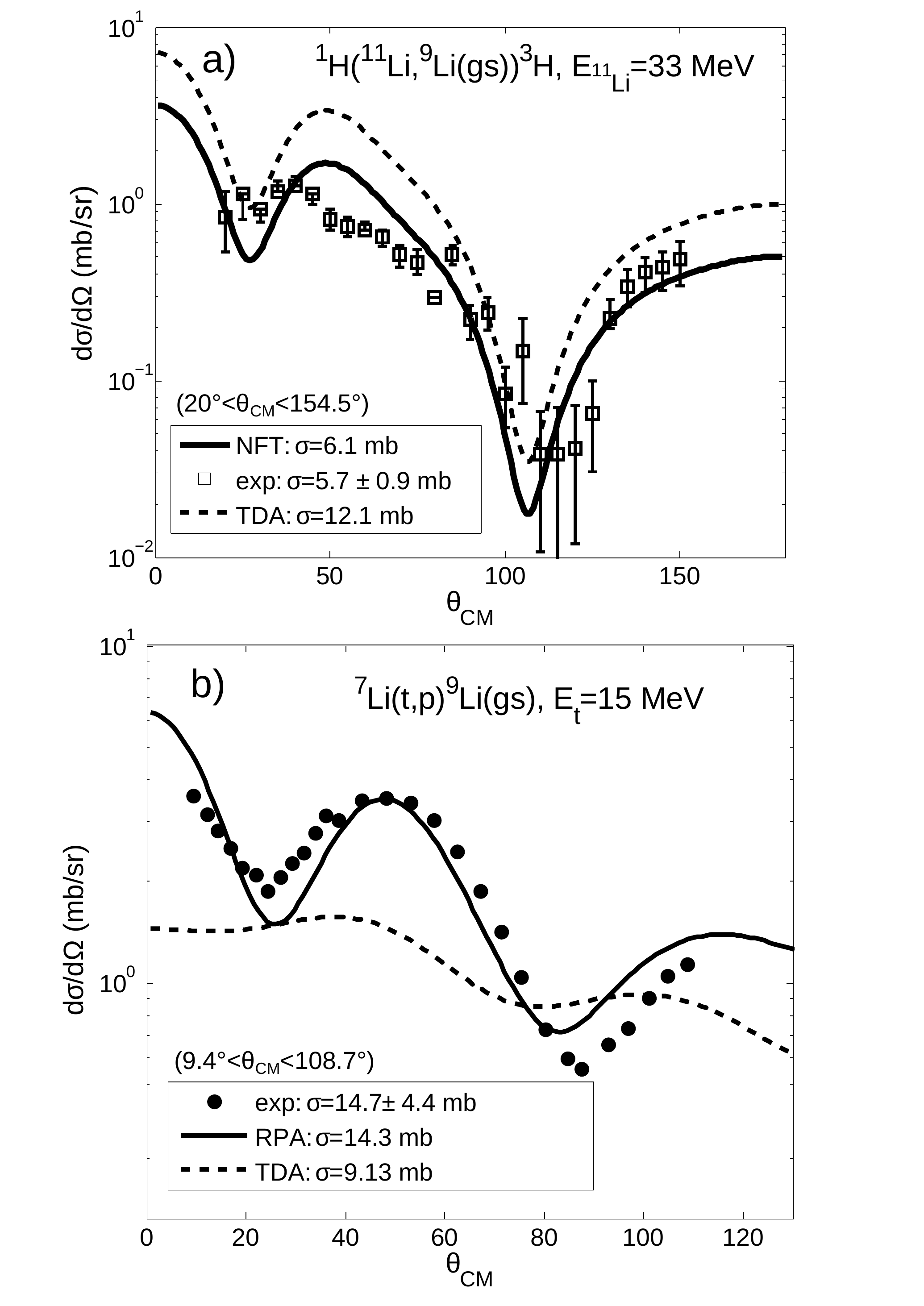}}
\caption{\begin{footnotesize} Absolute differential cross section of the pair addition and pair removal modes of $^9$Li in comparison with the experimental findings.
(a) The NFT results of the calculations of the absolute values of $d \sigma /d \Omega$ associated with the reaction ${}^{11}\textrm{Li}(p,t) \; {}^9\textrm{Li}(gs)$
(pair addition mode) reported in Fig. 6 are compared with those labeled Tamm-Dancoff approximation (TDA), in which the interweaving of single-particle and 
particle-hole like vibrational modes are neglected while the $|s^2>,|p^2>,|d^2>$ components of the two-neutron wavefunction are  normalized to one  (see text). (b) The absolute value
of the differential cross section
$d \sigma/d\Omega$ associated with the reaction  ${}^{7}\textrm{Li}(t,p) \; {}^9\textrm{Li}(gs)$  (pair removal mode)  and  calculated making use of the RPA 
two-nucleon transfer spectroscopic amplitudes ($X^r,Y^r$-values, Table 1) also reported in Fig. \ref{LiPairVibr}, is compared with that  obtained 
neglecting ground state correlations and labeled TDA (see text).    \end{footnotesize} }\label{fig7a}
\end{figure}

\begin{figure}
\centerline{\includegraphics*[width=.4\textwidth,angle=0]{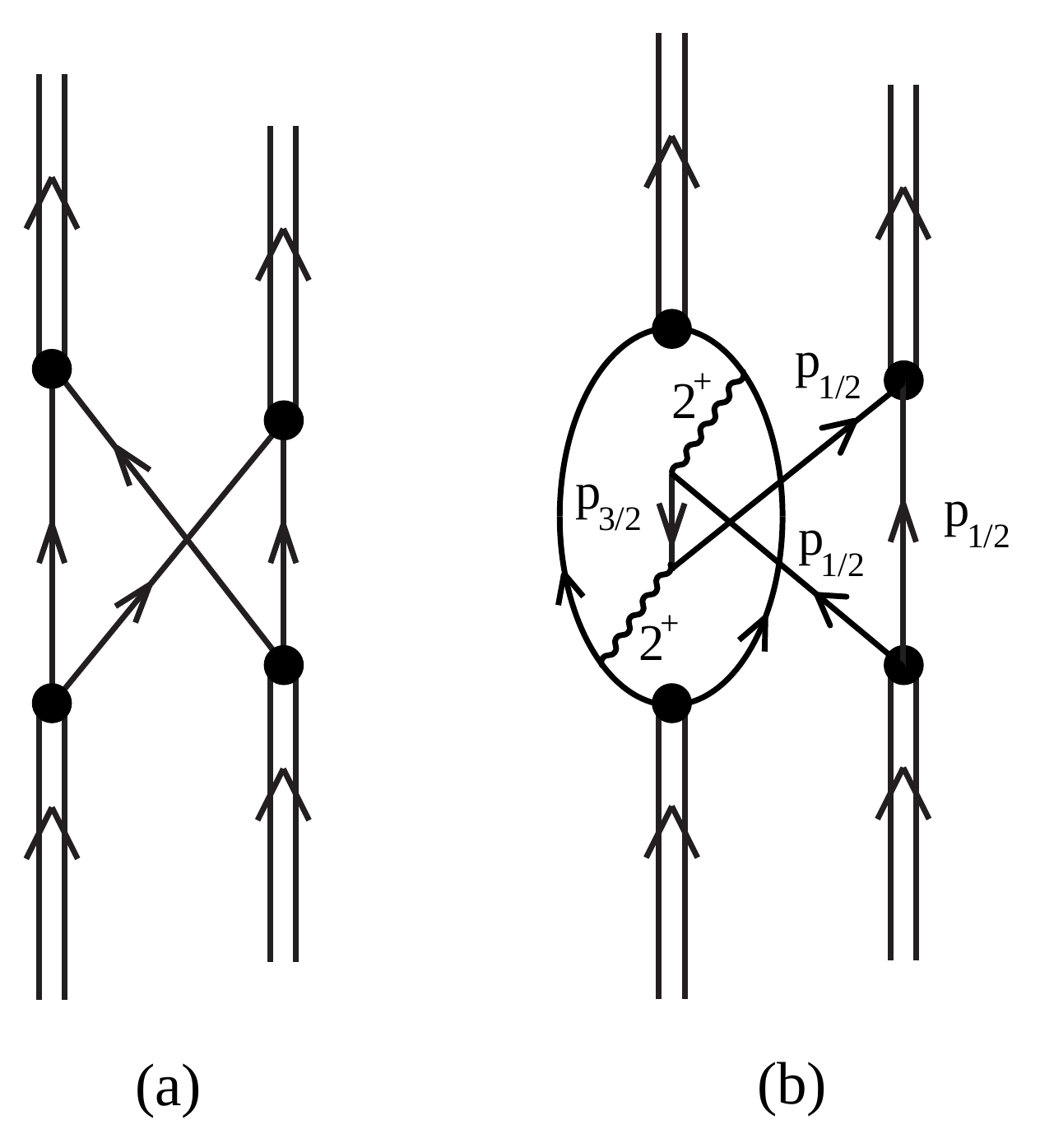}}
\caption{Pairing modes phonon--phonon interaction arizing from: (a) Pauli principle processes between pairing modes, and  (b) 
between single--particle and ph- phonon mediated induced pairing interaction (so called CO diagrams, cf. e.g. \cite{Mahaux:85}. 
See also Fig. \ref{fig9} (inset)).}\label{fig7}
\end{figure}

\begin{figure}
\centerline{\includegraphics*[width=.6\textwidth,angle=0]{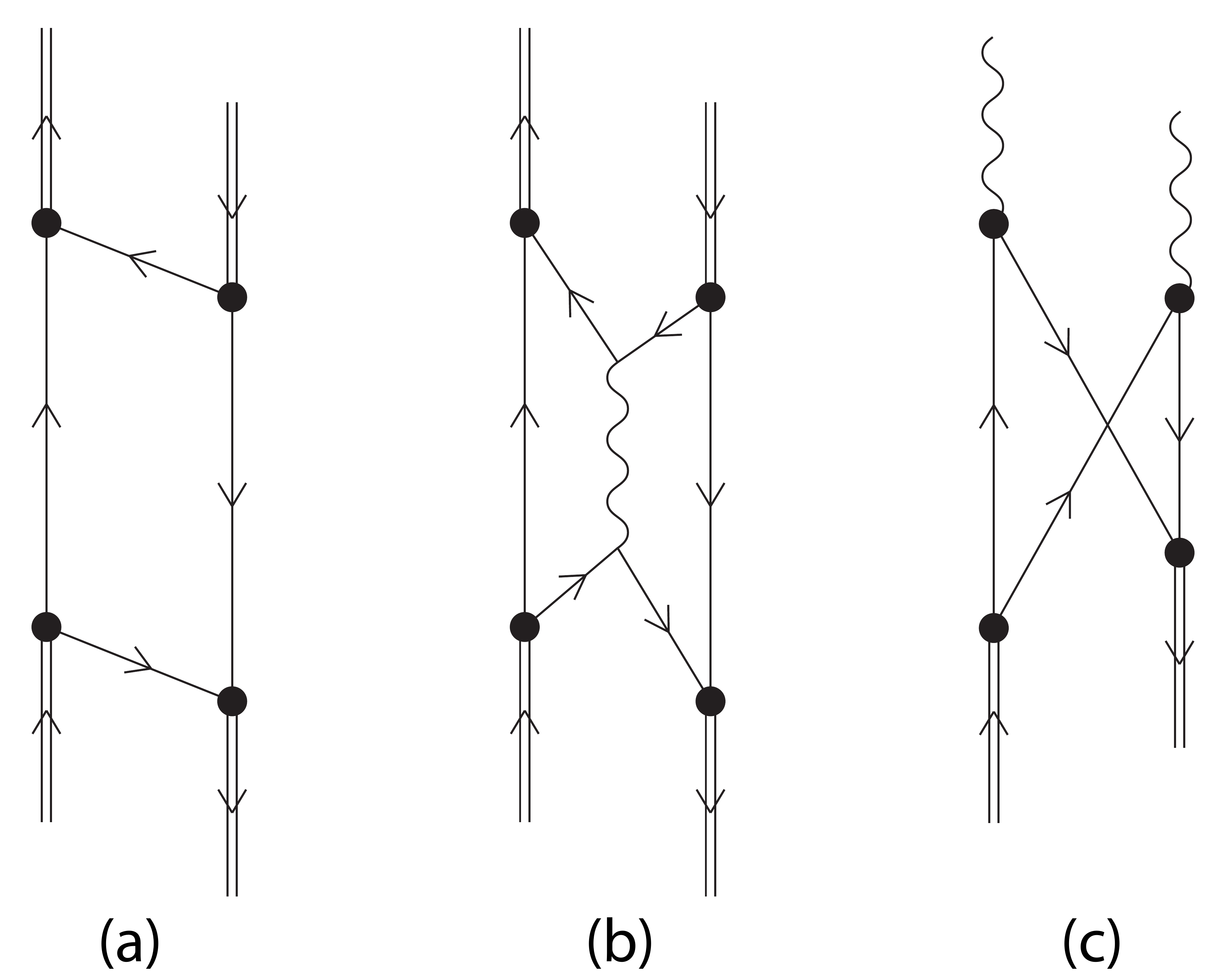}}
\caption{(a),(b) Examples of pair addition and pair removal modes interactions. (c) Interaction between the two-phonon pairing vibration state and the two-phonon particle-hole state.} \label{fig8}
\end{figure}

\begin{figure}
\centerline{\includegraphics*[width=.85\textwidth,angle=0]{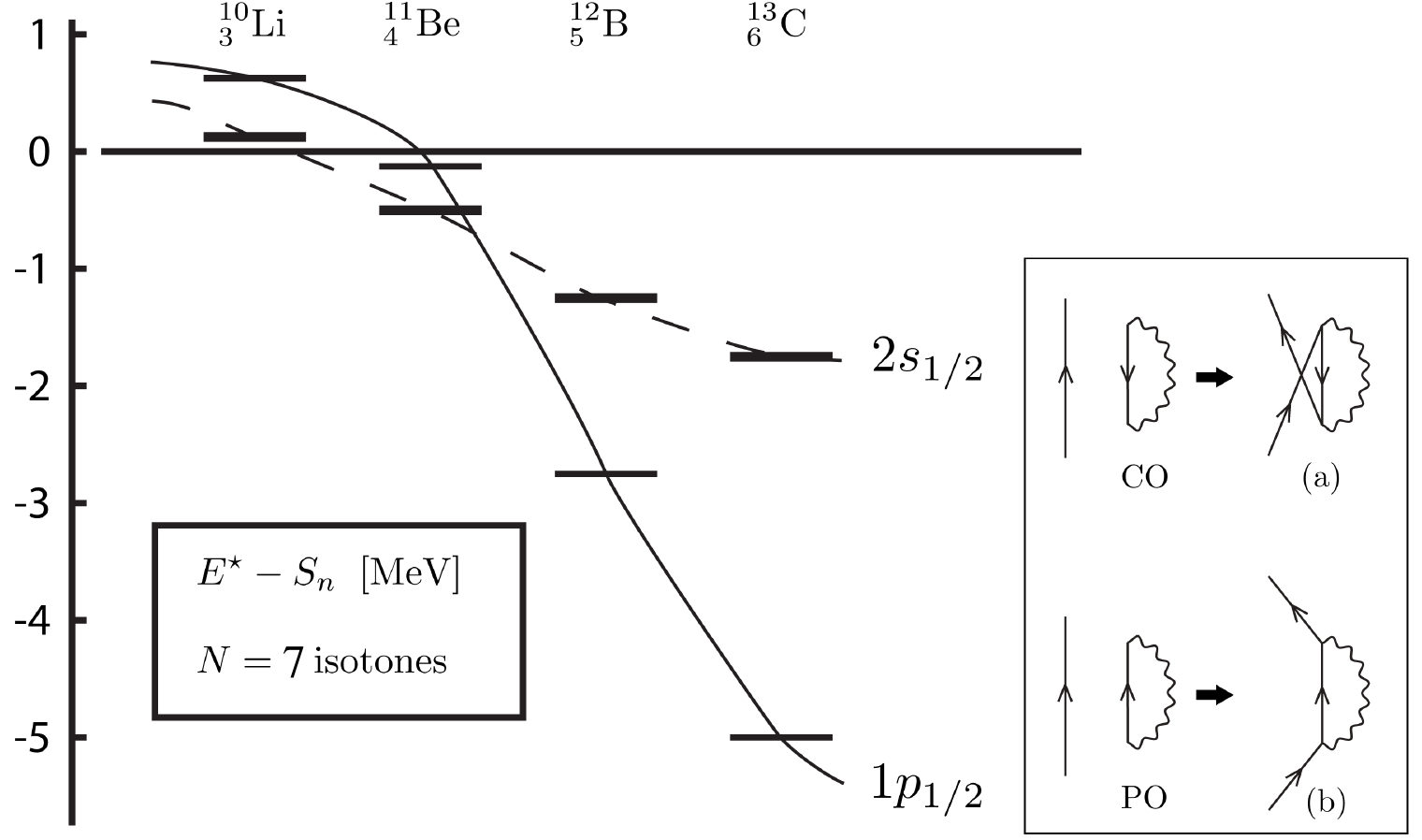}}
\caption{\begin{footnotesize} Single-particle states for 
$N=7$ isotones around $^{11}$Be associated with parity inversion. The thin horizontal lines represent the $1p_{1/2}$ single-particle state, while the thick ones the $2s_{1/2}$ orbital. In the case of $^{10}$Li one reports the centroid of the virtual and of the resonant states. $E^{*}$ stands for excitation energy and $S_n$ is the neutron separation energy. In the case of $^{10}$Li e.g. $S_n=0$, while $E^{*}_{s_{1/2}}=0.2$ MeV and  $E^{*}_{p_{1/2}}=0.5$ MeV. In the inset the correlation (CO) and polarization (PO) (virtual) contribution to the single-particle self energy are shown. An arrowed line pointing upwards represent a particle moving in a level with energy $\varepsilon_k>\varepsilon_F$, a downwards pointing line represent a hole state $\varepsilon_i<\varepsilon_F$, while a wavy line stands for a ph-like vibrational state. Their contribution to the real (single-particle ``legs'' propagating to $\pm \infty$ times) processes dressing the $1p_{1/2}$ and $2s_{1/2}$ neutron states of $^{10}$Li and $^{11}$Be are (a) and (b) respectively. In the first case the phonon corresponds essentially only to the $2^+$ vibration of the corresponding core ($^9$Li and $^{10}$Be respectively), and pushes the orbital upwards (Pauli principle, Lamb-shift-like process) making the dressed $p_{1/2}$ orbital more strongly unbound
than what it  was originally in the Saxon--Woods potential (see \cite{Bohr:69} Eqs. (2-180)--(2-182) pp. 238 and 239). In the case of the $2s_{1/2}$ orbital, it is mainly the process (b) which dresses the state making it almost bound (virtual state) as compared with the Saxon--Woods state. Within this context, it is of notice that in the binding of the two halo neutrons of $^{11}$Li to the $^{9}$Li core, it is essentially the pigmy resonance of $^{11}$Li which provides the largest contribution, the coupling to the $2^+$ vibration of the core $^9$Li giving a small shift in energy (nonetheless, it is this weak component of the self energy which is responsible for the excitation, in the
$^{11}$Li(p,t)$^9$Li reaction \cite{Tanihata:08}, of the $1/2^-, 2.69$ MeV state \cite{Potel:10}). In the case of $^{11}$Be the (p-h) vibrations are the $2^+$, $1^-$ and $3^-$ of the core $^{10}$Be, in keeping also with the fact that  $^{12}$Be does not display a pigmy $1^-$ resonance, not at least based on the ground state. 
It is of notice that graphs (a) and (b) give rise to an effective mass known as the $\omega$-mass. Associated with it are the $Z(\omega)=(m_\omega/m)^{-1}$ occupation factors (discontinuity at the Fermi energy; for details see ref. \cite{Mahaux:85} and refs. therein). \end{footnotesize}}\label{fig9}
\end{figure}

\begin{figure}
\centerline{\includegraphics*[width=.95\textwidth,angle=0]{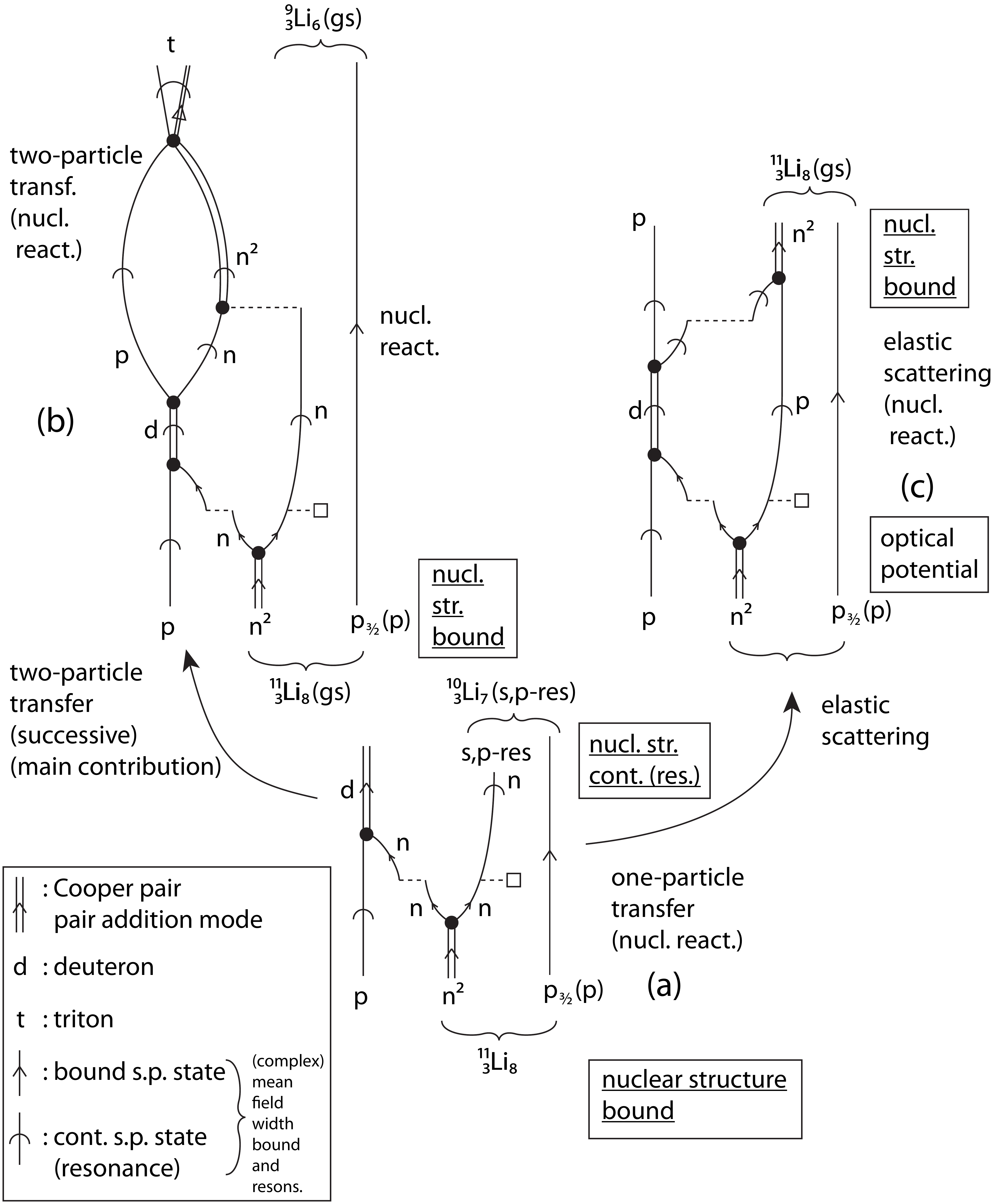}}
\caption{\begin{footnotesize} NFT diagrams summarizing the physics which is at the basis of the structure of $^{11}$Li \cite{Barranco:01} and of the analysis of the ${}^{11}\textrm{Li}(p,t) \; {}^9\textrm{Li}(gs)$ reaction \cite{Potel:10}. At variance with similar diagrams shown in Fig. \ref{fig2} (c) and Fig. \ref{fig3}, in the present figure emphasis is set on intermediate (like e.g. ${}^{10}$Li$+d$, see (a) and (b)) and elastic (see (c), see also Fig. \ref{fig11}) channels.\end{footnotesize}}\label{fig10}
\end{figure}

\begin{figure}
\centerline{\includegraphics*[trim = 1mm 120mm 1mm 20mm, clip, width=.95\textwidth]{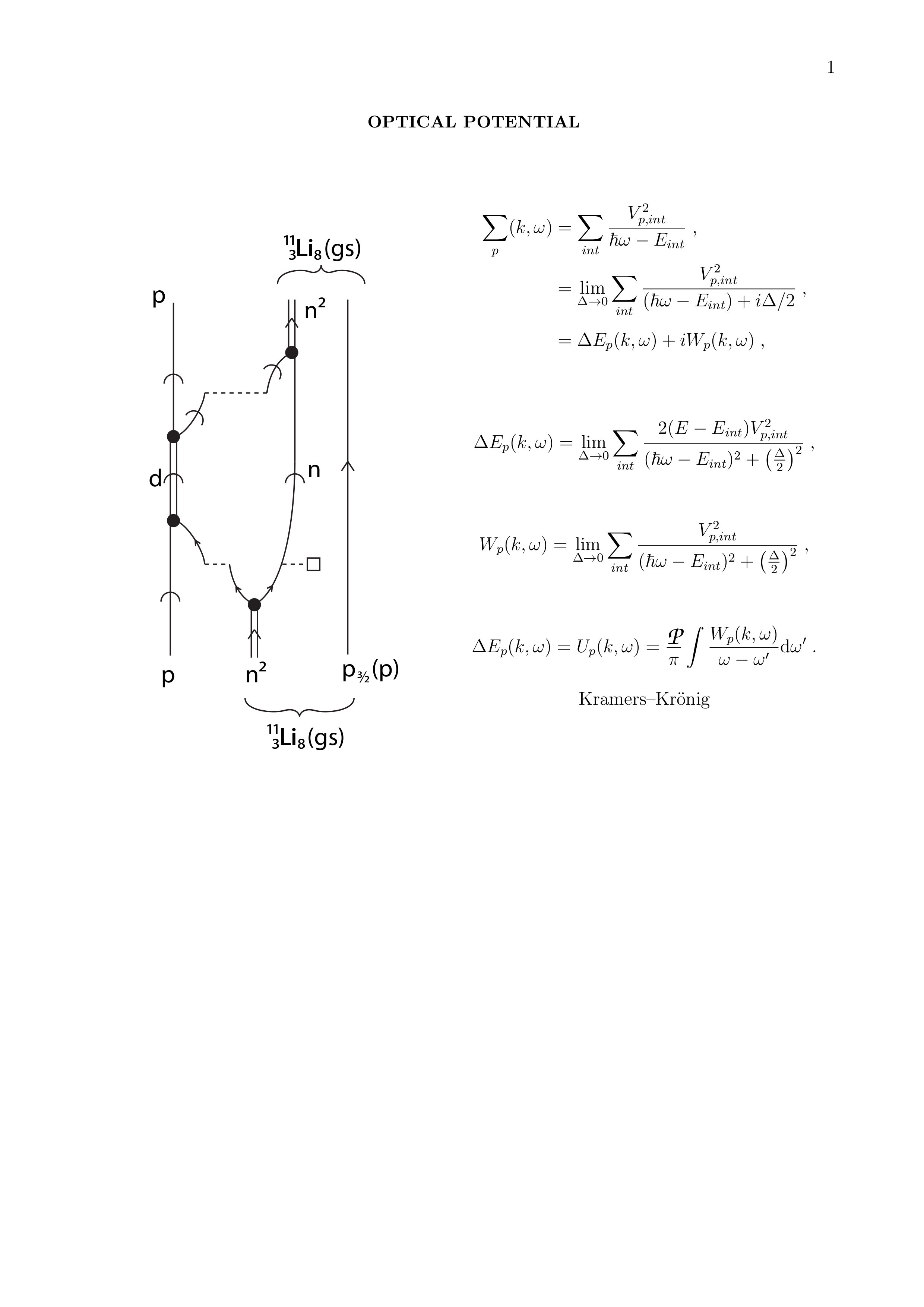}}
\caption{\begin{footnotesize} NFT diagrams and summary of the expression (see e.g. \cite{Mahaux:85} and refs. therein) entering in the calculation of one of the contributions (that associated with one-particle transfer and, arguably, the dominant one) to the ${}^{11}$Li$+p$ elastic channel. The self-energy function is denoted $\Sigma_p$, while the real and imaginary parts are denoted $\Delta E_p(=U_p)$ and $W_p$ respectively, the subindex $p$ indicating the incoming proton. These quantities are, in principle, a function of frequency and momentum. \end{footnotesize}}\label{fig11}
\end{figure}

\begin{figure}
\centerline{\includegraphics*[width=.95\textwidth,angle=0]{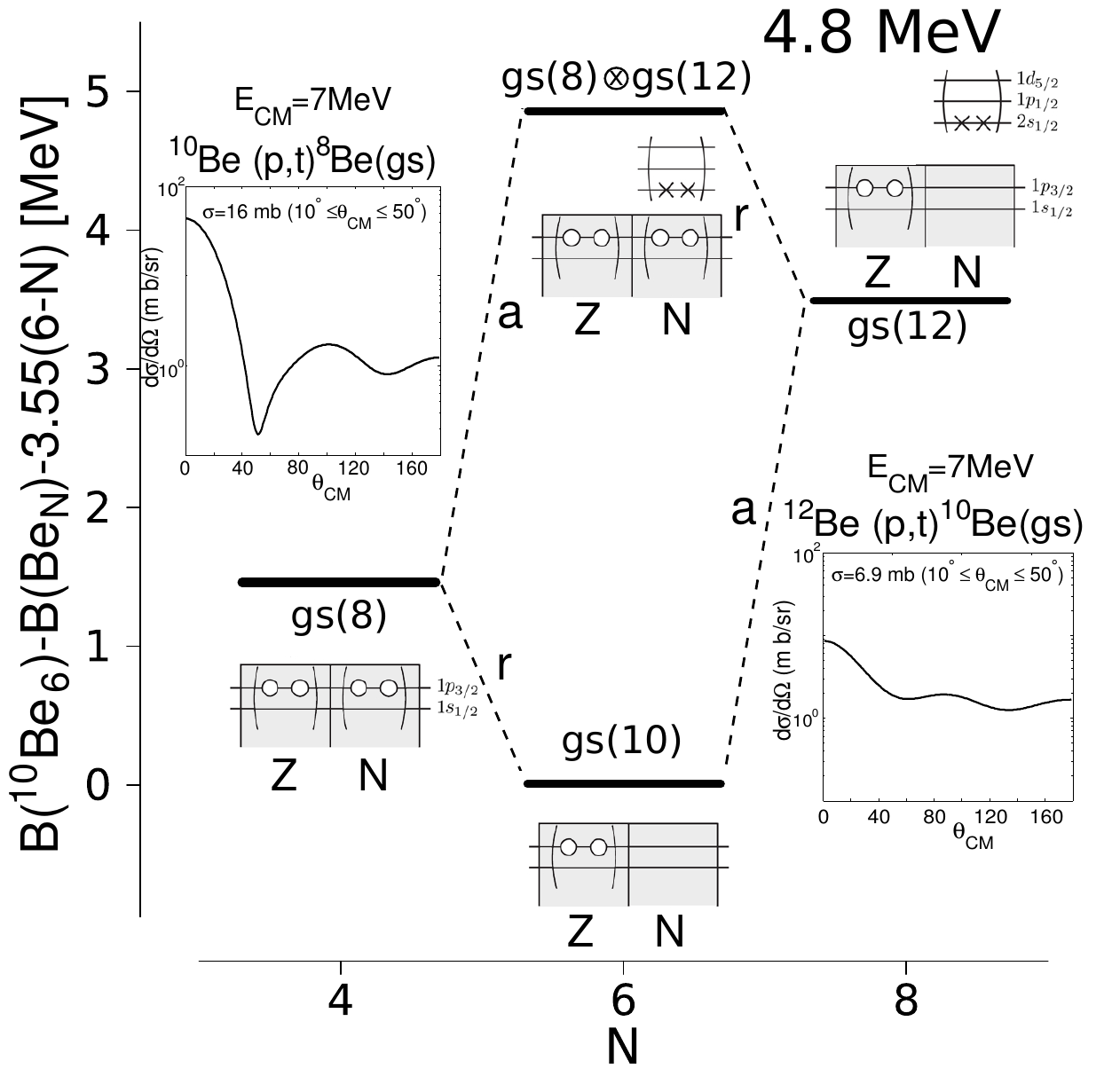}}
\caption{Pairing vibrational spectrum of Be$_{N}$ (see also Fig. \ref{fig21} below) and associated absolute two-nucleon transfer differential cross section calculated as explained in the text.} \label{fig12} 
\end{figure}

\begin{figure}
\centerline{\includegraphics*[width=.85\textwidth,angle=0]{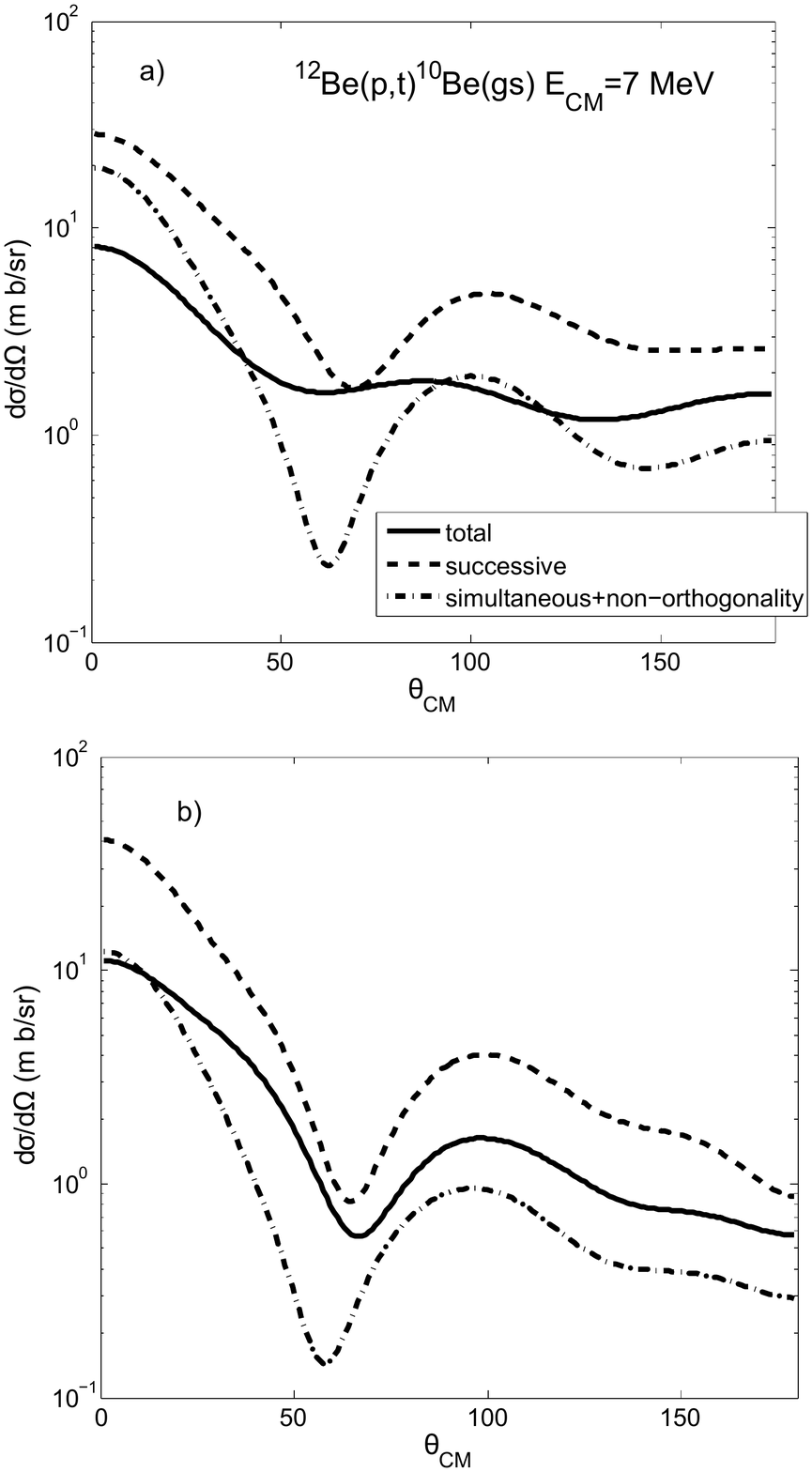}}
\caption{Absolute differential cross section associated with the reaction ${}^{12}\textrm{Be} (p,t) {}^{10}\textrm{Be} (gs)$ at $E_{CM}$ = 7 MeV, calculated making use of :  
(a) the  
wavefunction (\ref{Eq.waveBe}) (already shown in Fig. \ref{fig12}) and (b), the RPA wavefunction describing the $^{10}$Be  pair addition mode (see Table III).} \label{fig12a}
\end{figure}

\begin{figure}
\centerline{\includegraphics*[width=.95\textwidth,angle=0]{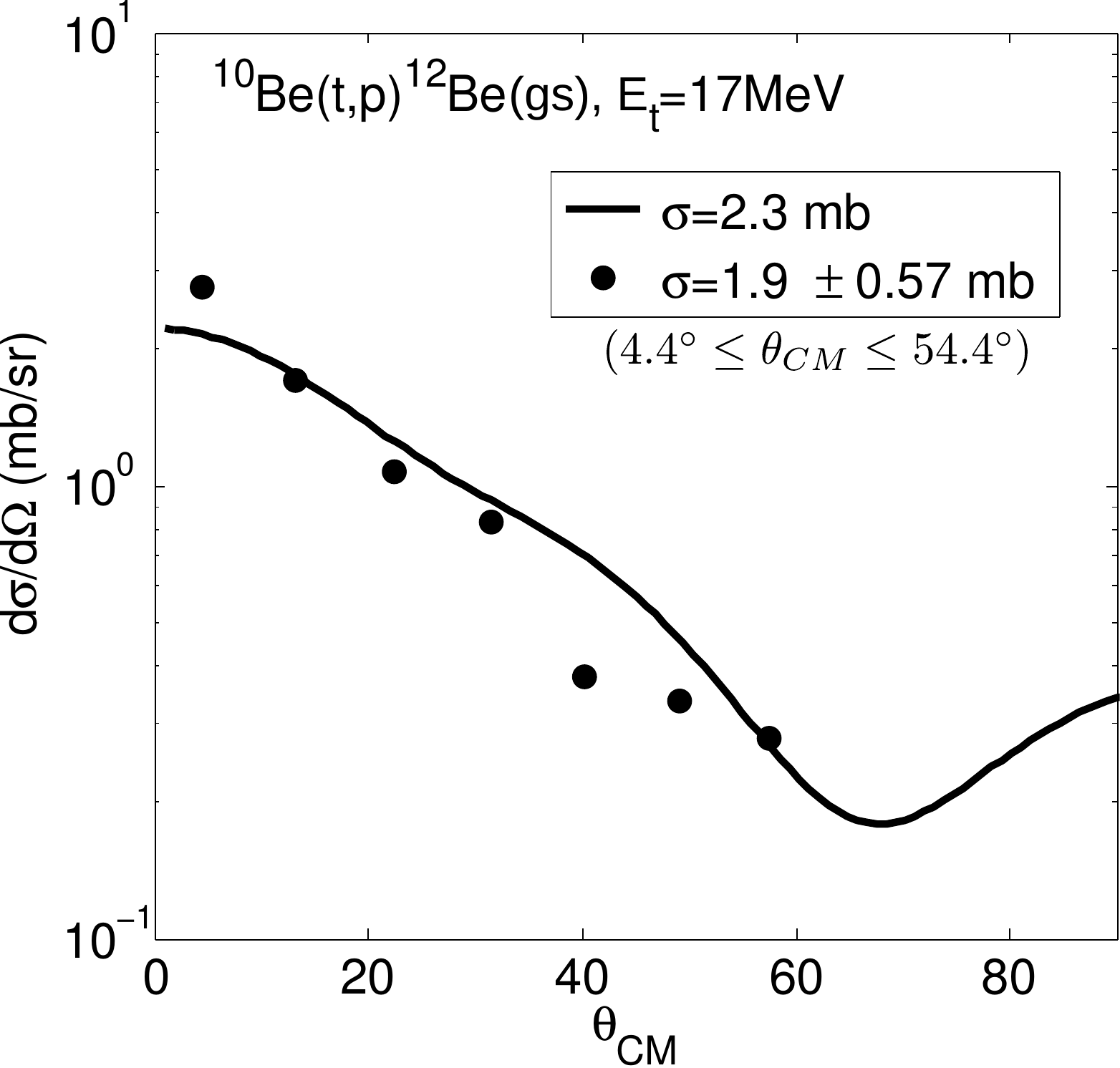}}
\caption{Absolute differential cross section measured \cite{Fortune:94} in the reaction 
${}^{10}\textrm{Be} (t,p) {}^{12}\textrm{Be} (gs)$  at 17 MeV triton bombarding energy (solid dots). The theoretical calculations (continuous solid curve) were obtained making use of the spectroscopic amplitudes associated with the wavefunction in Eqs. (\ref{Eq.waveBe})-(\ref{Eq.waveBe_b}), and the optical parameters of refs. \cite{An:06} and \cite{Fortune:94}
taking into account successive, simultaneous and non-orthogonality processes.} \label{fig19}
\end{figure}

\begin{figure}
\centerline{\includegraphics*[width=.95\textwidth,angle=0]{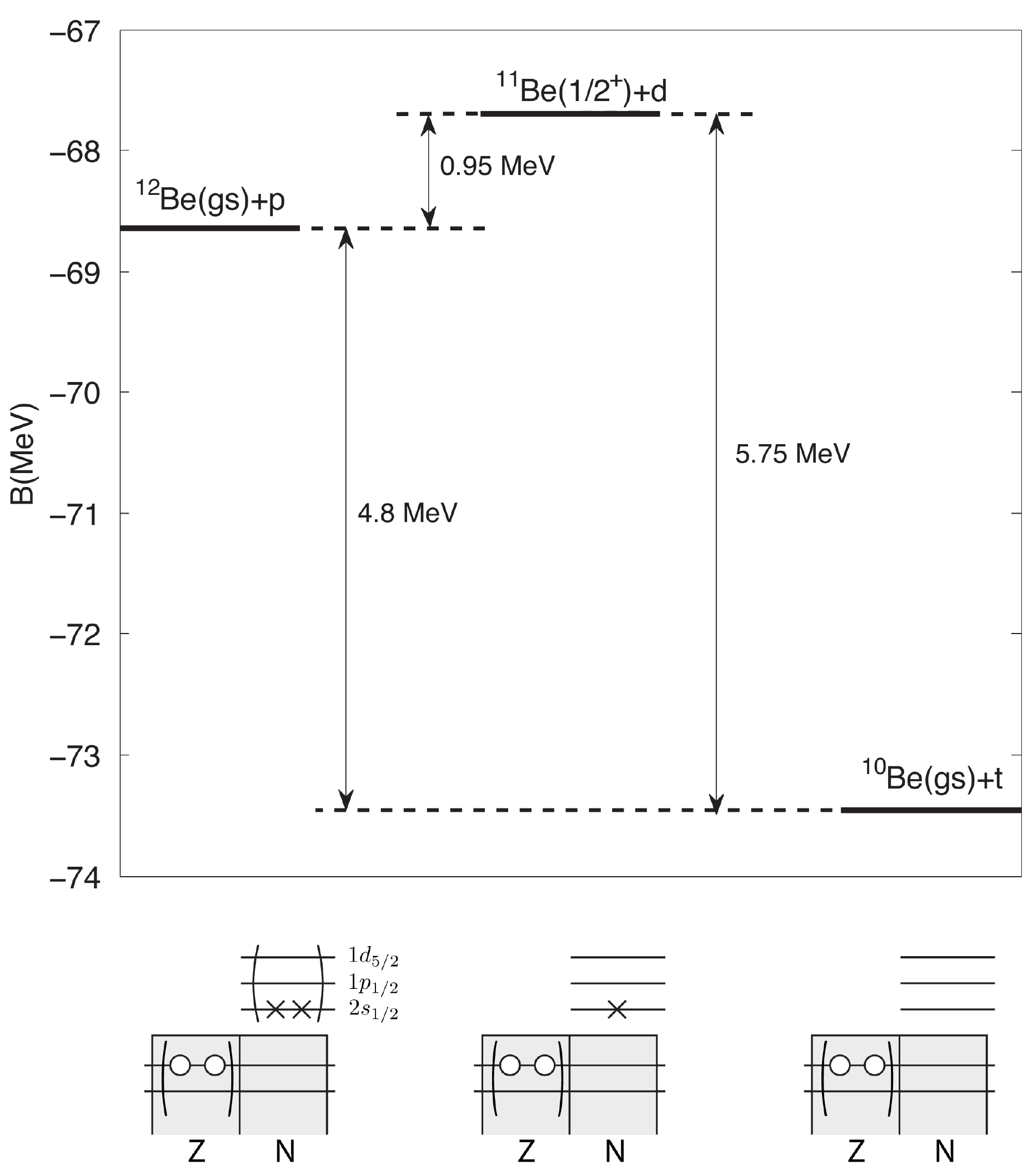}}
\caption{Q-values associated with the ${}^{12}\textrm{Be} (p,t) {}^{10}\textrm{Be}(gs)$, also for the successive transfer (see text).} \label{fig13}
\end{figure}

\begin{figure}
\centerline{\includegraphics*[width=.95\textwidth,angle=0]{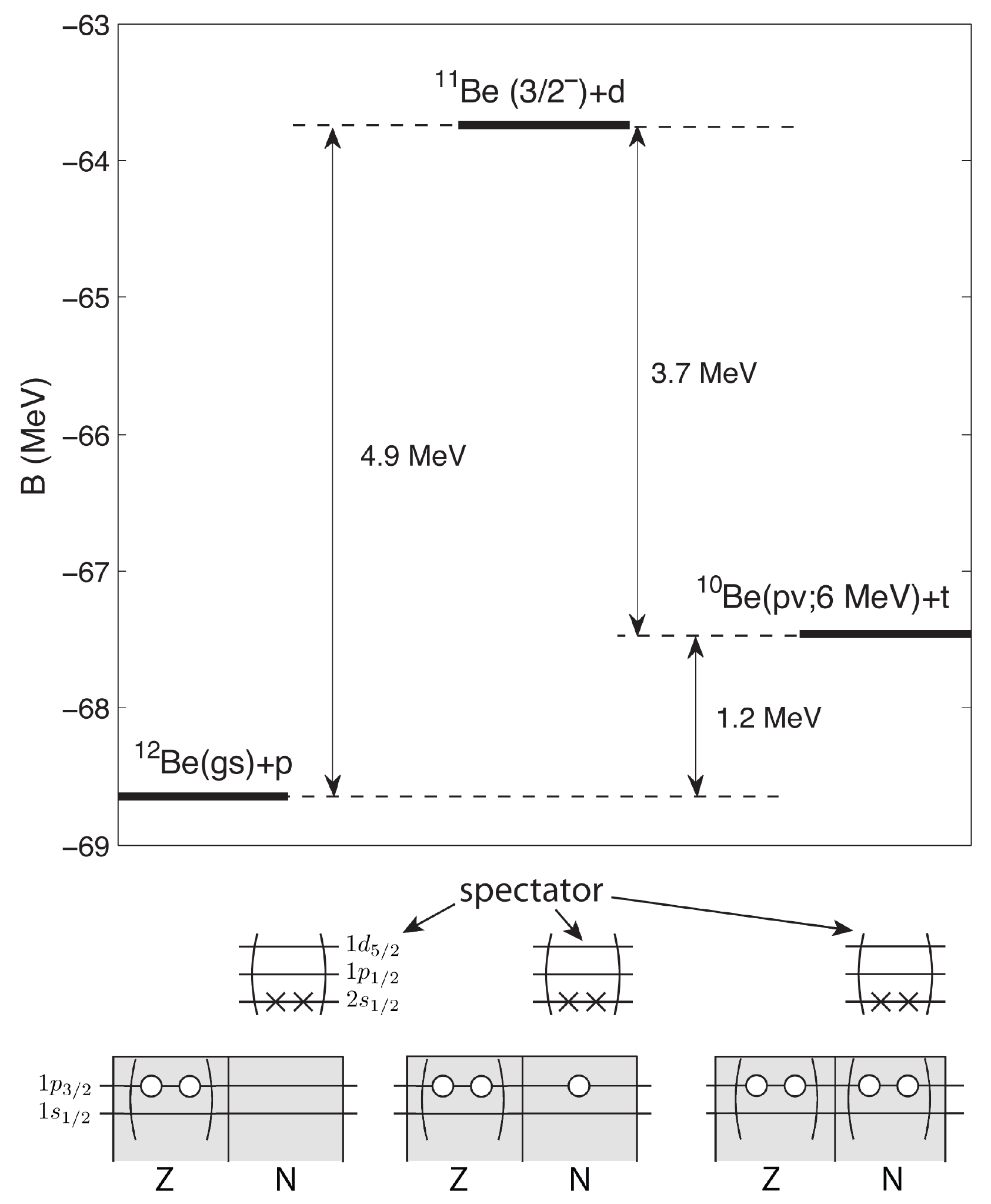}}
\caption{As for Fig. \ref{fig13} but for the reaction ${}^{12}\textrm{Be} (p,t) {}^{10}\textrm{Be}(pv; 6 \textrm{MeV})$.} \label{fig14}
\end{figure}

\begin{figure}
\centerline{\includegraphics*[width=.8\textwidth,angle=0]{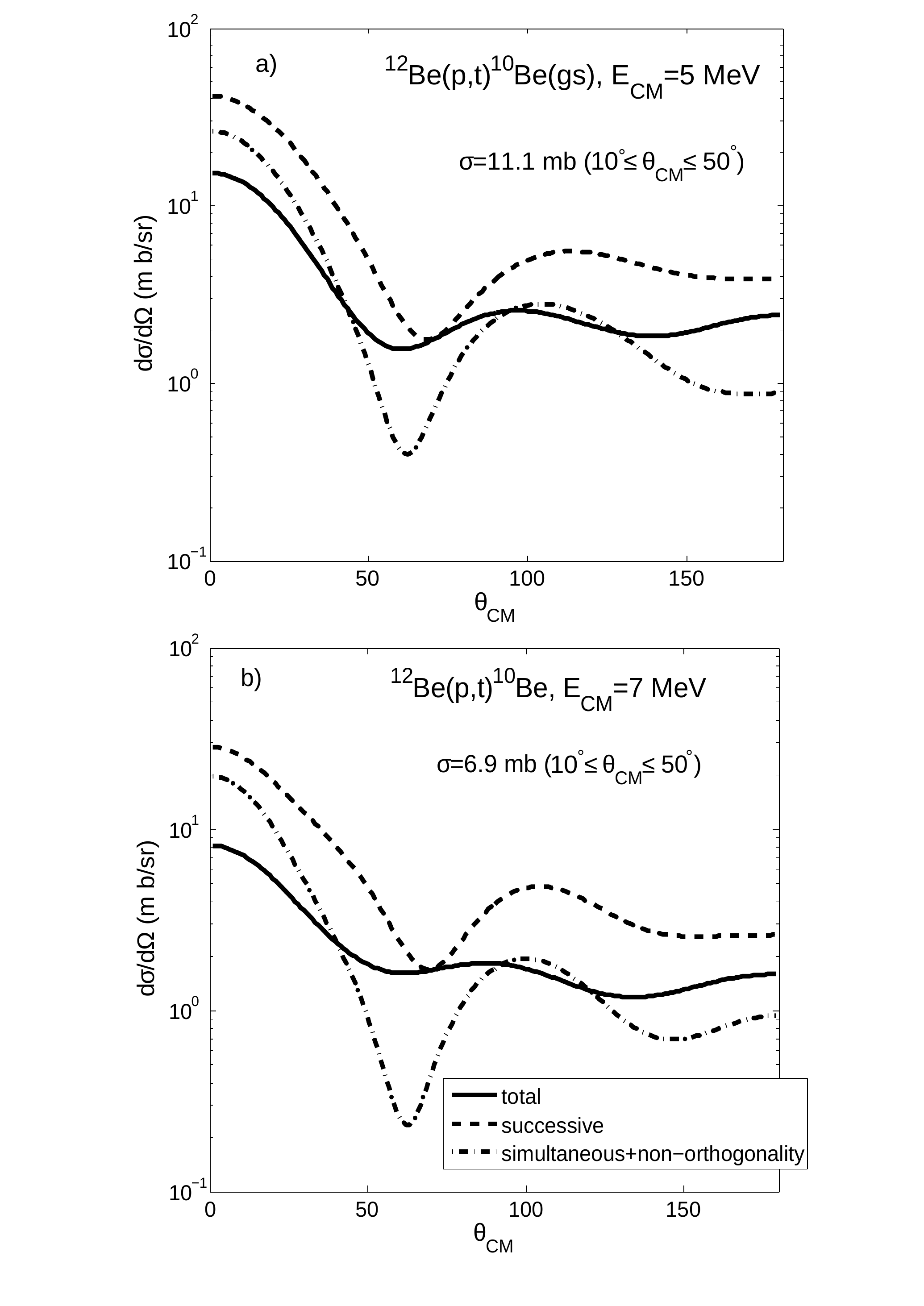}}
\caption{Absolute differential cross section associated with the reaction ${}^{12}\textrm{Be} (p,t) {}^{10}\textrm{Be} (gs)$ at two bombarding energies (a) has already been displayed in Fig. \ref{fig12}). 
Also given are  the integrated values of the cross section in the angular range $1 0^{\circ} \leq \theta_{CM} \leq 50^{\circ}$, as well as the different contributions to the total differential cross section.} \label{fig15}
\end{figure}

\begin{figure}
\centerline{\includegraphics*[width=.65\textwidth,angle=0]{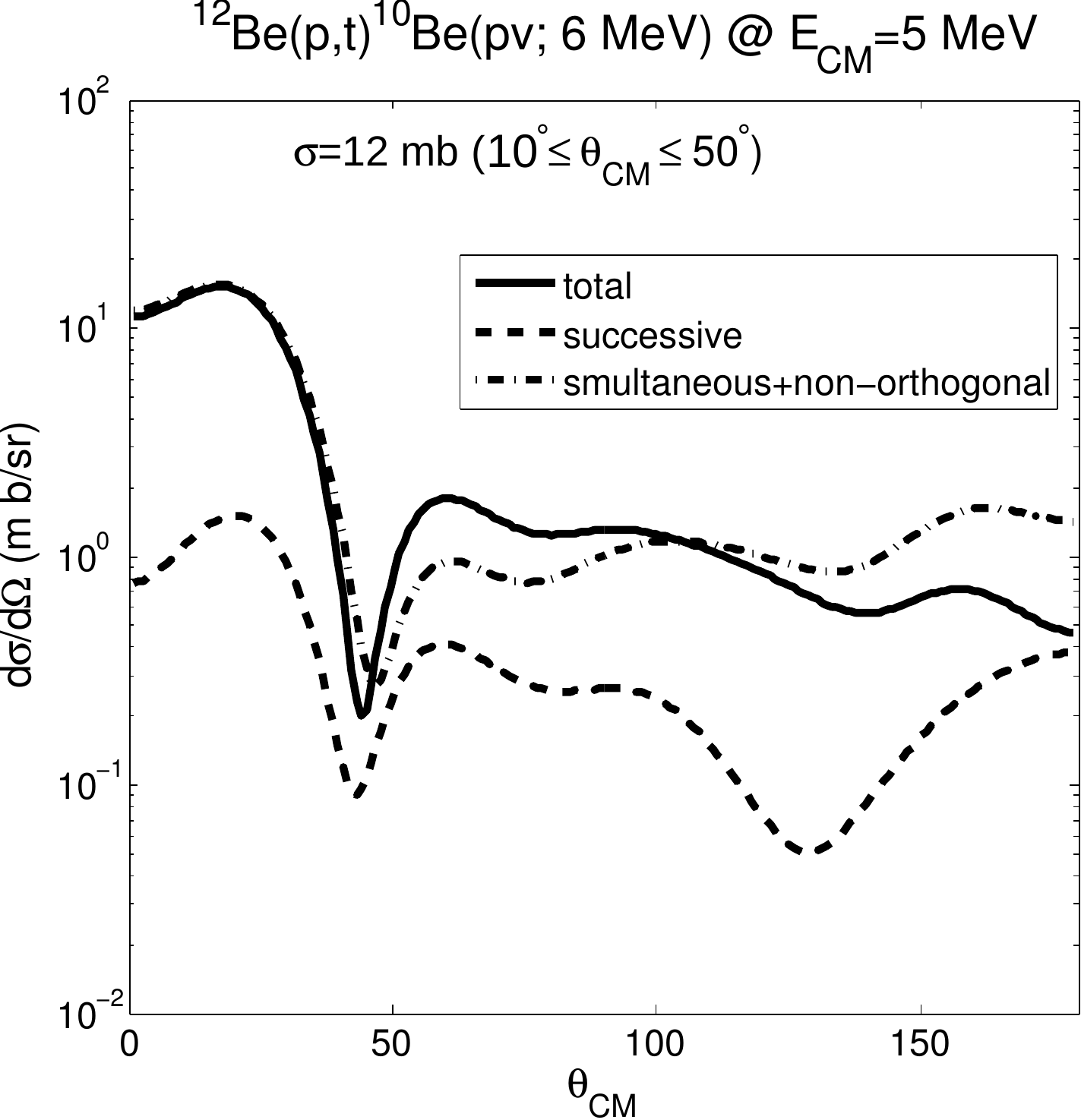}}
\centerline{\includegraphics*[width=.65\textwidth,angle=0]{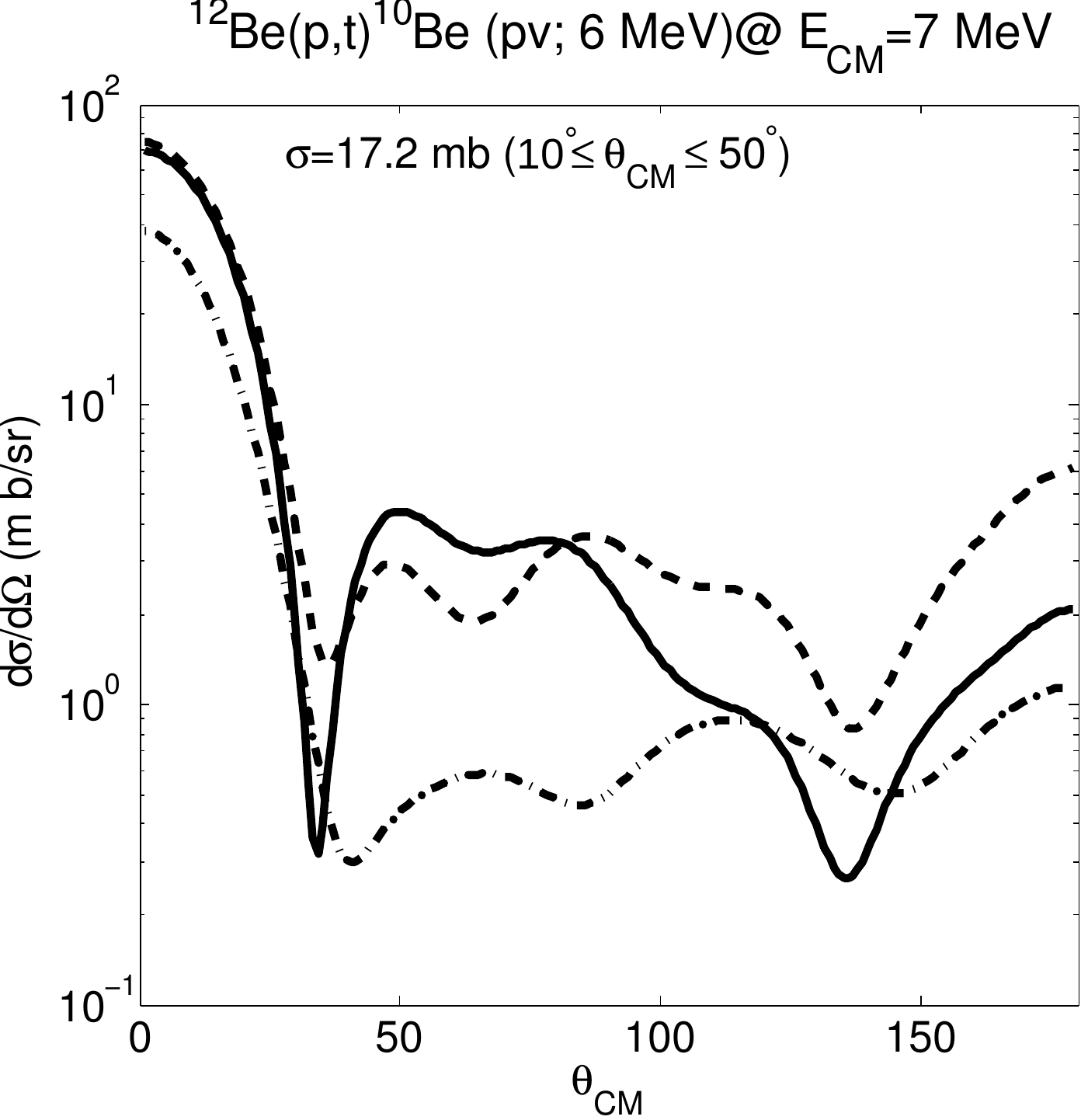}}
\caption{Absolute differential cross section associated with the reaction ${}^{12}\textrm{Be} (p,t) {}^{10}\textrm{Be} (pv; 6 \textrm{MeV})$ at two bombarding energies and  integrated values of the cross section in the angular range $10^{\circ} \leq \theta_{CM} \leq 50^{\circ}$.
Also displayed are the various contributions to the total differential cross section.} \label{fig16}
\end{figure}
\begin{figure}
\centerline{\includegraphics*[width=.95\textwidth,angle=0]{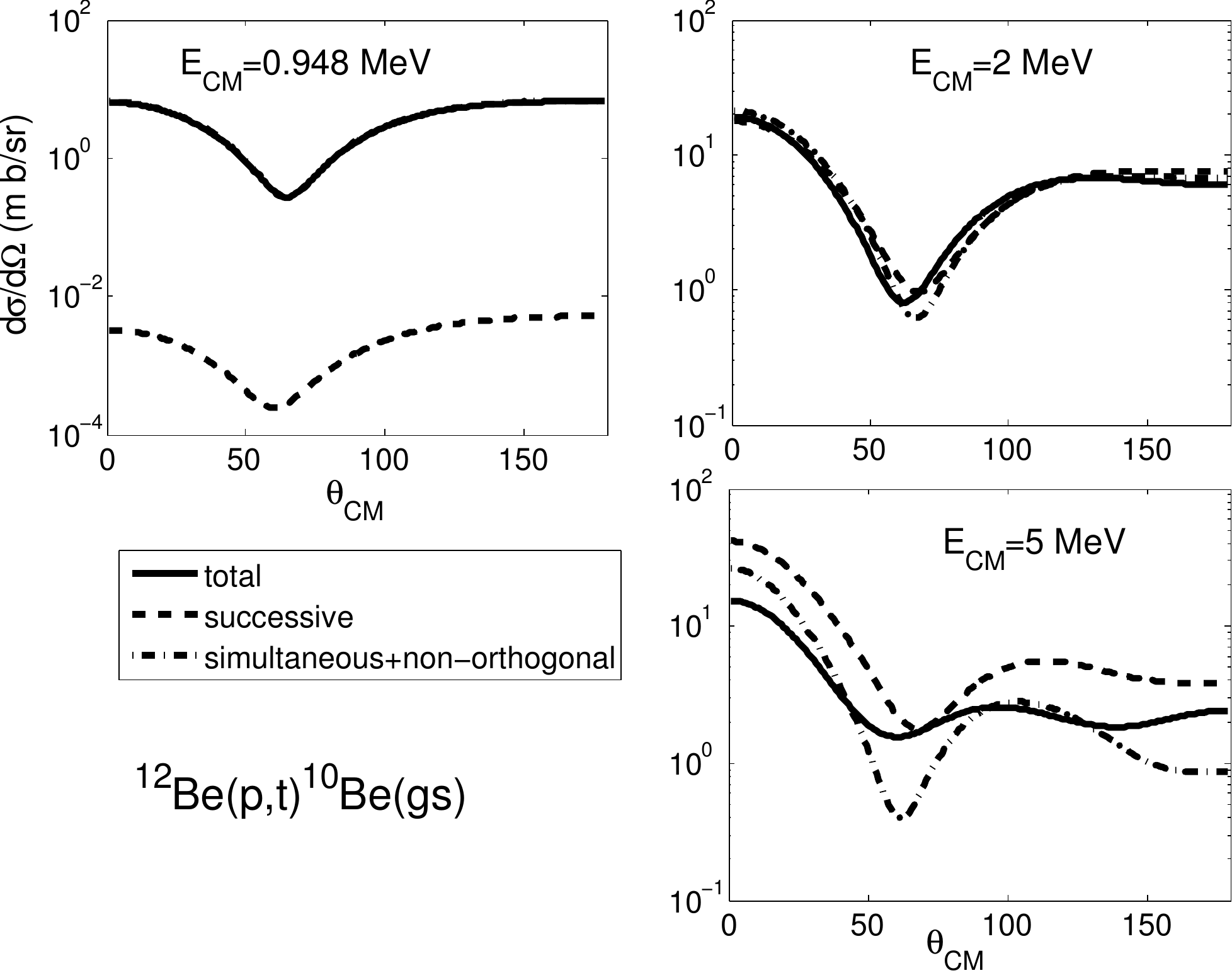}}
\caption{Absolute differential cross section associated with the reaction ${}^{12}\textrm{Be} (p,t) {}^{10}\textrm{Be} (gs)$ at three bombarding energies. Also displayed are the various 
contributions to the total differential cross section.} \label{fig17}
\end{figure}


\clearpage

\begin{table}
\begin{center}
\begin{tabular}{ c c c c c c }
\color{red} 
            &$1s_{1/2}$&$1p_{3/2}$&          &$1p_{1/2}$&$2s_{1/2}$ \\
\cline{2-3} \cline{5-6}
$\epsilon_i$ [MeV]& $-16.07$ & $-4.06$ &     $\epsilon_k$ [MeV]     & $0.025$ &$0.5$   \\
\hline 
$X^{r}$   &  0.058   &  1.049   & $Y^{r}$& 0.244   & 0.211      \\
\hline
\end{tabular}
\caption{\protect RPA wavefunction of the pair removal mode of ${}^{9}$Li. Single--particle energies were deduced 
from experimental binding energy differences, while $W_1(\beta= 2)=3.36$ MeV, $W_1(\beta=-2)=2.3$ MeV. The results obtained making use of a single $G$ or of two pairing 
coupling constants to take care of the difference of overlaps between core-core, core-halo and halo-halo single-particle wavefunctions are, in the present
case, essentially the same (see App. A, in particular Sect. $b$ of this appendix).   }
\label{Li9_PV}
\end{center}
\end{table}

\begin{table}
\begin{center}
\begin{tabular}{|c|c|}
 \multicolumn{2}{c}{$\sigma$($^{11}$Li(gs)  $\to$ $^9$Li (gs)) $(m$b)}\\
\hline
Theory & Experiment \\
\hline
  6.1 &  5.7 $\pm$ 0.9\\
\hline
\multicolumn{2}{c}{$\sigma$($^{7}$Li(gs)  $\to$ $^9$Li (gs)) $(m$b)}\\
\hline
 14.3 & 14.7 $\pm$ 4.4  \\
\hline
\end{tabular}
\caption{\protect\footnotesize Integrated two-neutron differential transfer cross sections, in the center of mass angular range \mbox{20$^{\circ}$--154.5$^{\circ}$} for the reaction $^{11}$Li$(^{1}H,^{3}H)^9$Li$(gs)$ and in the center of mass angular range \mbox{10$^{\circ}$--109$^{\circ}$}  for the reaction $^{7}$Li$(t,p)^9$Li$(gs)$ in which the measurements have been made, in comparison with the data (\cite{Tanihata:08,Young:71}).}\label{tab2}
\end{center}
\end{table}

\begin{table}
\begin{center}
\begin{tabular}{ c c c c c c c}

            &$1s_{1/2}$&$1p_{3/2}$&            &$2s_{1/2}$&$1p_{1/2}$&$1d_{5/2}$ \\
\cline{2-3} \cline{5-7}
$\epsilon_k$ [MeV] & $-19.55$  & $-6.81$ &$ \epsilon_i$ [MeV]& $-0.50$ &$-0.18$     & 1.28\\
\hline 
$X^{r}$   &  0.128   &  1.076   & $Y^{r}$  & 0.232   & 0.214         & 0.272 \\
\hline
$Y^{a}$   &  0.080   &  0.402   & $X^{a}$  & 0.727   & 0.588         & 0.543 \\
\hline
\end{tabular}
\caption{\protect RPA wavefunctions of pair removal and addition $0^{+}$ modes of ${}^{10}$Be, 
that is,  of the ground state of $^8$Be 
 and ${}^{12}$Be. The single--particle energies were deduced from experimental binding and excitation energies, and making use of the coupling constants $G_{cc} = 2$ MeV and $G_{hc} = G_{hh} = 0.68$ MeV (see App. A, in particular Sect. $b$). }
\label{Be10_PV}
\end{center}
\end{table}

\begin{table}
\begin{center}
\begin{tabular}{c | c c | c c | c | c | c }
& \multicolumn{2}{c|}{$\varepsilon_i$ [MeV]}& \multicolumn{2}{c|}{$\varepsilon_k$ [MeV]}& $E$ [MeV]& \; \; X \; \;    &\;  \; Y \;  \; \\
\hline
n \; & \; $p_{3/2}$   & $-8.6$          \;    & \; $p_{1/2}$ \;& $-3.6$              & 5.0                     &$0.90$ &$0.31$  \\
\hline 
p \; & \; $p_{3/2}$   & $-14.9$         \;    & \; $p_{3/2}$ \;& $-14.9$             & 7.6                     &$-0.56$&$-0.29$  \\
\hline
p \; & \; $p_{3/2}$   & $-14.9$         \;    & \; $p_{1/2}$ \;& $-7.8$              & 12.2                    &$ 0.24$&$ 0.16$  \\
\hline
n \; & \; $p_{3/2}$   & $-8.6$          \;    & \; $f_{7/2}$ \;& $16.5$              & 25.1                    &$-0.12$&$-0.10$  \\
\hline 
p \; & \; $s_{1/2}$   & $-29.0$         \;    & \; $d_{5/2}$ \;& $-1.6$              & 28.4                    &$-0.10$&$-0.08$  \\
\hline 
n \; & \; $p_{3/2}$   & $-8.6$          \;    & \; $f_{7/2}$ \;& $8.8$               & 17.5                    &$-0.10$&$-0.07$  \\
\hline 
n \; & \; $s_{1/2}$   & $-21.1$         \;    & \; $d_{5/2}$ \;& $2.2$               & 28.4                    &$-0.09$&$-0.07$  \\
\hline 
\end{tabular}
\caption{ Wavefunction of the lowest $2^{+}$  vibrational state (phonon) of $^{10}$Be (obtained from a QRPA calculation,
making use of a quadrupole separable interaction and a value of the proton pairing gap of $\Delta_p = 3.8$ MeV 
while setting  $\Delta_n =0$). The calculated energy 
 and the $B(E2)$ transition strength of the low lying $2^+$ 
are 2.5 MeV and 49.6 $e^2$fm$^4$ respectively. These results are to be compared with the experimental values of
 3.3 MeV and 52 $e^2$fm$^4$. The quantities $\varepsilon_i$ and $\varepsilon_k$ indicate the energy of the hole and  of the particle 
states  respectively for either protons (p) or neutrons (n). $E$  
denotes  the  associated two-quasiparticle energies, while $X$ and $Y$ are the QRPA amplitudes of the mode.}
\label{tab.Be10_2}
\end{center}
\end{table}

\clearpage

\appendix

\section{Pairing and surface vibrations for pedestrians}
\label{Appendix.A}
\begin{figure}
\centerline{\includegraphics*[width=.95\textwidth,angle=0]{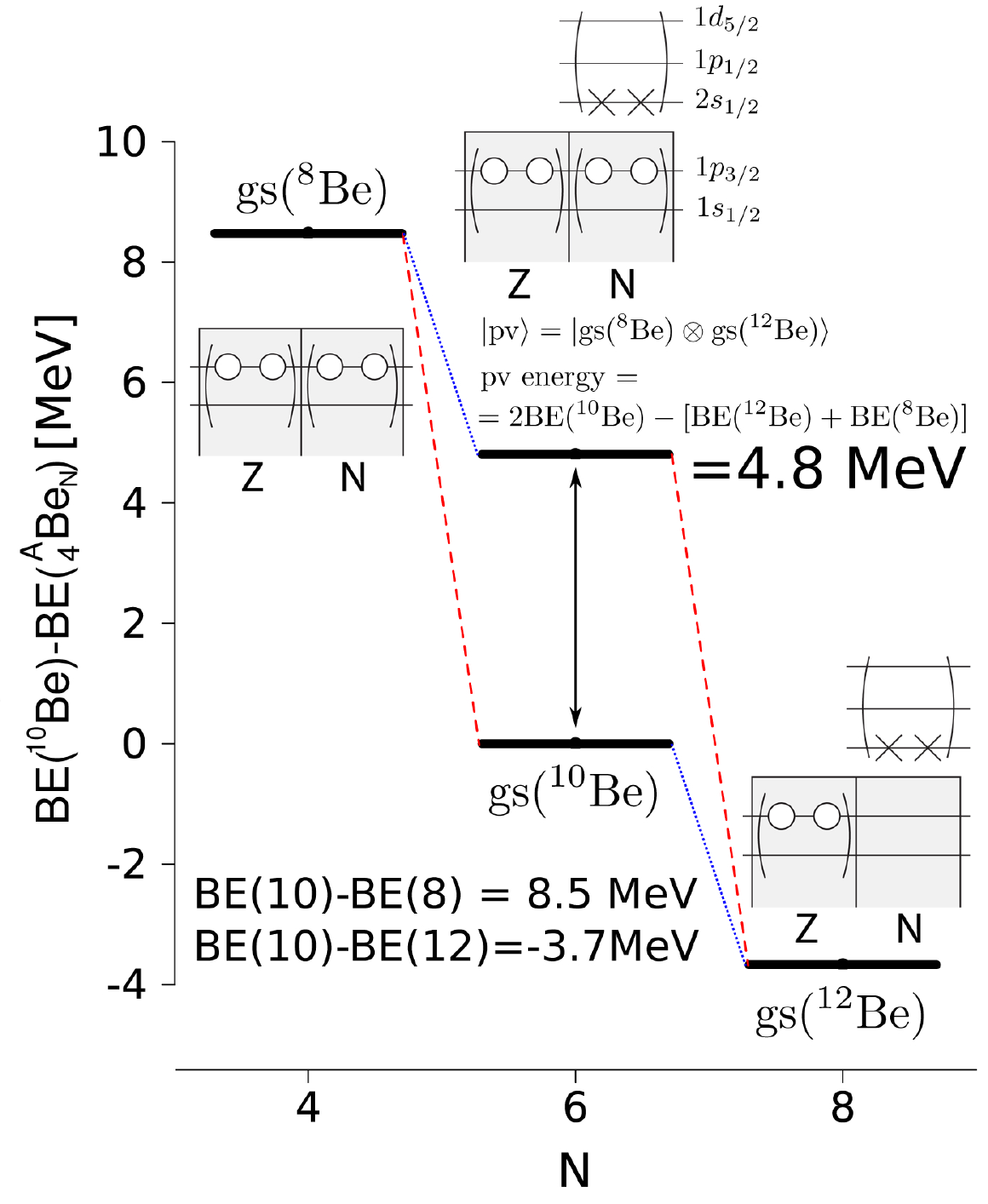}}
\caption{In the harmonic approximation, the two-phonon pairing vibration state of $^{10}$Be is the state $\vert gs({}^{8}$Be$) \otimes gs(^{12}$Be$)\rangle$, its energy being given by the sum of the relative binding energies (BE) of ${}^{8}$Be and ${}^{12}$Be as indicated (see also Fig. \ref{fig12}).} \label{fig21}
\end{figure}

A pairing vibration is a harmonic mode which changes the number of particles in $+2$ or $-2$, and can be observed, around closed shell nuclei, as strong transitions in two-particle transfer processes. For example around ${}^{208}$Pb, where the monopole, pair addition mode (a) is the ground state of ${}^{210}$Pb and the pair removal mode (r) is the ground state of ${}^{206}$Pb \cite{Broglia:73}. The two-phonon pairing vibration of ${}^{208}$Pb ($^{208}$Pb($pv;0^+,4.9$ MeV)) is thus a two-particle two-hole state, product of these two ground states, and thus expected, in the harmonic approximation, at an energy of 4.9 MeV above the ${}^{208}$Pb ground state (see \cite{Brink:05} and refs. therein, see also \cite{Bes:66}). Consequently, and within the harmonic approximation, in the reaction  ${}^{206}$Pb$(t,p){}^{208}$Pb($pv$) one 
excites the pair addition mode, 
with quite similar  ( Q-value, angular distribution, cross section, etc. ) observables to those 
associated with the ${}^{208}$Pb$(t,p){}^{210}$Pb$(gs)$ reaction. Of notice that the pair addition mode $\vert gs({}^{210}\textrm{Pb}) \rangle$ can be viewed as a correlated two-particle state, linear combination of $(j^{2}(0))^{2}$ 
 configurations, $j$ denoting 
valence orbitals
($2g_{9/2}$, $1i_{11/2}$, $3d_{5/2}$, ...) lying above the Fermi energy. 
Now, because of ground state correlations, the (RPA) wavefunction of the pair addition mode 
 contains  also components of the type $(j^{-2}(0))$, i.e. corresponding to the correlation of two-hole-states in the occupied valence orbitals lying below the Fermi energy ($3p_{1/2}$, $2f_{5/2}$, $3p_{3/2}$, ...). These correlations add coherently to the previous ones in binding the two neutrons (Cooper pair partners) of the pair addition mode
to the $^{208}$Pb core.  Similar arguments apply to the removal mode, but where the role of holes and particles  are exchanged. 
For an example see Table III, where the two-nucleon transfer spectroscopic amplitudes (see e.g. App. 2, ref. \cite{Broglia:73}), proportional to the RPA wavefunction amplitudes of ${}^{10}$Be, are collected.
In the case of ${}^{10}$Be one expects the two-phonon pairing vibrational 2p--2h state at an energy of 4.8 MeV above the ground state \footnote{It is of notice the totally accidental similarity between the harmonic predicted value of the two-phonon pairing vibrational state in $^{208}$Pb (4.9 MeV) and in $^{10}$Be (4.8 MeV).}, an energy which emerges directly from two differences in binding energies, namely $[(BE({}^{10}\textrm{Be})-BE({}^{8}\textrm{Be})] - [BE({}^{12}\textrm{Be})-BE({}^{10}\textrm{Be})] = 4.8$ MeV (see Fig. \ref{fig21}; see also Fig \ref{fig12} where the same spectrum is displayed, but subtracting to $BE({}^{10}\textrm{Be}))-(BE({}^{A}_4\textrm{Be}_N)$ a linear term in $N$, i.e. essentially eliminating the A-dependent term of Weisz\"{a}ker mass formula, so as to be able to compare the excitation energies of ${}^{8}\textrm{Be}(gs)$ and ${}^{12}\textrm{Be}(gs)$ with respect to ${}^{10}\textrm{Be}(gs)$.

In the reaction ${}^{210}$Pb$(p,t){}^{208}$Pb one would also excite, for example, the particle-hole quadrupole vibrational state of ${}^{208}$Pb, that is ${}^{210}$Pb$(p,t){}^{208}$Pb$(2^{+}; 4.07\textrm{MeV})$, $2^{+}$ which decays electromagnetically to the ground state of ${}^{208}$Pb with a B(E2) value corresponding to approximately 5 Weisskopf single-particle units. On one hand this state, which is very interesting by itself (in particular when exchanged between the two outer neutrons in the ${}^{210}$Pb ground state, it contributes to the pairing correlations of the pair addition mode), has nothing to do with the pairing vibrational spectrum, in the same way in which the $2^{+}$ state of ${}^{10}$Be
 is not related to the pairing spectrum around magic number $N=6$.
On the other hand, the residual interaction correlating the particle and the hole (linear combination of particle-hole excitations 
$(j_k,j^{-1}_{i})_{2^{+}}$ of the quadrupole vibration) pulls the hole close to the particle. This is reason why collective surface vibrations display both enhanced electromagnetic transition probabilities, i.e. B(E$\lambda$) values, and enhanced two-nucleon transfer cross section.
 The main difference between surface ((p-h)-like) vibrational modes and pairing vibrational modes lies,
as far as two-nucleon transfer processes are concerned, in the role played by the ground state correlations. While those associated with the pair addition and pair subtraction modes enhance the two-nucleon transfer cross sections, those associated with particle-hole excitations, while increasing the electromagnetic decay probabilities, decrease the two-nucleon transfer cross section (see e.g. \cite{Broglia:71a,Broglia:72b}). This competition between ((pp)-(hh))- and (ph)- ground state correlations in nuclei, is at the basis of the studies of pairing in nuclei in terms of the competition between deformed and spherical shapes (see e.g. \cite{Mottelson:76,Mottelson:62,Bes:69,Bayman:60b} and Fig. 22).

\begin{figure}
\centerline{\includegraphics*[width=.95\textwidth,angle=0]{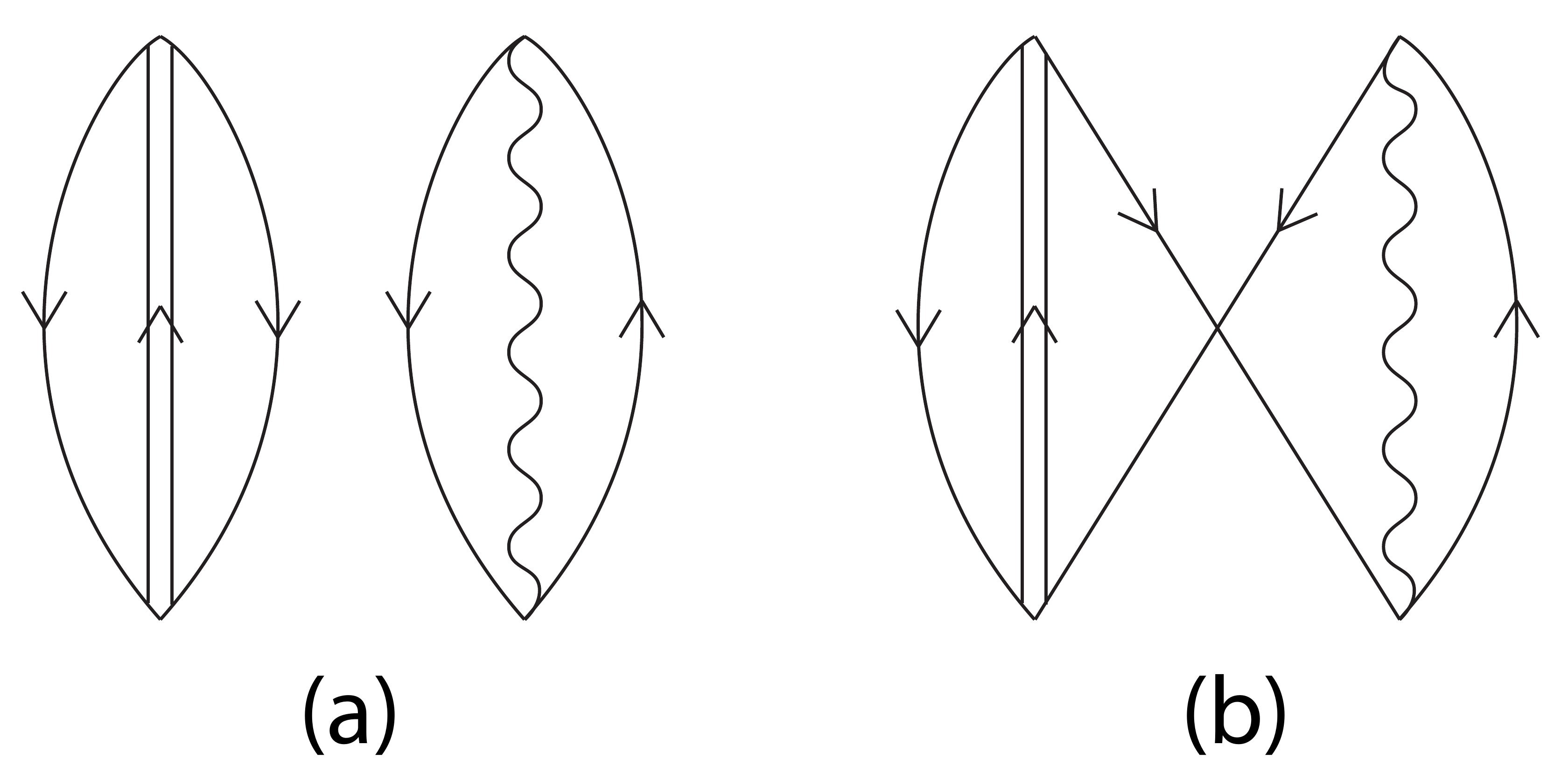}}
\caption{(a) Zero Point Fluctuations (ZPF) associated with a pair addition mode and a surface ((p-h)-like) vibration. (b) Pauli principle correction of (pp)-like ZPF in presence of (ph)-like ZPF.} \label{fig20}
\end{figure}

\subsubsection{Calculation of energy and wavefunctions of pairing vibrations}

The two-phonon, $0^{+}$ of pairing vibrational state, of the closed shell system $N_0$ can be written, in the harmonic approximation as (cf. \cite{Brink:05} Ch. 5, see also \cite{Broglia:73,Bes:66} and refs. therein)
\begin{equation}
 \vert pv \rangle = \vert gs(N_0+2) \rangle \; \vert gs(N_0-2) \rangle,
\end{equation}
where 
\begin{equation}
 \vert gs(N_0+2) \rangle = \Gamma_{1}^{\dagger} (\beta = +2)\vert 0 \rangle,
\end{equation}
and
\begin{equation}
 \vert gs(N_0-2) \rangle = \Gamma_{1}^{\dagger} (\beta = -2)\vert 0 \rangle,
\end{equation}
are the RPA ground state correlated states of the $N_0+2$ and the $N_0-2$ system respectively.
The pair addition and pair removal creation operator phonons are written as
\begin{equation}
 \Gamma_{n}^{\dagger} (\beta = +2)\vert 0 \rangle = \sum_k X^{a}_{n}(k)\Gamma^{\dagger}_{k} + \sum_i Y^{a}_{n}(i)\Gamma_{i},
\label{eq.A4}
\end{equation}
where
\begin{equation}
 \Gamma^{\dagger}_{k}=\left[a_k^{\dagger}a_k^{\dagger} \right]_{0},
\end{equation}
creates a pair of particles coupled to angular momentum zero in levels with energy $\varepsilon_k > \varepsilon_F$ ($k\equiv n_k l_k j_k$), while
\begin{equation}
 \Gamma_{i}=\left[a_i a_i \right]_{0},
\end{equation}
creates a pair of holes in the occupied orbitals, i.e. orbitals with energy $\varepsilon_i \leq \varepsilon_F$ ($i\equiv n_i l_i j_i$). The index $n$(=1, 2, ...) labels the lowest, the first excited, etc. states. In what follows we concentrate on the $n=1$ (ground) modes, otherwise explicitly mentioned.

\begin{figure}
\centerline{\includegraphics*[width=.95\textwidth,angle=0]{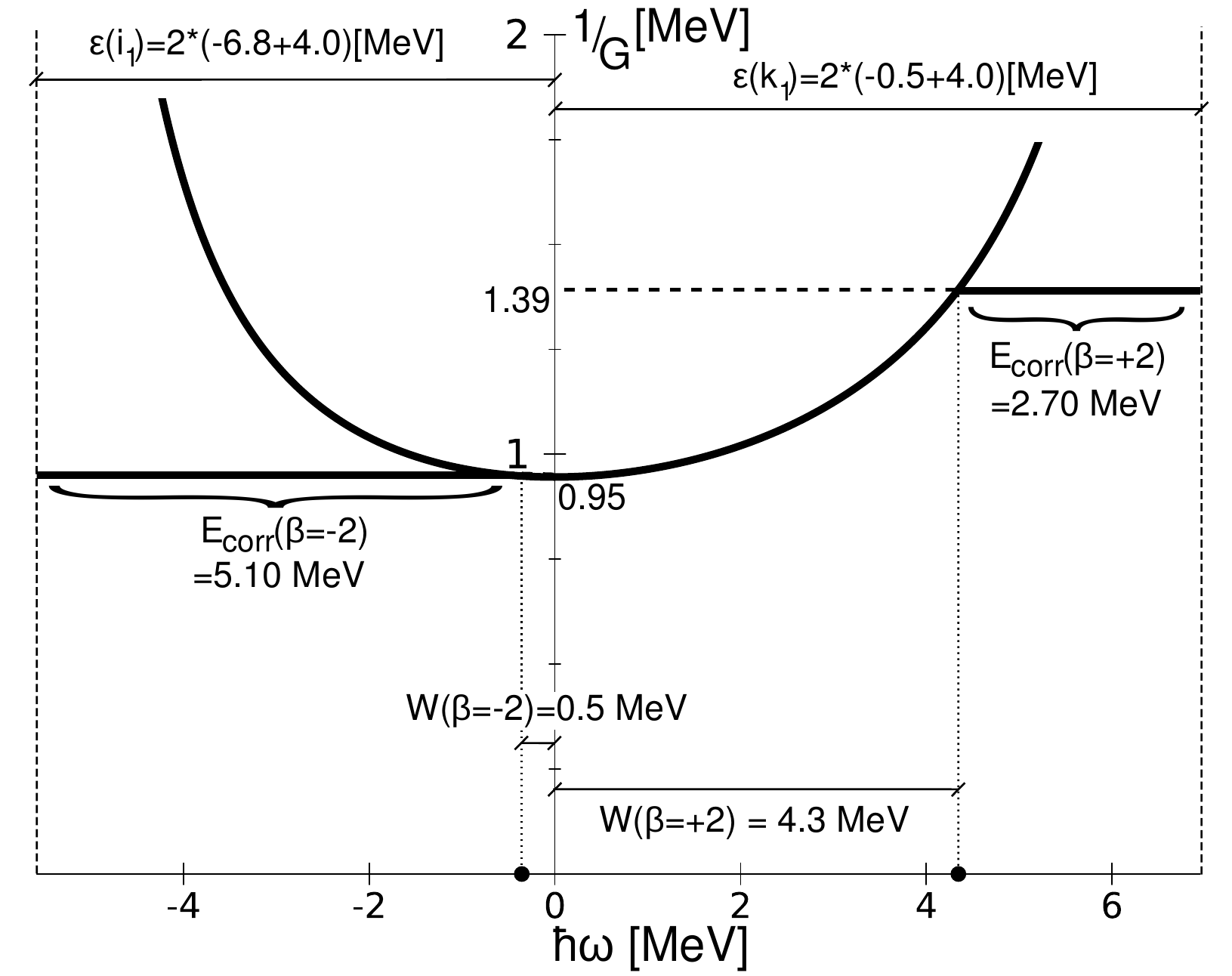}}
\caption{Dispersion relation and parameters characterizing the pairing vibrational spectrum of $^{10}$Be. The quantity $E_{corr}(\beta=+2)= (BE(^{12}\textrm{Be})-BE(^{10}\textrm{Be}))-2(BE(^{11}\textrm{Be})-BE(^{10}\textrm{Be})) = (68.7-65.0)-2(65.5-65.0) = 2.7$ MeV while $W_{1}(\beta=+2) = 2\varepsilon(k_1)-E_{corr}(\beta=+2)=2(-0.5+4.0)-2.7 =4.3$ MeV, 
with $\varepsilon(k_1)=S_n({}^{11}Be)-\varepsilon_F=-0.5+4.0$ MeV, $S_n({}^{11}Be)$ being the experimental neutron separation energy in $^{11}$Be. Thus $\varepsilon(k_1)$ is the energy of the first unoccupied state of $^{11}$Be, calculated respect to the Fermi energy $\varepsilon_F=-4.0$ MeV (for details see App. B). Similar expressions are valid for the pair removal mode, namely $E_{corr}(\beta=-2)=5.10$ MeV, $W_{1}(\beta=-2) = - 2\varepsilon(i_1)-E_{corr}(\beta=-2)=0.5$ MeV,
 where  $\varepsilon(i_1) = -6.8 + 4.0 $ MeV is 
the energy of the first occupied state respect to  the Fermi energy.
The inverse strengths $1/G(\pm 2)$ intersect the dispersion relation at values of $\hbar \omega$ leading to $W_{1}(\beta=\pm 2)$.}\label{fig22}.
\end{figure}

Assuming $\Gamma^{\dagger}_{k}$ and $\Gamma^{\dagger}_{i}$ display boson commutation relations, one can linearize the pairing Hamiltonian obtaining the dispersion relation for the lowest ($n=1$) modes (see Fig. \ref{fig22} for the case of the pairing vibrational spectrum around $^{10}$Be)
\begin{equation}
 S(\beta = \pm 2) = \frac{1}{G(\beta = \pm 2)},
\label{eq.A7}
\end{equation}
where (note that $\beta = \pm 2$ is simplified in what follows into $\pm 2$),
\begin{subequations}
\begin{equation}
 S(\pm 2) = S_{h} (\pm 2) + S_{c} (\pm 2),
\end{equation}
with
\begin{equation}
S_{h} (\pm 2)  = \sum_k \frac{\Omega_{k}}{2(\varepsilon_k - \varepsilon_F) \mp W_{1}(\pm 2)}
\end{equation}
\end{subequations}
and
\begin{equation}
 S_{c} (\pm 2)  = \sum_i \frac{\Omega_{i}}{2(\varepsilon_F - \varepsilon_i) \pm W_{1}(\pm 2)}
\end{equation}
while $\Omega = j + 1/2$ measure the pair degeneracy of the single particle orbital $j$.

The RPA amplitudes appearing in (\ref{eq.A4}) are defined as
\begin{subequations}
 \begin{equation}
   X^{a}_1(k)= \frac{ \frac{1}{2} \sqrt{\Omega_k}\; \Lambda_1(+2)}{2(\varepsilon_k-\varepsilon_F)-W_{1}(+2)} \quad, \quad
   Y^{a}_1(i)=-\frac{ \frac{1}{2} \sqrt{\Omega_i}\; \Lambda_1(+2)}{2(\varepsilon_F-\varepsilon_i)+W_{1}(+2)},
 \end{equation}
and
 \begin{equation}
   X^{r}_1(i)= \frac{ \frac{1}{2} \sqrt{\Omega_i}\; \Lambda_1(-2)}{2(\varepsilon_F-\varepsilon_i)-W_{1}(-2)} \quad, \quad
   Y^{r}_1(k)=-\frac{ \frac{1}{2} \sqrt{\Omega_k}\; \Lambda_1(-2)}{2(\varepsilon_k-\varepsilon_F)+W_{1}(-2)},
 \end{equation}
\end{subequations}

%

The particle-vibration coupling strengths $\Lambda_1(\beta=\pm2)$ are determined from the normalization conditions
\begin{equation}
 \sum_{k}\vert  {X^a_1}(k)\vert^{2}-\sum_{i} \vert Y^{a}_1(i)\vert^{2}  = 1,
\end{equation}
and
\begin{equation}
 \sum_{i}\vert X^{r}_1(i)\vert^{2}-\sum_{k} \vert Y^{r}_1(k)\vert^{2} = 1,
\end{equation}
respectively.

\subsubsection{Extension of the harmonic approximation to the case of more than one pairing coupling constant}

The relation between the pairing coupling constant $G$ and the matrix elements between pure two-particle configurations
$(j^2)_0$ coupled to angular momentum $J=0$, of a $\delta-$force of strength $V_0$ can be written as (see e.g. Eq. (2-24) p.41 ref. \cite{Brink:05}),
\begin{equation}
G = V_0 I,
\end{equation}  
where 
\begin{equation}
I = \int dr \;  r^2  {{\cal{R}}_f}^2(r) {{\cal{R}}_i}^2(r), 
\end{equation}
${\cal{ R}}$ being the radial single-particle wavefunction describing initial ($i$) and final ($f$) configurations. Making 
use of the approximation (constant value of ${\cal R}$ inside the nucleus of radius $R$),
\begin{equation}
{\cal{R}} = \sqrt{\frac{3}{R^3}} \Theta(r-R),
\end{equation}
where 
\begin{equation}
\Theta(r-R) =  \; 1  \quad (r \leq R) \quad ;  \quad  \Theta(r-R)  = \; 0  \quad (r > R),
\end{equation}
one can write 
\begin{equation}
I_{if} =  \frac{R_j^3}{3}\frac{3}{R_f^3} \frac{3}{R^{3}_i},
\label{iif}
\end{equation}
$R_j$ being equal to the smallest value of the radius $R_i$ and $R_f$.

A simple estimate of $V_0$ can be  worked out assuming $R_i = R_f = R$, and using  the standard expression 
for the pairing coupling constant $G \approx $ 25/$A$ MeV. One thus obtains  
\begin{equation}
4 \pi V_0 = \frac{25 {\rm MeV}}{\rho_0} \approx 150 {\rm MeV \; fm^3},
\label{vo}
\end{equation}
where $\rho_0 = A/\textrm{Vol} \approx 0.17$ {\rm fm$^{-3}$} ($\textrm{Vol}=4\pi/3 R^3$), corresponds  to the nuclear saturation density . 

Distinguishing  between core ($c$) and halo ($h$) single-particle wavefunctions and thus radii, one obtains two 
different expressions for (\ref{iif}), namely   
\begin{equation}
I_{cc} = \frac{4 \pi}{\textrm{Vol} (c)},
\end{equation}
and 
\begin{equation}
I_{hh} = I_{hc} = \frac{4 \pi} {\textrm{Vol} (h)}.
\end{equation}
Making use of Eq. (\ref{vo}) one can write 
\begin{equation}
G_{cc} = \frac{150 {\rm MeV  \; fm}^3}{\textrm{Vol} (c)}
\end{equation}
and
\begin{equation}
G_{hh} = G_{hc} =  \frac{150 {\rm MeV \;  fm}^3}{\textrm{Vol} (h).}
\end{equation}
Using the parametrization of the nuclear radius according to Eq. (2-181) of ref. \cite{Bohr:69}, one obtains $R_c = 1.27 \times  8^{1/3}$ fm = 2.54 fm.
leading to $\textrm{Vol} (c)  \approx 70$ fm$^{3}$. Thus 
\begin{equation}
G_{cc} \approx 2 \; {\rm MeV}.
\end{equation}
Let us now carry out a simple estimate of $G_{hh}$, assuming that $R_h$ is about 35\% larger than $R_c$, a value  in line with the observations 
in light halo nuclei. In this case, $\textrm{Vol} (h) \approx 170$fm$^{3}$.  Thus
\begin{equation}
G_{hh} = G_{hc} \approx 0.9 \;  {\rm MeV}.
\end{equation}

In the present case one can distinguish three terms in the pairing Hamiltonian, namely
\begin{equation}
 H_p = H_p(h)+H_p(c)+H_p(hc),
\end{equation}
where
\begin{equation}
 H_p (h) = -G_{hh} \sum_{k,k' \atop \varepsilon_k, \varepsilon_{k'} > \varepsilon_F} a^{\dagger}_{k} a^{\dagger}_{\tilde{k}} a_{k'} a_{\tilde{k'}},
\end{equation}
\begin{equation}
 H_p (c) = -G_{cc} \sum_{i,i' \atop \varepsilon_i, \varepsilon_{i'} \leq \varepsilon_F} a^{\dagger}_{i} a^{\dagger}_{\tilde{i}} a_{i'} a_{\tilde{i'}},
\end{equation}
and
\begin{equation}
 H_p (hc) = - G_{ch} \sum_{k,i} (a^{\dagger}_{k} a^{\dagger}_{\tilde{k}} a_{\tilde{i}} a_{i}+ a^{\dagger}_{i} a^{\dagger}_{\tilde{i}} a_{\tilde{k}} a_{k}).
\end{equation}
Linearizing these equations one obtains, instead of the dispersion relation (\ref{eq.A7}), the determinant (within this \textit{formal} context see e.g. App J of ref. \cite{Brink:05})
\begin{equation}
 \begin{vmatrix}
  1 - G_{hh} S_{h}(\pm 2) &     G_{hc} S_{c}(\pm 2) \\
      G_{hc} S_{h}(\pm 2) & 1 - G_{cc} S_{c}(\pm 2)
 \end{vmatrix} = 0,
\label{eq.A26}
\end{equation}
and in the amplitudes (A.10) the substitutions of $\Lambda_1(+2)$ and $\Lambda_1(-2)$ by
\begin{subequations}
 \begin{equation}
  G_{hh} A + G_{hc}B,
\label{eq.A27a}
 \end{equation}
and
\begin{equation}
 G_{cc} C + G_{hc}D,
\label{eq.A27b}
\end{equation}
\end{subequations}
respectively are to be made ($A = \sum_{k} X^{a}_{1}(k)$, $B = \sum_{i} Y^{a}_{1}(i)$, $C = \sum_{i} X^{r}_{1}(i)$ and $D$ \linebreak $=\sum_{k} Y^{r}_{1}(k)$). It is of notice that, in the case in which $G_{cc}=G_{hh}=G_{hc}=G$, (\ref{eq.A26}) becomes
\begin{equation}
 1-G(S_h(\pm 2) + S_c (\pm 2)) = 1- GS(\pm 2) = 0,
\end{equation}
while (\ref{eq.A27a}) and (\ref{eq.A27b}) coincide with $\Lambda_1(+2)$ and $\Lambda_1(-2)$ respectively.

While the results obtained by parametrizing the pair addition and pair removal modes in terms of $G_{hh} = G_{hc}$ and $G_{cc}$ are very different from those using a single G in the case of $^{10}$Be pair removal mode, they lead to rather similar results in the case of pair addition mode  (see Tables  \ref{Be10_PV} and \ref{Be10_PV-2}). This is connected with the fact that the pair addition mode is mainly built on halo-like orbitals.
The differences found between the results of the two calculations in the case of the pair addition and removal modes of $^{9}$Li (namely $^{11}$Li($gs$) and $^{7}$Li($gs$)) is quite small. This is keeping with the fact that in this case the $d_{5/2}$ orbital plays a rather minor role in the halo component of the corresponding wavefunctions. In other words, the basic halo subspace more than doubles going from Li ($s_{1/2},p_{1/2}$) to Be($s_{1/2},p_{1/2},d_{5/2}$). Returning to the pair removal mode of $^{10}$Be, it is seen from Fig. \ref{fig23}, that the predicted absolute cross section associated with the reaction $^{10}$Be($p,t$)$^{8}$Be($gs$) ($E_{CM} = 7$ MeV) and the integrated between $10^{\circ}$ and $50^{\circ}$ increases, from the value of 16 mb, to an (unrealistic) value close to the geometric cross section ($\sigma = \pi R^2 (^{10}\textrm{Be}) \sim \pi (2.5180 \pm 0.0114 fm)^2 \approx 200 \textrm{mb}$).

\begin{table}
\begin{center}
\begin{tabular}{ c c c c c c c}

            &$1s_{1/2}$&$1p_{3/2}$&            &$2s_{1/2}$&$1p_{1/2}$&$1d_{5/2}$ \\
\cline{2-3} \cline{5-7}
$\epsilon_k$ [MeV] & $-19.55$  & $-6.81$ &$\epsilon_i$ [MeV]& $-0.50$ &$-0.18$     & 1.28\\
\hline 
$X_{r}$   &  0.237   &  1.993   & $Y_{r}$  & 0.969   & 0.892         & 1.137 \\
\hline
$Y_{a}$   &  0.053   &  0.264   & $X_{a}$  & 0.699   & 0.563         & 0.517 \\
\hline
\end{tabular}
\caption{\protect RPA amplitudes associated with the pair addition and removal modes of $^{10}$Be. The calculation were carried making use of a single pairing coupling constant for the pair addition mode ($G=0.72$ MeV) and for the pair removal mode ($G=1.053$ MeV).}
\label{Be10_PV-2}
\end{center}
\end{table}

\begin{figure}
\centerline{\includegraphics*[width=.85\textwidth,angle=0]{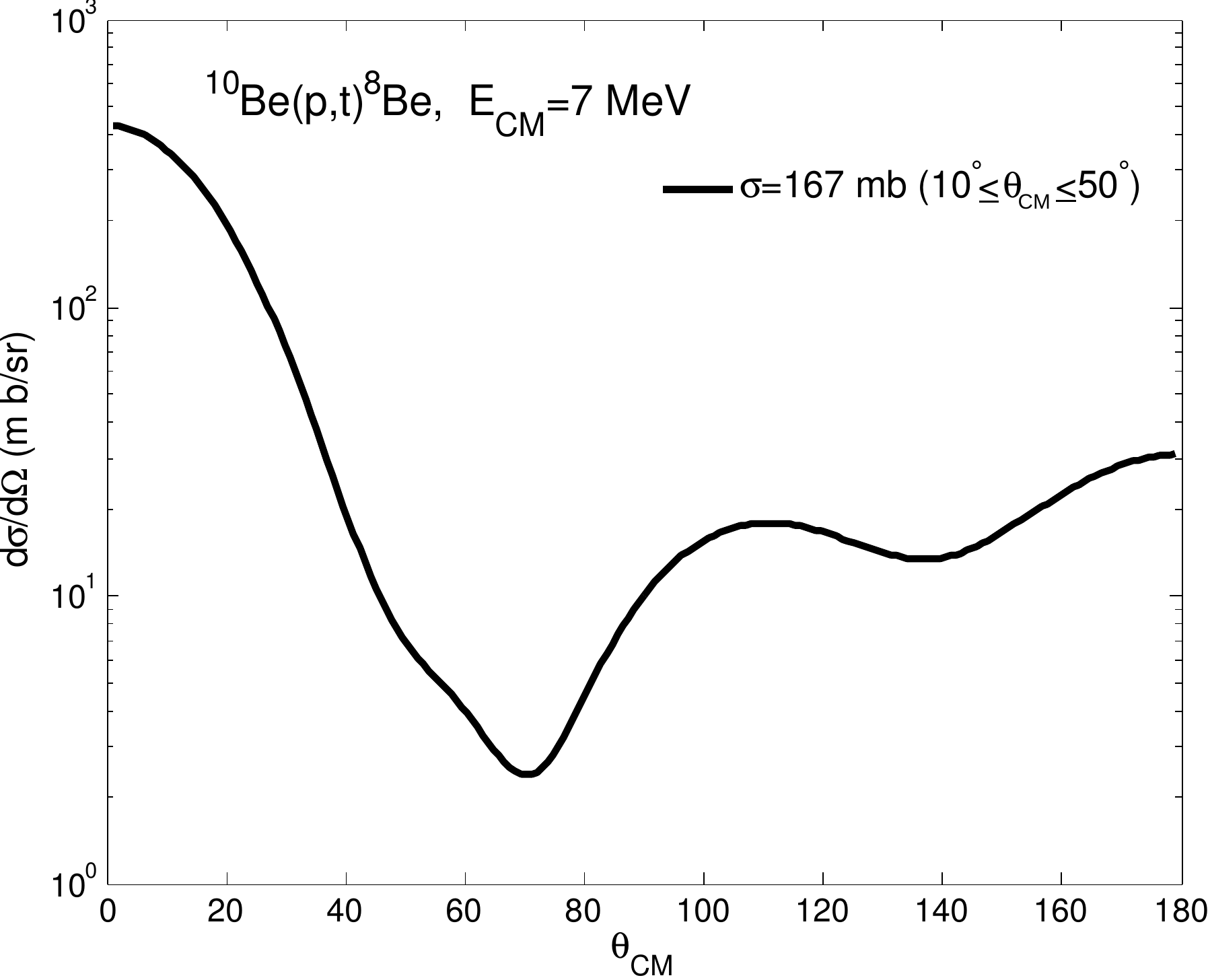}}
\caption{Absolute differential cross section associated with the reaction $^{10}$Be$(p,t)^{8}$Be$(gs)$ at $E_{CM}=7$ MeV making use of the spectroscopic amplitudes calculated making use of a single pairing coupling constant (see Table \ref{Be10_PV-2}).} \label{fig24}
\end{figure}

\section{Fermi energy and Fermi level (chemical potential)}
The Fermi energy $\varepsilon_F$ is the energy of the highest occupied state in a macroscopic ("infinite") system of non-interacting fermions. Under these conditions it is the increase in the ground state energy when one particle is added to the system. At zero temperature it coincides with the chemical potential (Fermi level).

The situation is somewhat more subtle in the case of a finite, many-body system, like the atomic nucleus (see e.g. \cite{Mahaux:85,Dickhoff:05} and refs. therein), as spatial quantization leads to a discrete spectrum. In this case it is useful to refer to the properties of the one-body Green function (e.g. in the Lehmann representation)
\begin{equation}
 G(\vec{r},\vec{r}\; ';\omega)=\sum_i \frac{\varphi_{i}(\vec{r})\varphi^{*}_{i}(\vec{r}\; ')}{\omega-\omega_i-i\eta} + \sum_k \frac{\varphi_{k}(\vec{r})\varphi^{*}_{k}(\vec{r}\; ')}{\omega-\omega_k-i\eta},
\end{equation}
where
\begin{equation}
 \varphi_{i}(\vec{r})=\langle \psi_{i}^{(A-1)} \vert a(\vec{r}) \vert \psi_0^{(A)}\rangle,
\end{equation}
and
\begin{equation}
 \varphi_{k}(\vec{r})=\langle \psi_{0}^{(A)} \vert a(\vec{r}) \vert \psi_i^{(A+1)}\rangle.
\end{equation}
Here $\psi_{i}^{(A-1)}$ is an eigenstate of the Hamiltonian describing the (A-1) particle system, with energy $E_{i}^{A-1}$, $\psi_{k}^{A-1}$ being eigenstate of the $(A+1)$-particle system with energy $E_{k}^{(A+1)}$ and
\begin{equation}
  \omega_k= E^{(A+1)}_{k}-E_{0}^{(A)}, \; \omega_i= E^{(A)}_{0}-E_{i}^{(A-1)},
\label{eq.b4}
\end{equation}
where $E_{0}^{(A)}$ is the energy of $\psi_{0}^{(A)}$. Closure relations yield
\begin{equation}
 \sum_k \varphi_{k}(\vec{r}) \varphi_{k}^{*}(\vec{r}\; ')+\sum_i \varphi_{i}(\vec{r}) \varphi_{i}^{*}(\vec{r}\; ')=\delta(\vec{r}-\vec{r}\; ').
\label{eq.B5}
\end{equation}

\begin{figure}
\centerline{\includegraphics*[width=.75\textwidth,angle=0]{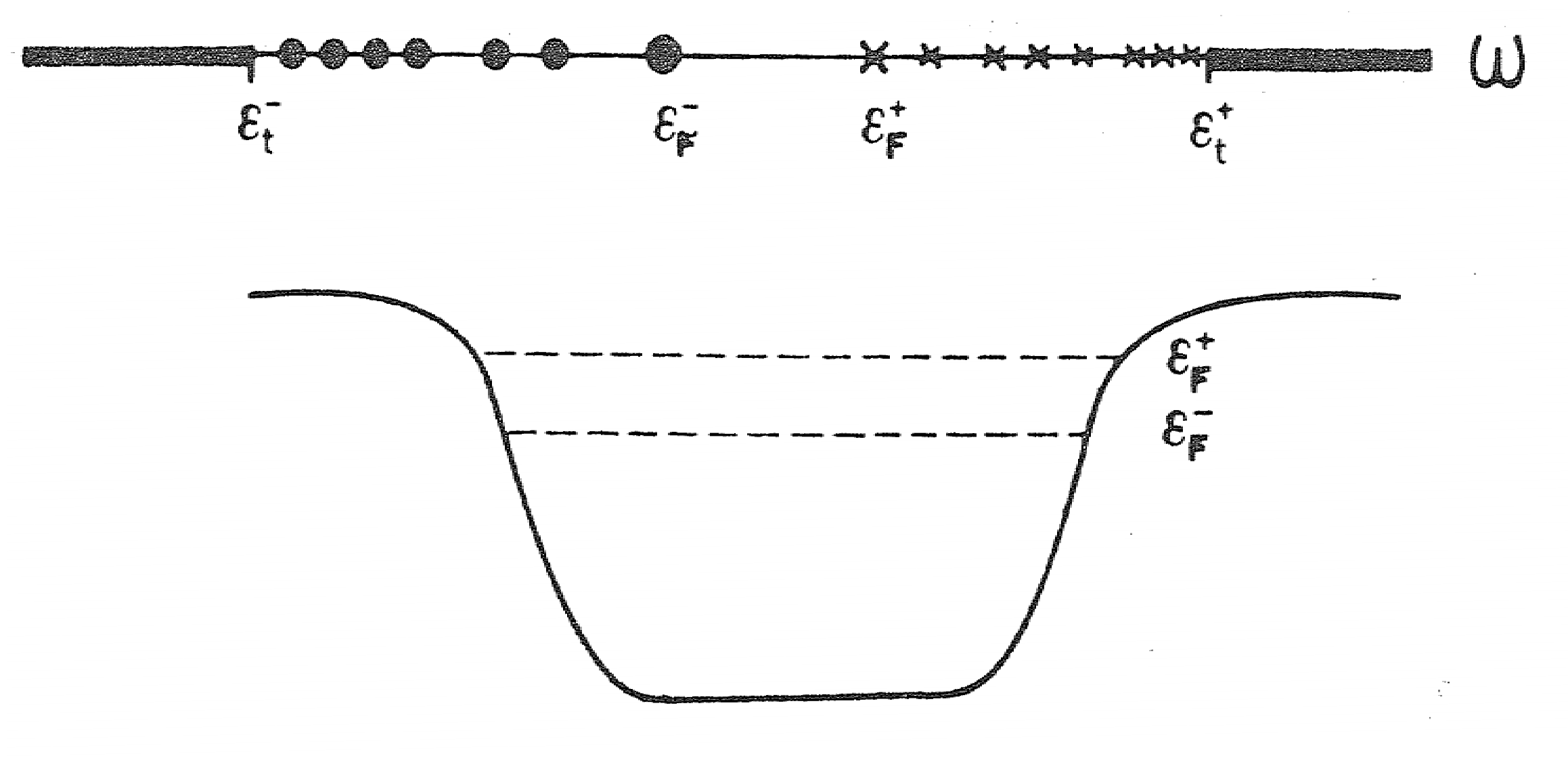}}
\caption{Schematic representation of the singularities of the one-body Green function $G(\vec{r},\vec{r}\;',\omega)$ in the complex $\omega$-plane.}\label{fig23}
\end{figure}

The analytic properties of $G(\vec{r},\vec{r}\;';\omega)$ in the complex $\omega$-plane are schematically represented in Fig. \ref{fig23}. The one-body Green function has a right-hand cut which runs below the real $\omega$-axis between $\omega=\varepsilon_{t}^{+}$ and $\omega=+\infty$. Here, $\varepsilon_{t}^{+}$ is the energy ($\varepsilon(k_2)-\varepsilon_F$) of the lowest excited state $k_2$ of the $(A+1)$ system measured with respect to $\varepsilon_F=(\varepsilon_{F}^{+}+\varepsilon_{F}^{-})/2=(E_{0}^{(A+1)}-E_{0}^{(A-1)})/2$ (see (\ref{eq.B7}) and (\ref{eq.B9}) below), that is, 
 it is the threshold energy  of the first inelastic channel. It is of notice that $\varepsilon(k_2)$ is equal to $-3.25$ MeV and $-0.18$ MeV in the case of ${}^{208}$Pb and ${}^{10}$Be respectively. The one-body Green function has a left-hand cut which runs above the real $\omega$-axis and extends from $-\infty$ to $\varepsilon_{t}^{-}$ with
\begin{equation}
 \varepsilon_{t}^{-}=E_{0}^{(A)}-E_{0}^{(A-2)}.
\label{eq.B6}
\end{equation}
The left-hand cut corresponds to the possibility of the decay of the $(A-1)$ into the $(A-2)$ nucleus. The one-body Green function has discrete poles $\omega_i$ between $\varepsilon_{t}^{+}$ and $\varepsilon_{F}^{-}$, where
\begin{equation}
 \varepsilon_{F}^{-}=E_{0}^{(A)}-E_{0}^{(A-1)}.
\label{eq.B7}
\end{equation}
These poles correspond to bound states of the $(A-1)$ nucleus with excitation energy
\begin{equation}
 E_{x}^{(A-1)}=\varepsilon_{F}^{-}-\omega_i.
\end{equation}
The Green function has also poles $\omega_k$ on the real axis between $\omega=\varepsilon_{F}^{+}$ and $\omega=0$, where
\begin{equation}
 \varepsilon_{F}^{+}=E_{0}^{(A+1)}-E_{0}^{(A)}.
\label{eq.B9}
\end{equation}
These correspond to bound states of the $(A+1)$-nucleus with excitation energy
\begin{equation}
 E_{x}^{(A+1)}=\omega_k-\varepsilon_{F}^{+},
\end{equation}
see Eq. (\ref{eq.b4}).

Let us illustrate this description with two examples. In the case of neutrons in ${}^{208}$Pb one has 
\begin{equation}
 \begin{array}{l l}
  \varepsilon_{t}^{+}= \quad 2.40 \textrm{MeV}, \quad &   \varepsilon_{t}^{-}= -14.09 \textrm{MeV}, \\
  \varepsilon_{F}^{+}= \;-4.03 \textrm{MeV}, \quad &   \varepsilon_{F}^{-}= -7.27 \textrm{MeV}.
 \end{array}
\end{equation}
In the case of the neutrons of ${}^{10}$Be one has
\begin{equation}
 \begin{array}{l l}
  \varepsilon_{t}^{+}= \quad 3.48 \textrm{MeV}, \quad &   \varepsilon_{t}^{-}= - 8.48 \textrm{MeV}, \\
  \varepsilon_{F}^{+}= \;-0.50 \textrm{MeV}, \quad &   \varepsilon_{F}^{-}= -6.82 \textrm{MeV}.
 \end{array}
\end{equation}
Of notice that when dealing with molecular systems, the quantities $\varepsilon_{F}^{-}$ and $\varepsilon_{F}^{+}$ are referred to as Highest Occupied Molecular Orbital (HOMO) and Lowest Unoccupied Molecular Orbital (LUMO).

While in the nuclear case, the prescription (see discussion following Eq. (\ref{eq.B5}))
\begin{equation}
 \varepsilon_F=\frac{1}{2}(\varepsilon_{F}^{+}+\varepsilon_{F}^{-}),
\label{eq.B13}
\end{equation}
has been used for closed shell nuclei \cite{Mahaux:85} (e.g. $\varepsilon_F({}^{208}\textrm{Pb})=-5.6$ MeV, $\varepsilon_F({}^{10}\textrm{Be})=-3.7$ MeV)), a more satisfying definition exists.
In particular when studying nuclear superfluidity. In this case the BCS variational (Lagrange) multiplier used to fix the average number of nucleons, and usually denoted $\lambda$, is the chemical potential of the system (Fermi level) which in this case is equal to the Fermi energy $\varepsilon_F$.

It can be shown that $\lambda/\hbar = \dot{\varphi}$, $\varphi = i \partial/\partial N$ being the variable conjugated to the particle number operator $\hat{N} = \sum_\nu a^{\dagger}_{\nu} a_{\nu}$, while $\dot{\varphi}$ is the rotational frequency in gauge space of the (ground state) pairing rotational band associated with gauge symmetry restoration (Anderson-Nambu-Goldstone mode, see \cite{Brink:05} Ch. 4, and refs. therein). In keeping with the fact that gauge symmetry breaking is intimately correlated with the mixing of occupied ($i$) and empty ($k$) states, i.e. with particle number violation, and that a rotation can be viewed as large amplitude vibration in which the restoring force goes to zero while inertia remains finite, particle number (gauge) symmetry restoration implies 
$W_1(\beta=+2)=W_1(\beta=-2)=0$.
Making use of this prescription one obtains $\varepsilon_F=-5.8$ MeV and 
$\varepsilon_F= -4.0$ MeV, for ${}^{208}$Pb and ${}^{10}$Be respectively (minimum of the dispersion relation, see Fig. 5 of ref. \cite{Bes:66} and Fig \ref{fig22}).

In a mean field approximation the energy of the last occupied single-particle level measured with respect to the continuum gives the nucleon separation energy. That is, 
the energy needed for promoting a  bound nucleon into the lowest state of the continuum, the difference between the binding energy of a given nucleus and of its nearest neighbor. 
In other words, the energy gained adding (or removing) a nucleon.

In the case of ${}^{208}$Pb the mean field definition of the neutron separation energy $S_n$ is the energy needed to take a neutron moving in the last occupied $p_{1/2}$ orbit and bring it to become barely unbound. In other words, it is the energy of the $p_{1/2}$ state with respect to the continuum, i.e. $S_n({}^{207}$Pb$)=6.737$ MeV. Now the difference of binding energy between ${}^{208}$Pb and ${}^{207}$Pb measures the energy that is gained by adding a neutron to ${}^{207}$Pb, that is $BE({}^{208}$Pb$)-BE({}^{207}$Pb$)=7.27$ MeV.
In the case of ${}^{10}$Be the mean field approximation neutron separation energy is $S_n({}^{9}$Be$)=1.665$ MeV while $BE({}^{10}$Be$)-BE({}^{9}$Be$)=6.81$ MeV.
These results  testify to the fact that a pure mean field description of nuclei at large, and of  $^{10}$Be and $^{208}$Pb 
in particular,  is insufficient to account for all the effects defining the ground state of the system and that a more complete description is needed.


\section{The Copenhagen-Buenos Aires-Milan Nuclear Field Theory program}

Let us elaborate on this need expressed making use of the $N=6$ shell closure,
associated with the so called parity-inversion phenomenon,
of which isotones $^9$Li and $^{10}$Be constitute paradigmatic embodiments. Within the framework of NFT, parity-inversion is associated with the inordinately large state dependent dressing of the $1p_{1/2}$ and $2s_{1/2}$ Saxon-Woods orbitals (to which a starting $k$-mass $\approx 0.7 m$ has been adscribed, see e.g. \cite{Barranco:01,Gori:04} in keeping with the low density of the fermion system under consideration) describing the motion of a neutron around the core $^9$Li ($^{10}$Li resonant states) and $^{10}$Be ($^{11}$Be bound states) (see diagrams (a) and (b) inset of Fig. 10). In other words, the self-energy corrections described by these diagrams eventually propagated to all orders of perturbation (and which can be parametrized in terms of the $m_\omega$ effective mass) lead, in the case of $^{10}$Li and $^{11}$Be, to energy shifts larger than ever seen in systems lying along the stability valley with one nucleon on top of a closed shell system (see e.g. $^{209}$Bi and $^{209}$Pb \cite{Bortignon:77,Bes:71,Bes:71b,Bes:71c} and refs. therein).
In keeping with this fact it is then only natural that the associated spectroscopic amplitudes ($Z_\omega=(m_\omega/m)^{-1}$) are quite far from the single-particle values, and that the interaction between two nucleons surrounded by a cloud of e.g. dipole pigmy resonances, leads to a Cooper pair which is so extended  as to make the $^{11}$Li radius 
about as large as 2/3 of the  $^{208}$Pb radius. Now, this (halo) Cooper pair is the pair addition pairing vibrational mode of $^9$Li. Because this "object" is so fragile, and at variance with for example $\vert {}^{210}$Pb$(gs) \rangle$, a spectator, e.g. the pair removal mode of $^9$Li, i.e. $\vert gs(^{7}$Li$)\rangle$, or the presence of a two-phonon p-h state (see Fig. 9(c)), can strongly perturb it. This is again at variance with what one observes in connection with the two-phonon, 4.9 MeV pairing vibrational mode of $^{208}$Pb (see \cite{Broglia:73,Bes:66} and refs. therein).

Summing up, the strongly renormalized $^{10}$Li and $^{11}$Be single-particle states, i.e. dressed nucleons displaying effective $\omega-$masses and spectroscopic factors $Z_\omega=(m_\omega/m)^{-1}$ very different from those of the original single-particle levels associated with the ``bare'' potential, interact among them, when moving in time reversal states, also very differently than bare nucleons do.
 Halos are just but a manifestation of this extreme interweaving of single-particle and collective motion which is at the basis of the structure of nuclei and which, in the present case, actually changes the very nature of the original Mayer and Jensen model \cite{Mayer:55}, starting by the magic numbers, let alone the occupation numbers.
This is because, as observed by Bohr and Mottelson \cite{Mottelson:62,Mottelson:76,Bohr:75,Bohr:76} the distribution of single-particle levels of the mean field which determines, in first approximation, the motion of nucleons, emerge from exactly the same nuclear properties which determine the collective motion of the nucleus as a whole. In fact, in a sense, the most collective nuclear motion is associated with the fact that a nucleon at the Fermi energy moves essentially independent of the rest of the other nucleons (long mean free path as compared to the nuclear radius), feeling their pushing or pulling only when trying to leave the nucleus, and being forced to bounce elastically off the nuclear surface.

A consequence of this fact, among other things, is that the distribution of valence orbitals around the Fermi energy determines the particle-hole excitations which are at the basis of e.g. collective surface vibrations. Thus, the fact that low-lying modes are, as a rule, of quadrupole and octupole nature, as well as the associated spontaneous symmetry breaking phenomena (quadrupole and octupole rotational bands). The associated zero point motion can modify most of the nuclear properties, like the mean square radius
(see e.g. \cite{Esbensen:83,Barranco:87a} and refs. therein), as well as the position and occupation of single-particle levels. In extreme cases, like the ones discussed above, even the magic numbers. Of notice that this is a dynamic Jahn-Teller type of effect, 
the one most commonly observed being  the effect  associated with the (static) phenomenon of spontaneous symmetry breaking in which the restoring force of, for example, a quadrupole vibration vanishes, its inertia remaining finite. The fingerprints of such phenomena are important rearrangements of single-particle levels (Nilsson potential \cite{Nilsson:75}), 
 and the presence of rotational bands with enhanced quadrupole transitions.

These effects can be so extreme in heavy nuclei that they lead e.g. to the phenomenon of exotic decay, the first observed case being that of $^{223}$Ra$\rightarrow{ }^{209}$Pb$+{}^{14}$C (see e.g. \cite{Brink:05} Ch. 7 and refs. therein), as extreme as the fact that $^{11}$Li is only marginally bound even if it is associated with a Mayer and Jensen magic number ($N=8$), magic number which because of the strong polarizability of the system (quadrupole vibration and pigmy resonance), shifts dynamically levels around leading to parity inversion and transforming $^{11}_{3}$Li$_{8}$ from a single-closed shell system, into the exotic, halo, single Cooper pair weakly bound systems by the exchange of vibrations as recently experimentally demonstrated (see e.g. \cite{Potel:10,Tanihata:08,Ball:11}), pair addition mode of the $N=6$ neutron closed shell.

Summing up,
the Mayer and Jensen (independent particle) shell model \cite{Mayer:55}, and the liquid drop model \cite{Bohr:36} are just two extreme aspects of the dynamical shell model
(see e.g.  \cite{Mahaux:85} and refs. therein)  in which the observed nuclear properties emerge from the interweaving (dynamical and static) of single-particle and collective degrees of freedom.

This scenario is at the basis of the Bohr-Mottelson model. Within this context, the systematic use of NFT to work out the variety of couplings and processes dressing nucleons and collective modes (Feynman-Goldstone-like diagrams as well as B\`{e}s-BRST) provides the theoretical framework to carry, arguably, the Copenhagen-Buenos Aires-Milan 
program (for both structure \cite{Bes:76a,Bes:76b,Bes:76c,Bes:75,Broglia:76,Mottelson:76,Bortignon:77,Bohr:76,Bes:90} and reactions \cite{Broglia:05c})
to completion, renormalizing not only the elementary modes of excitation, but also their interactions, effective operators and related scattering processes and probes.

The extreme consequences of such renormalizations observed in e.g. the nuclei $^{10}$Li, $^{11}$Li, $^{11}$Be, $^{12}$Be, 
force  one to recognize that such a program is not an option, but that  it is a prerequisite to be able to compare quantitatively theory with experiment.

\section{Nuclear Reactions}
\subsubsection{Elastic scattering}
Consider an incident beam of particles of mass $m$ moving along the z-axis with a velocity $v$ and impinging on a scattering center. Such a system can be described in terms or a plane wave
\begin{equation}
 \Psi_{inc} = \Psi_\alpha e^{ikz},
\end{equation}
where $k=mv/\hbar$, $v$ being the velocity corresponding to the incident projectile energy, while $\Psi_\alpha$ describes the intrinsic structure of the incoming particles along the $z$-axis. the associated incident current can be determined from the relation
\begin{equation}
 \vec{I} = \frac{\hbar}{2im} (\Psi^{*} \vec{\nabla} \Psi - (\vec{\nabla} \Psi^{*})\Psi ) = \Im m\{ \Psi^{*} \vec{\nabla} \Psi \},
\end{equation}
leading to
\begin{equation}
 \vec{I}_{inc} = | \Psi_\alpha |^2 v \hat{z}.
\end{equation}


The scattering center interacts with the incident particles to produce an outgoing (elastic) scattering (spherical) wave, with axial symmetry with respect to the incident beam. At a large distance from the scatter this wave has the form
\begin{equation}
 \Psi_{scatt} = \Psi_\alpha f(\theta) \frac{e^{ikr}}{r}.
\end{equation}
The associated scattered current is equal to
\begin{equation}
 \vec{I}_{scatt} = | \Psi_\alpha |^2 v \frac{|f(\theta)|^2}{r^2} \hat{r}
\end{equation}
The flux outgoing particles is given by the projection of $\vec{I}_{scatt}$ on the unit area $d\vec{A}$ parallel to $\hat{r}$, that is,
\begin{equation}
 \vec{I}_{scatt} \cdot d\vec{A} = |\Psi_\alpha|^2 v \frac{dA}{r^2}.
\end{equation}
It is of notice that $\frac{ d\vec{A} }{r^2}=d\Omega \hat{r}$ where $d\Omega$ is the unit of solid angle, thus
\begin{equation}
 \vec{I}_{scatt} \cdot d\vec{A} = |\Psi_\alpha|^2 v d\Omega.
\end{equation}
The differential cross section is defined as the scattered flux divided by the incoming flux, namely,
\begin{equation}
 d\sigma = \frac{\vec{I}_{scatt} \cdot d\vec{A}}{\vec{I}_{inc} \cdot \hat{z}} = |f(\theta)|^2 d\Omega,
\end{equation}
and
\begin{equation}
 \frac{d\sigma}{d\Omega} =  |f(\theta)|^2
\end{equation}

%

\subsubsection{Inelastic and transfer reactions}
Let us consider the reaction $A(a,b)B$, where $\alpha = a + A$ and $\beta = b + B$ label the elastic and a general inelastic or transfer channel respectively.
\begin{figure}[h!]
\centerline{\includegraphics*[width=.33\textwidth,angle=0]{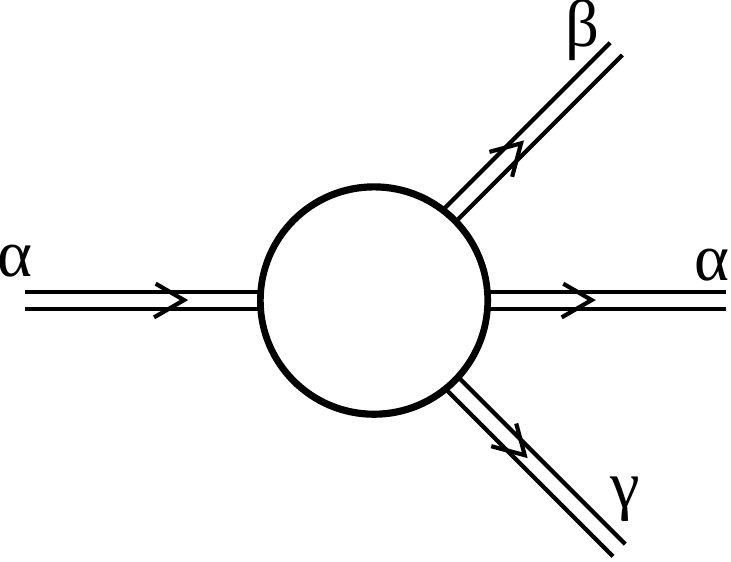}}
\end{figure}
The wavefunction describing the above process can be written as
\begin{equation}
 \Psi = e^{i \overrightarrow{k}_{\alpha} \cdot \overrightarrow{r}_{\alpha}} \Psi_{a}(\xi_{a})\Psi_{A}(\xi_{A})+\sum_{\beta} \Psi_{scatt,\beta} 
\end{equation}
where $\Psi_{a}(\xi_{a})$ and $\Psi_{A}(\xi_{A})$ are the intrinsic wavefunctions in channel $\alpha$ ( $\int \vert \Psi_{a}(\xi_{a}) \vert ^2 \textrm{d}\xi_a = \int \vert \Psi_{A}(\xi_{A}) \vert ^2 \textrm{d}\xi_A = 1$ ) while
\begin{equation}
 \Psi_{scatt,\beta} = f_{\beta}(\theta) \frac{e^{ik_{\beta}r_{\beta}}}{r_{\beta}}\Psi_{b}(\xi_{b})\Psi_{B}(\xi_{B}),
\end{equation}
with $\int \vert \Psi_{b}(\xi_{b}) \vert ^2 \textrm{d}\xi_b = \int \vert \Psi_{B}(\xi_{B}) \vert ^2 \textrm{d}\xi_B = 1$.

In analogy with the elastic process, one can define the reaction differential cross section $\textrm{d}\sigma_{\alpha \rightarrow \beta}$ as 
\begin{equation}
 \textrm{d}\sigma_{\alpha \rightarrow \beta} = \frac{v_{\beta}\vert f_{\alpha \rightarrow \beta}(\theta) \vert ^2}{v_{\alpha}} \textrm{d} \Omega,
\end{equation}
that is,
\begin{equation}
 \frac{\textrm{d}\sigma_{\alpha \rightarrow \beta}}{\textrm{d} \Omega} = \frac{v_{\beta}}{v_{\alpha}}\vert f_{\alpha \rightarrow \beta}(\theta) \vert ^2.
\end{equation}

For elastic scattering we have $v_{\alpha}=v_{\beta}$. When the scattering is non-elastic, the factor ($v_{\alpha}/v_{\beta}$) enters because the cross section refers to particle flux whereas the wave amplitude $f$ describe particle density.

\subsubsection{Detailed Balance}
\begin{figure}
\begin{center}
\includegraphics[width=0.8\textwidth]{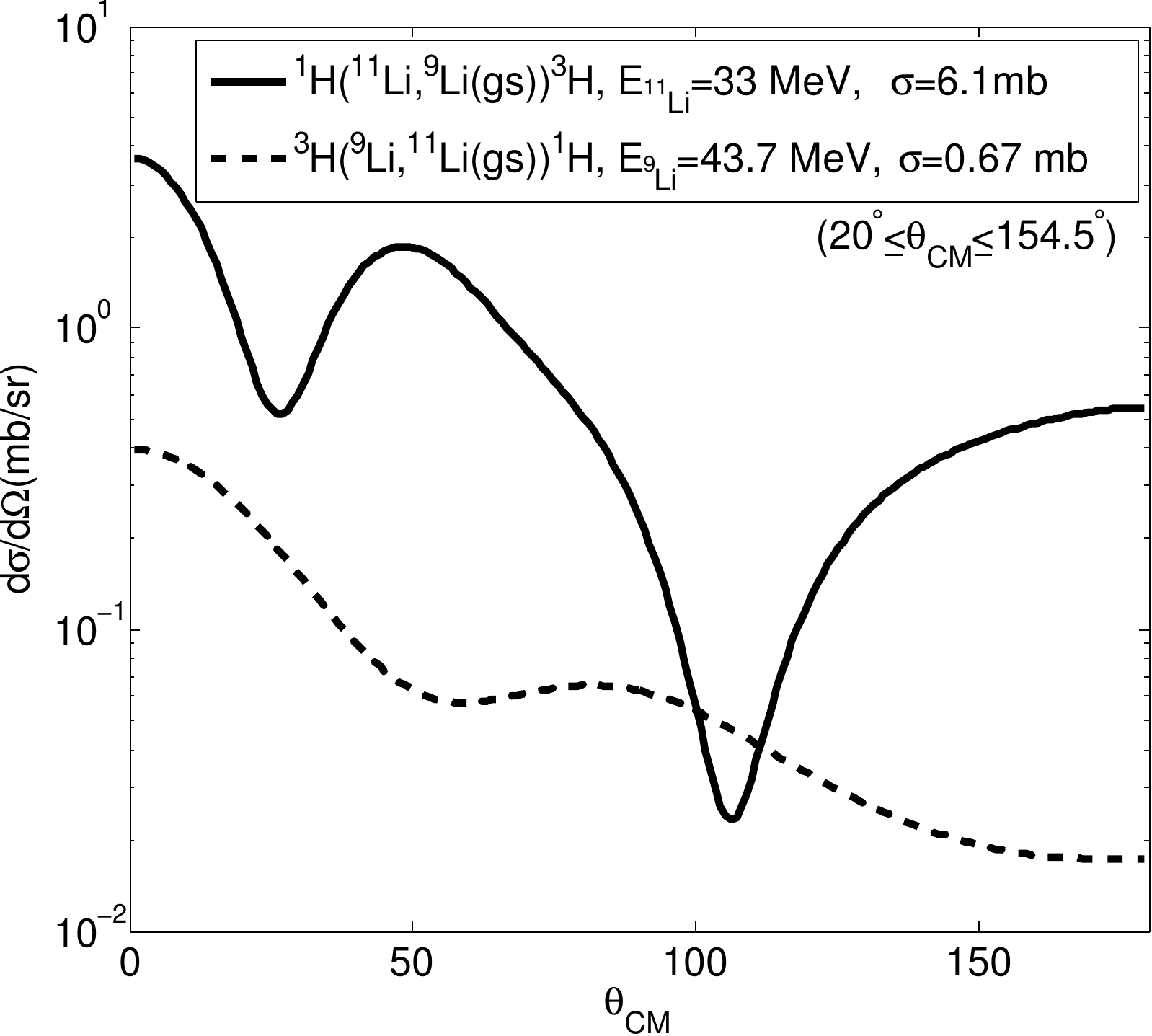}
\end{center}
\caption{Two nucleon transfer reaction on Lithium with inverse kinematics: (continuous curve), absolute differential cross section for the reaction $^1$H$(^{11}$Li$,^{9}$Li$(gs))^3$H at a bombarding energy of $E_{^{11}\textrm{Li}}=33\textrm{ MeV}$ ($Q=8.18$ MeV) corresponding to a center of mass energy of $E_{CM} = 2.75$ MeV; (dashed curve) $^3$H$(^{9}$Li$,^{11}$Li$(gs))^1$H at a $^{9}$Li bombarding energy of $E_{^{9}\textrm{Li}}=43.7\textrm{ MeV}$ ($E_{CM} = 10.93$ MeV).}
\label{kanungo}
\end{figure}

Of course
\begin{equation}
 \vert f_{\alpha \rightarrow \beta}(\theta) \vert ^2 = \vert f_{\beta \rightarrow \alpha}(\theta) \vert ^2,
\end{equation}
in keeping with detailed balance. However, the corresponding cross sections may be very different. For example in the case of the reaction at $E_{^{11}\textrm{Li}} = 33$ MeV ($E_{CM} = 2.75$ MeV) 
\begin{equation}
 ^{11}\textrm{Li}(p,t)^{9}\textrm{Li}(gs),
\end{equation}
where the total cross section is $\sigma \approx 6 \textrm{mb}$, $(v_{\beta}/v_{\alpha}) = v_{^{9}\textrm{Li}-t}/v_{^{11}\textrm{Li}-p} \approx 3$.
Consequently,  the inverse reaction $^9$Li(t,p)$^{11}$Li(gs) (for which $k_i = 0.347$ fm$^{-1}$ and $k_f=1.085$ fm$^{-1}$) at identical bombarding conditions in the center of mass system
is predicted to display a total cross section of $\approx$ 0.6 mb (see Fig. \ref{kanungo}). 

\section{Correlation length and Cooper pair transfer}

The two halo neutrons in $^{11}$Li have a separation energy of about 380  keV. Consequentely, one can consider this system as a 
textbook example to study the nuclear embodiment of a Cooper pair, namely two-fermions on top of a Fermi sphere with which it communicates solely in terms of the Pauli principle, all other possible medium polarization effect being included in the strongly renormalized interaction acting among the pair partners (within this context see \cite{Bardeen:55}). Making use of  
the fact that the observed mean square radius of $^{11}$Li is $<r^2> ^{1/2} = 3.55 \pm 0.1$ fm and that $R \approx (5/3)^{1/2} <r^2>^{1/2}$, one can argue that, making use of the  parametrization $R = r_0 A^{1/3}$, with $r_0 =1.2$ fm, for the nuclear
radius, the exotic $^{11}$Li nucleus  behaves as a normal system lying  along the stability valley of mass number $A \approx 60$, i.e. a nucleus displaying a radius of
4.6 fm, to be compared with that of $^9$Li, $R = 1.2 \times  9^{1/3}$ fm $\approx $ 2.5 fm.

\begin{figure}
\begin{center}
\includegraphics[width=0.6\textwidth]{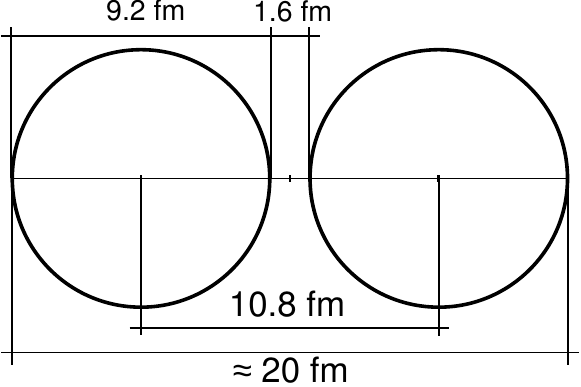}
\end{center}
\caption{Schematic representation of the distance of closest approach configuration 
of a two-particle transfer process in the reaction $^{11}$Li + $^{11}$Li at distances  of the order of $2(R+a)$, 
where $a$ is a length of the order of the $^{11}$Li nuclear diffusivity. For distances smaller than this distance, 
absorption will essentially depopulate  any particular direct channel, e.g. two-particle transfer channel. At distances  considerably larger,  
the two-particle transfer formfactor will be too small to lead to measurable transfer cross sections. Thus, one can assume that 
$2a$ is the optimal surface-surface relative distance for Cooper pair tunneling.}      
\label{schematic}
\end{figure}

While one cannot produce a chunk of matter made out of $^{11}$Li separated by about two nuclear diffusivities ($2 a \approx 1.6$ fm), and thus observe neutron supercurrents (nuclear superfluidity), as it is possible in the case of e.g. alkali doped $C_{60}$ fullerides, namely Van der  Waals solids (cf. \cite{Pickett:94,Lieber:94,Gunnarsson:04,Chancey:97,Dresselhaus:96,Broglia:04b} and refs. therein), 
it is in principle conceivable to study the Cooper pair tunneling  in which the partner neutrons are  correlated over a distance of 
20 fm. This is, in principle, to be done with the help of the    $^{11}$Li + $^{11}$Li reaction.
From the associated differential cross section of pair tunneling, and thus the building blocks of (neutron) superflow in a \textit{gedanken} non-overlapping $^{11}$Li solid (of unit cell as shown in Fig. \ref{schematic}), namely (hopping) pair tunneling, one may estimate how large the contribution to it arises 
from situations in which tunneling  takes place when the halo neutrons are separated by a  full correlation length  ($\xi \approx $ 20 fm),
as schematically shown in Fig. \ref{schematic}.


\end{document}